\documentstyle[12pt,a4wide,epsfig,psfig,twoside,openbib,cite]{article}

\begin{document}
\renewcommand{\topfraction}{1}
\renewcommand{\bottomfraction}{1}
\renewcommand{\textfraction}{0}
\thispagestyle{empty}
\section*{}
\hfill \parbox[t]{3cm}{WUB 99-02}\\[10mm]
\noindent
{\Large Flavor singlet phenomena in lattice QCD} 
\\[5pt]

\noindent
{\large Stephan G\"usken}

\noindent
{\small\em Physics Department, University of Wuppertal, D-42097 Wuppertal,
Germany}
\vskip 2.0cm
\begin{center}
{\bf Abstract}
\end{center}
Flavor singlet combinations of quark operators 
${\cal{O}}_S^{\Gamma} = \bar{u}\Gamma u  + \bar{d}\Gamma d + \bar{s}\Gamma s$ 
contribute to many important physical observables in the low energy
region of QCD. Experimentally one finds the values
of some of these observables to be in sharp contrast to the naive
(perturbative) theoretical expectations. This indicates that non
perturbative vacuum properties might play a major role in 
the comprehension of these phenomena. An example of such a vacuum 
contribution is the axial anomaly, which appears in the divergence
of the flavor singlet axial current and which is connected to the
topological properties of QCD.

From a field theoretical point of view flavor singlet matrix
elements differ from non singlet amplitudes in the occurrence
of so called disconnected insertions. These are correlations
of hadron propagators with quark-antiquark loops or correlations
between quark-antiquark loops, which are mediated by vacuum fluctuations.
According to their respective flavor composition, the disconnected
insertions cancel largely in non singlet processes, but add in flavor
singlet amplitudes.

The lattice approach provides an ideal tool to study 
flavor singlet phenomena. Being a  first principle
method it should be capable to uncover non perturbative 
vacuum contributions and to yield, on the long run, reliable       
results for the size of such contributions in QCD. 
  
The present article reviews the status of flavor singlet matrix
element calculations
in lattice QCD with respect to methods, results and reliability.
Special emphasis is paid to the discussion of state of the art
calculations of the pion nucleon sigma term $\sigma_{\pi N}$, the flavor
singlet axial coupling of the proton $G_A^1$, and the $\eta'$ mass.
\newpage
\renewcommand{\contentsname}{{\noindent\small\em Contents:\\[-1cm]}}
{\footnotesize\tableofcontents}
~\\[\baselineskip]
\newpage
\noindent
\newpage

\section{Introduction\label{sec_intro}}

Quantum field theories are today by far the most successful candidates in
describing the fundamental forces of nature, and there is still no
evidence that the standard model -- a Lorentz and gauge
invariant quantum field theory with gauge group 
$SU(3)_c \times SU(2)_w \times U(1)_Y$ -- fails in the description
of electroweak and strong phenomena.

One of the most fascinating consequences  of such theories is
that the physical vacuum is not vacuous. It rather
behaves like a Lorentz invariant 
medium, which is made up from ground state fluctuations of quantum
fields. This property, which is absent in classical field
theories and also in quantum theory, implies a tremendous spectrum 
of new and partly curious physics phenomena, whose range has still
not been fully explored yet.

A particle brought into such a vacuum, interacts with it.
Excitation, polarization and anti-polarization effects occur
and give rise to quite singular phenomena such as the decay of 
particles into classically forbidden channels, e.g. the decay
$\Phi \rightarrow 3\pi$, the running of the
classically constant coupling with the distance of two
particles or the property of confinement
of strongly interacting particles\cite{ref_confinement}.

Due to its internal dynamics, the vacuum itself may
exhibit a non-trivial structure. 
Symmetries, which are present on the classical level, can
be destroyed by quantum fluctuations. 

A qualitatively well understood consequence of such a spontaneous
symmetry breaking  is the appearance of massless (Nambu-Goldstone)   
bosons in case of a broken global symmetry\cite{nambu,goldstone},
and of (massive) gauge bosons\cite{higgs,englert,kibble,bernstein}
in case of a broken (local) gauge symmetry.
The former mechanism, applied to the chiral $SU(3)$ symmetry, explains
the smallness of the masses of the pseudo  scalar meson octet members.
The latter allows to include massive $W$ and $Z$
gauge bosons into the theory of electroweak interactions. 
 
A third and much more hidden way of symmetry breaking occurs,
if the quantum fluctuations form an anomaly, i.e. if the conservation
of a classical current gets spoiled by vacuum effects. In the
standard model, this is
known to happen to the axial vector current\cite{adler_pi2g,ABJ_anomaly}.
On the one hand the implications of
such symmetry breaking can be disastrous, if the anomalous
current couples to a gauge field. In that case the renormalizability
of the theory, and thus its physics content,
gets lost\cite{ABJ_renormalization} unless the gauge properties and the particle content of the
theory are arranged  such that the anomalous contributions cancel each other.
Fortunately, the standard model is constructed in this way.
On the other hand, if the anomalous current does not couple to a gauge field,
this `dynamical' symmetry breaking supplies a new handle to
understand quite a number of physics phenomena. Besides the
successful calculation of the decay $\pi_0 \rightarrow \gamma\gamma$,
\cite{adler_pi2g}, it has been shown\cite{thooft_eta}
that the anomaly of the flavor singlet axial vector current 
is related to a non-trivial topological structure of the vacuum, which in turn
can be used to explain qualitatively the large mass of the $\eta'$
meson\cite{witten_eta,veneziano_eta}. Moreover, the unexpectedly small value
of the flavor singlet axial vector coupling of the proton $G_A^1$,
which lead to the so called proton spin crisis, is also closely
connected 
to the influence of this anomaly\cite{ga_anomaly1,ga_anomaly2,ga_anomaly3}.  

We do not attempt to discuss all the implications of the
quantum vacuum here.
We rather want to concentrate on a certain class of phenomena 
in the deeply non-perturbative regime of the quantum field theory of
strong interactions (QCD), which are expected to be determined to
a large extend by vacuum properties: The flavor singlet phenomena.

\subsection{The Subject\label{sec_intro_subj}}

\begin{figure}
\epsfxsize=12cm
\epsfbox{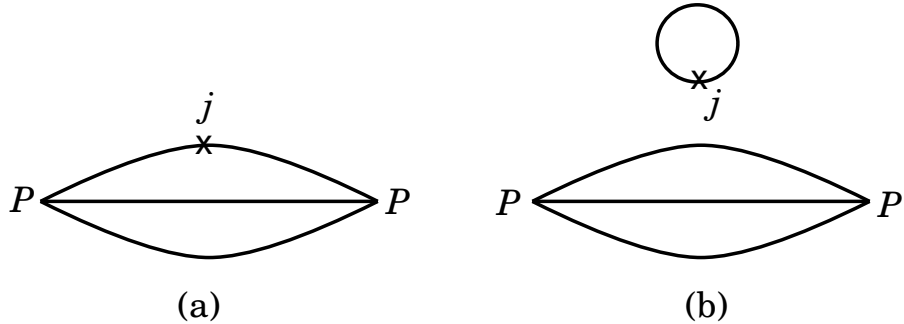}
\caption{\label{fig_schema_con_dis}{\it Connected (a) and disconnected (b)
insertions of a proton interacting with a current $j$. Please note that
all quark lines, including the quark loop, are connected by infinitely
many gluon lines and virtual quark loops. 
 }} 
\end{figure}

Suppose a (composite) particle, say a proton, interacts with a
current $j$. Then quantum field theory tells us that, 
in addition to the direct coupling, a second process contributes,
which is the interaction of the current $j$ with a vacuum
quark-antiquark loop in the field of the proton.
Both contributions, the connected (direct) and the
disconnected\footnote{Note that these `disconnected' insertions are
still connected in the field theoretical sense. There are however no
valence quark lines which connect the quark-loop and the Proton
propagator.} (vacuum) insertion, are shown schematically 
in fig.\ref{fig_schema_con_dis}. 

One would expect however, that the disconnected insertion is small
compared to the connected one if the current $j$ has a flavor non-singlet
structure. Consider for example the  flavor triplet and flavor octet
currents 
\begin{equation}
j_3 = \bar{u}\Gamma u - \bar{d}\Gamma d \; ,\; j_8 = \bar{u}\Gamma u +
\bar{d}\Gamma d - 2 \bar{s}\Gamma s \;,
\end{equation}
where $\Gamma$ denotes a combination of Dirac matrices, and the
electromagnetic current
\begin{equation} 
j_{em}^{\mu}= 2/3 \bar{u}\gamma^{\mu} u  
-1/3\bar{d}\gamma^{\mu} d - 1/3\bar{s}\gamma^{\mu} s \;.
\end{equation}
Assuming an approximate flavor symmetry of the vacuum, the
contributions from each quark flavor $u,d,s$ to the disconnected insertion
would largely cancel each other.

The situation changes drastically if we choose $j$ to be a
flavor singlet current 
\begin{equation}
j_1 =\bar{u}\Gamma u + \bar{d}\Gamma d + \bar{s}\Gamma s \;.
\end{equation}
In this case, the single quark
contributions simply add up, and there is no a priori reason why the
disconnected part should be small any more. Thus, flavor singlet
processes are promising candidates to study vacuum polarization and
vacuum structure effects. 

So far we have introduced flavor singlet phenomena solely as  
generic constructs. Of course, to be able to extract physics
information from such processes one has to know 
to which physical situations they are related, whether
they can be characterized by observables which are accessible 
to experiment and, last not least, how the diagrams
in fig.\ref{fig_schema_con_dis} can be calculated reliably in QCD.
Only then, by comparison of the theoretical
expectation with the experimental measurement, flavor singlet
phenomena can help to scrutinize the quantum field theoretical
picture of the vacuum.\\ 

The question of how and to what precision flavor singlet matrix
elements can be calculated is one of the major issues 
of this review.
As mentioned above, we will concentrate on flavor singlet phenomena
in the deeply non-perturbative regime of QCD, i.e. at momentum
transfers below $O(\Lambda_{QCD})$. The only known technique, which
allows to
solve QCD in this regime without additional assumptions is the
method of lattice QCD. Thus we will use this method here.\\

Flavor singlet processes contribute to many physical
observables, the most important ones being related to the chiral
structure of the strong interaction. 
A scalar current ($j_1(\Gamma=1)$), multiplied by a (quark) mass 
has the
structure of a mass term in the QCD Lagrange density. Indeed, it turns
out that $j_1(\Gamma=1)$, when inserted between hadron ground states, just
measures the
explicit breaking of the chiral symmetry in QCD. The quantity 
$m_q \langle N|\bar{u} u + \bar{d} d|N \rangle$ is known as
the pion-nucleon-sigma term
$\sigma_{\pi N}$. Its value can be extracted from $\pi\,N$ scattering
data\cite{sigma_experiments}, although the analysis is rather
indirect and requires much theoretical input\cite{gasser_nsigma}. 

A very interesting situation
arises if we choose $j_1$ to be a axial vector current 
($\Gamma= \gamma_{\mu}\gamma_5$). The
conservation of this current is spoiled by the axial anomaly, and one
would expect to track the effect of this anomaly in the 
disconnected insertion of fig.\ref{fig_schema_con_dis}. In fact, 
the corresponding amplitude, connected + disconnected, just
defines the axial flavor singlet coupling of the proton $G_A^1$, which
is extracted from the first moment of the spin dependent proton
structure function $g^P_1$. Naively, $G_A^1$ can be
interpreted as the fraction of the the proton spin carried by
quarks. It was first measured from polarized muon proton scattering 
by EMC \cite{EMC_exp}. The result, which has been
consolidated by several succeeding 
experiments\cite{SMC_exp,SLAC_exp1,SLAC_exp2,SMC_exp_new,SLAC_exp_new}
revealed a much smaller singlet coupling than 
expected\cite{ellis_jaffe_sr}.
One might wonder of course whether this
result can be derived directly from QCD by a calculation of the diagrams
in fig.\ref{fig_schema_con_dis}.

The flavor singlet tensor charge of a nucleon, 
$Q_T = \langle N|j_1(\Gamma=\sigma_{\mu \nu}\gamma_5)| N \rangle$
can be extracted from  the first moment of the chirality violating
structure function $h_1$ of a nucleon.
Since the chiral anomaly is absent here, one would not expect
an anomalous result as found for $G_A^1$. 
Although experimental measurements of $Q_T$
are not available yet, it is interesting to investigate 
by a QCD calculation whether this expectation is justified.

\begin{figure}
\epsfxsize=12cm
\epsfbox{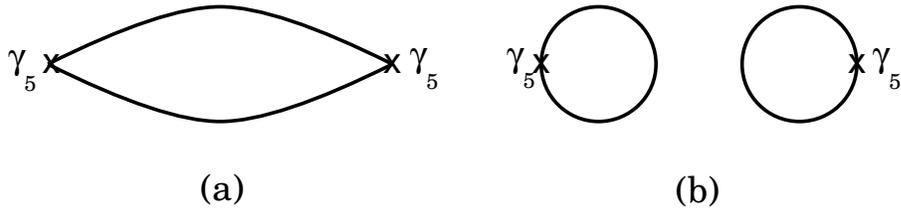}
\caption{\label{fig_schema_eta}{\it Connected (a) and disconnected (b)
contributions to the $\eta'$ propagator.}} 
\end{figure}

Closely related to the axial vector current is the pseudo scalar
current ($\Gamma=\gamma_5$), as the divergence of the former
is connected to the latter
by the axial Ward identity. Thus we expect to perceive the influence of
the axial anomaly also if we choose $j_1$ to be a pseudo scalar current.
The prominent application of this choice is the $U(1)$ problem of QCD, i.e.
large mass of the `would be ' Goldstone boson $\eta'$.
We mentioned above that a deep
qualitative understanding in terms of the axial anomaly in conjunction
with a topological structure of the vacuum has been achieved already,
but of course one would like to calculate the $\eta'$ mass directly
from QCD. The graphs which have to be evaluated for this purpose are
slightly different from those given in fig.\ref{fig_schema_con_dis}.
As can be seen from fig.\ref{fig_schema_eta}, the connected
contribution is simply the standard propagator of a pseudo scalar 
particle. The disconnected part, which contributes
in case of the flavor singlet combination, consists of the
correlation of two (pseudo scalar) vacuum loops. Clearly, one expects that the
contribution of this correlation gives rise to the large $\eta'$ mass.

To summerize, flavor singlet phenomena are closely connected with the
quantum structure of the vacuum. Thus, a strong
motivation to study these processes is to learn about this structure,
and to contribute to our knowledge about whether and how
the standard model concept of the
quantum vacuum is realized in nature.  \\

In the following chapters we will discuss the application of lattice
QCD to the determination of $\sigma_{\pi N}$, $G_A^1$,
and to the $U(1)$ problem\footnote{The
results of a pioneering lattice calculation of the tensor charge $Q_T$ can
be found in ref.\cite{aoki_qt}.}.

\subsection{The Method\label{sec_intro_meth}}

Although there has been tremendous progress in the
qualitative understanding of vacuum properties over the last 30 years,
a precise quantitative matching of experimental results with
theoretical calculations is still missing in many cases. The reason
is mainly not a  lack of experimental results, but rather lies
in the fact that most of the QCD vacuum phenomena cannot be treated within
the framework of standard perturbation theory. 
The only method known today which is capable to treat non-perturbative
phenomena from first principles, i.e. from the definition of the
theory itself, is the method of lattice gauge
theory\cite{lattice_textbooks}.
We describe here only
some important properties of lattice gauge theory which will enable
the reader to follow the line of presentation of this 
review, and to develop a feeling for the reliability of the results.

Lattice QCD was invented by K.~Wilson\cite{lattice_wilson} about
3 decades ago. 
The basic intention was to setup a gauge invariant and non
perturbative regularization scheme for QCD.
This can be achieved if one puts the path
functional\cite{path_functional}, which 
in principle contains all information of a given quantum field theory,
on a space-time grid with a lattice constant $a$ and a lattice extension
$L\times a$. The reduction from continuous to discrete space-time
instantly introduces an
ultraviolet cutoff, which removes possible singularities due to 
high momentum limits. Since the action density can be formulated
gauge invariantly on that grid, this procedure defines a gauge 
invariant regularization scheme. 

The second important ingredient, which makes quantum field theory
look like statistical mechanics and which opens the door to a
numerical treatment is the Wick rotation, i.e. the
change from Minkowski metric of the space-time grid to Euclidian
metric by a rotation of 90 degrees of the time axis in the complex plane. 
Fortunately all (time independent) properties of the theory remain 
untouched by this manipulation. The great advantage is, that the
time dependence of n-point correlation functions is changed from
an oscillatory to an exponential behavior. This enables for the numerical
evaluation of the path integral.\\

With these settings, n-point functions of (arbitrary) operators
$\sl{O}_1,\sl{O}_2,\dots\sl{O}_n$ can be expressed by
\begin{eqnarray}
\lefteqn{\langle \sl{O}_1(x_1)\sl{O}_2(x_2) \dots \sl{O}_n(x_n) \rangle
=} \label{eq_path_lattice}  \\ 
& & Z^{-1} \int d[U] d[\psi] d[\bar{\psi}]
 \left\{ \sl{O}_1(x_1)\sl{O}_2(x_2) \dots \sl{O}_n(x_n) 
e^{-S_f(U,\psi,\bar{\psi},M,g_0) - S_g(U,g_0)}\right\}\;, \nonumber
\end{eqnarray}
where
\begin{equation}
Z^{-1} = \int d[U] d[\psi] d[\bar{\psi}]
e^{-S_f - S_g}  \;. 
\end{equation}
The fermion fields $\psi$ and $\bar{\psi}$ are anti-commuting
Grassmann variables. The gluonic degrees of
freedom are represented by the gauge-links $U_{\mu}(x)$. Those are
$SU(3)$ color matrices, defined as the parallel transporters, 
which gauge-invariantly connect the fermion fields at lattice point
$x$ with the fermion fields at the neighboring point in the direction
$\mu$. $S_f(U,\psi,\bar{\psi},M,g_0)$ and $S_g(U,g_0)$
denote the fermionic and gluonic parts of the
(lattice discretized) QCD action.

Note that there are only two free
parameters in eq.(\ref{eq_path_lattice}), the bare coupling $g_0$ and
the bare quark mass\footnote{For clarity we consider here only one
(light) quark mass. Although the extension to more (non-degenerate)
quark masses is
in principle straightforward, one faces severe problems in the
implementation of the Monte Carlo process. Most (unquenched) lattice
calculations are therefore done with (two or four) degenerate quarks.}$M$.
All dimensionful quantities are given in
units of the lattice spacing $a$. We emphasize that $a$ is not
a free parameter: At a given choice of $g_0$, $M$ can be tuned such
that the lattice value of a dimensionless quantity, say the ratio
$M_{\pi}/M_{\rho}$, corresponds to its physical value. The size of $a$ is
then determined by the matching condition\footnote{Lattice values 
are indicated by capital letters, physical values by small letters.}
$M_{\rho} = m_{\rho}\,a$.\\

The first step in the evaluation of the path integral can be
done analytically. Since $S_f$ has the generic
form
\begin{equation}
 S_f = \bar{\psi} M(U)  \psi\;,
\label{eq_fermion_matrix}
\end{equation}
 the fermionic degrees of freedom
can be integrated out\cite{analytic_path}.
Eq.(\ref{eq_path_lattice}) then reads
\begin{equation}
\langle \sl{O}_1(x_1)\sl{O}_2(x_2) \dots \sl{O}_n(x_n) \rangle
= Z^{-1} \int d[U] \det(M) 
F(U,M^{-1}(U)) 
e^{- S_g} \, .
\label{eq_path_lattice_det}
\end{equation}
$F$ is a known function, which depends on the choice of
the operators $\sl{O}_i$. It can
be expressed in terms of the inverse fermionic matrix $M^{-1}$
and the gauge links $U$. For example, for the quark propagator,
$\sl{O}_1(x_1)\sl{O}_2(x_2)=\psi(x_1,a,\alpha)\bar{\psi}(x_2,a',\alpha')$,
$F$ is given by $M^{-1}_{x_1,b,\beta;x_2,b',\beta'}$, where 
$b,b'$, $\beta,\beta'$ denote color and Dirac indices.  \\

To proceed further in the evaluation of eq.(\ref{eq_path_lattice_det})
one relies on numerical importance sampling 
methods\cite{mc_quenched,mc_full}, which generate
 `gauge configurations' [U] with a probability
\begin{equation}
 P[U] \propto \det(M[U])\times exp(-S_g[U])\;.
\end{equation}
The expectation value  
$\langle \sl{O}_1(x_1)\sl{O}_2(x_2) \dots \sl{O}_n(x_n) \rangle$ then
is just the average of $F[U]$ over all gauge configurations $[U]$.
Note that all results achieved with such a Monte Carlo procedure 
pick up a statistical uncertainty, since
one is always limited to a finite sample of gauge configurations.
Of course, this uncertainty can be reduced systematically by 
increasing the size of the statistical sample.\\

The most computer time consuming part of this stochastic approach
is the generation of a statistically representative
sample of configurations. Especially the calculation
of $\det(M[U])$, which is a highly non local quantity, slows
down the Monte Carlo process considerably, and it is only in these days
that computers become powerful enough to create statistically
significant ensembles of [U] on acceptably large lattices.     
Therefore, in the past, the bulk of lattice results has been achieved
in the quenched approximation, i.e. by setting the fermion determinant
to the fixed value $\det(M)=1$. In terms of physics, `quenching'
corresponds to a neglect of fermionic vacuum loops
in the process of generating gauge configurations.

Naively one might expect that most of the quenching effects can be
accounted for by a redefinition of the coupling $g_0$. However, this
expectation could be misleading when one studies flavor
singlet phenomena. As explained above, fermionic vacuum loops play a crucial
role in this case. Allowing for such loops in the construction of 
n point functions, but neglecting them in the creation of vacuum
configurations could induce a systematic bias. The size of this
bias can only be
estimated by a comparison of quenched and unquenched results. This will
be one important topic of this review.  

We did not yet discuss 
the explicit forms of $S_f$ and $S_g$. Clearly there is a lot a
freedom in the definition of these quantities, as they have to coincide 
with the QCD action only in the continuum limit, i.e. at
$a\rightarrow 0$ and $L \rightarrow \infty$. Of course one tries to
construct the discretized action density such that as many continuum
symmetries as possible are preserved  and discretization errors 
are kept as small as possible. However, due to the so called fermion
doubling problem\cite{fermion_doubling1,fermion_doubling2,nielsen_ninomiya},
this turned out to be a highly non-trivial task.
Two different
prescriptions, one proposed by 
Wilson\cite{lattice_wilson,wilson_action_fermionic}
and one by Kogut and Susskind\cite{ks_action}, have been mainly 
used for $S_f$ over the recent years. With the Wilson action,
discretization errors occur, together with an explicit breaking of the chiral
symmetry,  at $O(a)$  in the lattice spacing. The Kogut-Susskind method
preserves the $U(1)$ chiral symmetry, but mixes internal (spin) and
external (space) degrees of freedom. In that scheme discretization
errors  contribute at $O(a^2)$.

Much effort has been invested to 
reduce the discretization errors of the Wilson action from $O(a)$ to
$O(a^2)$. The most promising candidates to achieve this goal are by now
the semi-perturbatively\cite{sw_tadpole_imp} and the
non-perturbatively\cite{sw_luescher_imp}  improved versions 
of the Sheikholeslami-Wohlert action\cite{SW_action}. 

The lattice definition of $S_g$ is much less problematic. The Wilson
prescription\cite{lattice_wilson}, which exhibits discretization
errors at $O(a^2)$ is still mostly used, but improved 
versions\cite{wilson_action_gluonic_imp} ($O(a^4)$) are also available.\\
 
In order to make contact with continuum physics, the parameters 
$g_0$ and $M$ have to be tuned such that the cutoff
$a^{-1}$ is removed while keeping the quark-mass $m$ at its
physical value. In QCD this means that, because of asymptotic freedom,
one has to take
the limits $g_0 \rightarrow 0$, $M= m\,a \rightarrow 0$. Of course, to
keep the physical size $L\,a$ of the lattice finite, one has to
increase $L \propto 1/a$ in this procedure.

In practice, one takes these limits by an extrapolation of
the results from several sets of parameters $(g_0,M,L)$ to the continuum.

The question of how close to the continuum limit these sets of parameters
{\it can} be chosen, is, after all, a question of computer power. 
In terms of statistical physics, the continuum limit is located 
at a phase transition point of second (or higher) order. Close to
this point all correlation lengths of the system start to grow
exponentially and the Monte Carlo process of generating 
decorrelated configurations slows down more
and more.

Current
resources have enabled for Monte Carlo simulations with lattice
specifications 
$a \approx 0.05\,fm\,,\, L/a \approx 3\,fm\,,\,m\approx 1/2 m_{strange}$
in the quenched case, and 
$a \approx 0.1\,fm\,,\, L/a \approx 2\,fm\,,\,m\approx m_{strange}$
for full QCD\cite{yoshie_ref}. \\

The question of how close to the continuum limit the parameter sets
{\it have to be} chosen to allow for a reliable extrapolation,
is difficult to answer in general. Although the functional
dependence of the lattice results on $M$ and $a$  is well known
close to their respective limits\footnote{Close to the continuum limit
the $M$ dependence  can then be
calculated by chiral perturbation
theory\cite{gasser_chiral,others_chiral,quenched_chiral} and the 
$a$ dependence is proportional to the order in $a$ at which discretization
errors occur.} it is not a priori clear, at which values of $M$ and
$a$ the asymptotic behavior sets in. Therefore one has to check the
onset of asymptotics carefully with the lattice data in each case.
Clearly, the reliability of the continuum extrapolation 
improves successively as more and more lattice results become available.
\newpage
\section{The Pion-Nucleon-Sigma Term \label{sec_PNST}}  

The Pion-Nucleon-Sigma term $\sigma_{\pi N}$ is defined as the
double commutator of the axial charge with the Hamilton density,
sandwiched between nucleon states at momentum zero:
\begin{equation}
\sigma_{\pi N} = \frac{1}{2M} 
\langle N | [Q^5(0),[Q^5(0),{\cal H}(0)]|N \rangle\;,
\label{eq_pi_n_sigma_def}
\end{equation}
where $M$ is the nucleon mass and $\cal H$ is the QCD Hamilton density.
The axial charge is given by
\begin{equation}
Q^5(x_0) = \int A^0(x_0,\vec{x}) d^3x \;\;,\;\mbox{with}\;
A^0 =  \bar{q} \gamma_0 \frac{\lambda_3}{2}q\;\;, \mbox{and} \; q=(u,d,s)\;.
\end{equation} 
Since $\sigma_{\pi N}$ contains the time derivate
$\dot{Q^5} = [Q^5,{\cal H}]$ of the axial charge,
it supplies a direct measure of the explicit 
chiral symmetry breaking of QCD. If the chiral symmetry would be
conserved, the axial charge would be also conserved ($\dot{Q^5}=0$)
and $\sigma_{\pi N}$ would vanish. 

In order to exploit eq.(\ref{eq_pi_n_sigma_def}) one uses the
assumption that the chiral symmetry breaking term of the Hamilton
density is given by\cite{sb_gellmann_oakes_renner,sb_glashow_weinberg}
\begin{equation}
{\cal H}' = m_u \bar{u}u + m_d \bar{d}d + m_s \bar{s}s\;. 
\end{equation}
Neglecting, in addition, $SU(2)$ flavor symmetry breaking
effects, i.e. assuming $m_u = m_d$, eq.(\ref{eq_pi_n_sigma_def})
takes the form
\begin{equation}
\sigma_{\pi N} = \frac{1}{2M}\frac{m_u+m_d}{2}
\langle N|\bar{u}u + \bar{d}d |N \rangle \;.
\label{eq_pi_n_sigma_use}
\end{equation}
Note that the matrix element on the r.h.s. 
exhibits a ($SU(2)$) flavor singlet structure, as discussed in the
previous section.
   
The determination of the physical value of $\sigma_{\pi N}$ from
$\pi N$ scattering experiments requires a lot of theoretical input.
The isospin even $\pi N$ scattering amplitude \footnote{$\bar{D}^+$ is the
scattering amplitude with the born
term removed.} $\bar{D}^+$ is related to $\sigma_{\pi N}$
at the (unphysical)
Cheng-Dashen point\footnote{The Cheng-Dashen point is defined by
$q^2=q'^2=2m_{\pi}^2$, $t=2m_{\pi}^2$, $\nu=\nu_B=0$. $q,q'$ are the 
momenta of the incoming and outgoing pion, $t,\nu,\nu_b$ are
given by $t=(q_{\mu}-q'_{\mu})^2$,
$\nu=\frac{PQ}{M}$, $\nu_B=\frac{t-q^2-q'^2}{4M}$. $Q_\mu$ and
$P_\mu$ denote the pion and
nucleon momentum sums: 
$Q_\mu=\frac{1}{2}(q_\mu - q'_\mu)$ and 
$P_\mu=\frac{1}{2}(p_\mu - p'_\mu)$.} by \cite{brown_pardee}
\begin{equation}
\Sigma= F_\pi^2 \bar{D}^+(2m_\pi^2) \; ,\;\Sigma \rightarrow
 \sigma_{\pi N} \;\; \mbox{for}\;\; m_u,m_d \rightarrow 0 \;.
\end{equation}
Since $\sigma_{\pi N}$ is defined at pion momentum transfer $t=0$,
a number of successive  
extrapolations is necessary to extract its value from the experimental
data: One first has to extrapolate  $\bar{D}^+$ from the physical
region $t \leq 0$ to the Cheng-Dashen point\cite{hoehler}. The most
recent analysis of experimental data leads to 
$\sigma_{\pi N}(2m_\pi^2) \simeq 60$MeV\cite{gasser_nsigma}, where a
(small) correction term $\Delta_R = \Sigma - \sigma_{\pi N}(2m_\pi^2)$
due to the non vanishing quark masses has been included.

The second
step is to calculate the shift 
$\Delta_\sigma = \sigma_{\pi N}(2m_\pi^2) - \sigma_{\pi N}(0)$ of
$\sigma_{\pi N}$ when going from the Cheng-Dashen point to $t=0$. 
Gasser, Leutwyler and Sainio\cite{gasser_nsigma} have performed a 
careful analysis of this shift by means of dispersive relations and
chiral perturbation theory. They find $\Delta_\sigma \simeq 15$MeV, and,
correspondingly
\begin{equation}
\sigma_{\pi N} \simeq 45 \mbox{MeV}\;.
\label{eq_sigma_exp}
\end{equation}
A similar analysis has been done in the framework of heavy baryon
chiral perturbation theory
by the authors of ref. \cite{meissner_nsigma}. They find 
$\sigma_{\pi N} = 48\pm 10$MeV, which is consistent with
eq.\ref{eq_sigma_exp}. \\
   
Additional information about $\sigma_{\pi N}$ can be extracted 
from the baryon octet mass splittings. On the one hand, the octet
quantity 
\begin{equation}
\sigma_0 = 
\frac{m_u+m_d}{2} \langle N|\bar{u}u + \bar{d}d - 2\bar{s}s|N \rangle
\end{equation}
is related to $\sigma_{\pi N}$ by
\begin{equation}
\sigma_{\pi N} = \frac{\sigma_0}{1 - y}\;,\;
y=
\frac{2\langle N|\bar{s}s |N \rangle}{\langle N|\bar{u}u+\bar{d}d |N
  \rangle} \;.
\label{eq_ypar_def}
\end{equation}
Thus, under the assumption that the strange quark contribution to
the nucleon mass is small, $\langle N|\bar{s}s |N \rangle \simeq 0$,
one would expect $\sigma_{\pi N} \simeq \sigma_0$.

On the other hand, $\sigma_0$ is directly related to the nucleon mass
shift caused by explicit  $SU(3)$ flavor symmetry breaking. To first order
in $m_s - 1/2(m_u+m_d)$ one finds
\begin{equation}
\Delta M_N = \frac{1}{3}\left( \frac{m_u+m_d}{2} -m_s \right)
\langle N|\bar{u}u + \bar{d}d - 2\bar{s}s |N\rangle \;,
\end{equation} 
or, for the baryon octet mass splittings in general
\begin{equation}
\Delta M_B = \langle B|c_8 u_8 |B\rangle 
= \alpha tr(\bar{B}u_8B) + \beta tr(\bar{B}Bu_8)\;.
\end{equation}
$B$ on the r.h.s. denotes the baryon matrix, and the octet quantities
$c_8$ and $u_8$ are defined as
\begin{equation}
c_8= \frac{1}{\sqrt{3}}\left( \frac{m_u+m_d}{2} - m_s\right) \quad ,
\quad
u_8=\frac{1}{\sqrt{3}}(\bar{u}u + \bar{d}d - 2\bar{s}s)\,.
\end{equation}
From the analysis of the mass splittings one naively extracts
$\sigma_0 \simeq 25$MeV. It has been pointed out 
however\cite{gasser_2nd_order}, that the corrections due to terms
$\propto ( m_s - 1/2(m_u+m_d))^2$ enhance this value on a
$(20-30)\%$ level. Including these terms one finds\cite{gasser_s0}
\begin{equation}
\sigma_0 = 35\pm 5 \mbox{MeV}\;.
\end{equation}
Comparing eq.(\ref{eq_sigma_exp}) with these findings one concludes that
\begin{equation}
y \simeq  0.2 - 0.4\;.
\label{eq_y_exp}
\end{equation}

Let us try to interpret this result in terms of connected and
disconnected insertions. The term $\langle N|\bar{s}s |P \rangle$
receives a contribution only from the disconnected part of
fig. \ref{fig_schema_con_dis}, since the nucleon does not
contain a valence strange quark. Thus, the result $y \ne 0$ is direct
evidence, that disconnected diagrams really contribute to
$\sigma_{\pi N}$. 

To get an idea of what eq.(\ref{eq_y_exp}) can tell about
connected and disconnected insertions, we reformulate $y$ in terms 
of the ratio of disconnected
to connected contributions 
\begin{equation}
R_{dc} = \frac{\langle N|\bar{u}u + \bar{d}d| N\rangle_{disc}}
{\langle N|\bar{u}u + \bar{d}d| N\rangle_{con}}
\label{eq_rdc_def}
\end{equation}
and of the flavor symmetry breaking parameter $\alpha_{sb}$ 
of the disconnected insertions
\begin{equation}
\label{eq_alpha_def}
\alpha_{sb} = \frac{\hspace{-0.4cm} \langle N|\bar{s}s| N\rangle}
{\langle N|\bar{q}_lq_l| N\rangle_{disc}}\;,
\end{equation} 
where $q_l$ denotes a light quark.
This yields
\begin{equation}
y = \alpha_{sb} \left( \frac{R_{dc}}{1+R_{dc}}\right) \;.
\label{eq_y_alternative}
\end{equation}
Thus, naively assuming flavor symmetry of the disconnected (vacuum)
parts ($\alpha_{sb}=1$), one would need $R_{dc}\simeq 0.5$ to achieve 
$y \simeq 0.3$, i.e. the connected contribution to $\sigma_{\pi N}$
should dominate by a factor of 2.
On the other hand, if one would assume connected and disconnected
parts to contribute equally to $\sigma_{\pi N}$, one would need a
strong symmetry breaking, $\alpha_{sb} \simeq 0.6$, to explain the 
experimental estimate of eq.(\ref{eq_y_exp}).

Of course, these assumptions are speculative and an accurate
determination of the matrix elements in 
eq's.(\ref{eq_rdc_def},\ref{eq_alpha_def}) is highly needed.
It will be interesting to see, to what extent state of the art 
lattice calculations can contribute to this task.

\subsection{Operators and Correlation Functions\label{sub_oper}}

We are interested in a computation of matrix elements 
$\langle N|\bar{q}q |N\rangle$. The path integral approach, 
c.f. eq. (\ref{eq_path_lattice}), tells us 
how to determine correlation functions of operators. In this paragraph
we explain which correlation functions are needed and how they can be
be combined to obtain the required information.\\

Consider the correlation functions
\begin{equation}
G_{NN}(x_0) = 
\sum_{\vec{x}}\langle N(\vec{0},0)^{\dagger} N(\vec{x},x_0) \rangle
\label{eq_proton_prop}
\end{equation} 
and
\begin{equation}
G_{NqN}(x_0) = 
\sum_{\vec{x}}\sum_{\vec{y},y_0}
\langle N^{\dagger}(\vec{0},0) \bar{q}(\vec{y},y_0)q(\vec{y},y_0)
N(\vec{x},x_0) \rangle \;.
\label{eq_scalar_insertion_prop}
\end{equation}
Here, $N$ denotes the creation operator of a nucleon-like particle,
i.e. a particle with the nucleon quantum numbers, which is 
however not necessarily in its ground state. Such an operator is most
easily expressed in terms of the fermion fields\footnote{We write down
here the expression for the proton. For the neutron, $u$ and $d$ have
to be interchanged.} $u$ 
and $d$ \cite{def_proton_op}\footnote{For the time being explicit
forms of operators and correlation functions in terms of fermion
fields and fermion propagators will be valid only for the Wilson
quark action.}
\begin{equation}
N_{\gamma} = \sum_{a,b,c} (u^{a}C\gamma_5 d^b) u_{\gamma}^c 
\epsilon^{a,b,c} \;.
\label{eq_proton_op}
\end{equation}  
$C$ is the charge conjugation operator, Greek letters denote Dirac
indices, and small Latin letters are used for color indices.
The time ($x_0$) dependence of the momentum zero nucleon propagator
$G_{NN}$ is well known. It has the form\footnote{For the sake of
clarity we will always assume a lattice with infinitely large
extension in time. The corrections due to finite time extension are
well known, but they will not be discussed here.}
\begin{equation}
G_{NN}(x_0) = \sum_{n} |\langle 0| N^{\dagger} |n \rangle|^2 \;e^{-M_n x_0}
\stackrel{x_0 \rightarrow \infty}{\longrightarrow}
 |\langle 0| N^{\dagger} |N \rangle|^2 \;e^{-M_N x_0} \;.
\label{eq_proton_prop_time_dep}
\end{equation} 
Thus, although $G_{NN}$ contains all (radial) excitations
of the nucleon, the ground state is isolated in the limit of large
$x_0$.

We now turn to the time dependence of the `insertion propagator'
$G_{NqN}$. Since the term $m_q\bar{q}q$ is contained is the QCD
action, $G_{NqN}$ can be constructed by deriving $G_{NN}$ with
respect to the quark mass. Because of 
\begin{equation}
\langle N^{\dagger}(\vec{0},0) N(\vec{x},x_0) \rangle
= Z^{-1} \int d[U] d[\psi] d[\bar{\psi}]
N^{\dagger}(\vec{0},0) N(\vec{x},x_0)
e^{-S_f(m_q\bar{q}q) - S_g} \\
\end{equation}
it follows 
that

\begin{eqnarray}
\lefteqn{\langle N^{\dagger}(\vec{0},0) \sum_{\vec{y},y_0}(\bar{q}q)(\vec{y},y_0)
N(\vec{x},x_0) \rangle_c} \hspace{3.0cm} \nonumber \\
 & & = G_{NqN}(x_0) - \langle N^{\dagger}(\vec{0},0) N(\vec{x},x_0)
\rangle \langle \sum_{\vec{y},y_0}(\bar{q}q)(\vec{y},y_0) \rangle
\nonumber \\
 & & =
-\frac{\partial}{\partial m_q}
\left\{ Z^{-1} \int d[U] d[\psi] d[\bar{\psi}]
N^{\dagger}(\vec{0},0) N(\vec{x},x_0)
e^{-S_f - S_g} \right\}  \nonumber \\
 & &=  - \frac{\partial}{\partial m_q}
\langle N^{\dagger}(\vec{0},0) N(\vec{x},x_0) \rangle \;.
\label{eq_3point_from_2point}
\end{eqnarray}

Using eq. (\ref{eq_proton_prop_time_dep}) one finally gets in the
limit of large $x_0$\footnote{This summation method has been introduced
by Maiani et al.\cite{maiani_ratio}.}

\begin{equation}
R^{SUM}(x_0) \equiv \frac{G_{NqN}(x_0)}{G_{NN}(x_0)} - \langle
\sum_{\vec{y},y_0}(\bar{q}q)(\vec{y},y_0) \rangle = 
A + \frac{\partial M_N}{\partial
  m_q}x_0 \; , 
\label{eq_scalar_ratio}
\end{equation} 
with $A=-\partial/\partial m_q (\ln |\langle 0 |N^{\dagger}|N \rangle|^2)$.
Note that it 
is the connected correlation
function, in the field theoretical meaning
\begin{equation}
\langle N (\bar{q}q)N^{\dagger} \rangle_c =
\langle N (\bar{q}q)N^{\dagger} \rangle - 
\langle N N^{\dagger}\rangle \langle \bar{q}q\rangle \:,
\end{equation}
which contributes to eq. (\ref{eq_scalar_ratio}). Thus, the 
lattice value of chiral condensate $\langle \bar{q}q\rangle$
is explicitely subtracted\footnote{As the latter diverges cubicly  
with the cutoff $a^{-1}$ \cite{vladikas_chiral_cond},  this could lead
to numerical problems for
lattice simulations close to the continuum limit.}.

The next step is to determine the derivative $\partial M_N/\partial
m_q$ in terms of the matrix element $\langle N|\bar{q}q|N\rangle$.
The Feynman-Hellmann theorem\cite{feynman_hellmann} states that
\begin{equation}
\frac{\partial E}{\partial \lambda} = 
\langle \frac{\partial {\cal H}}{\partial \lambda}\rangle\;,
\label{eq_feynman_hellmann}
\end{equation}
for any parameter $\lambda$ on which the  Hamiltonian $\cal H$ depends.
Thus, since the explicitly quark mass dependent part of $\cal H$ is given by 
$m_q \bar{q}q$ one finds
\begin{equation}
\frac{\partial M_N}{\partial m_q} = \langle N|\bar{q}q|N\rangle \;.
\label{eq_MN_sigma}
\end{equation}
One procedure to determine the relevant matrix element is therefore
to calculate the correlation functions $G_{NN}$ and $G_{NqN}$ by
solving the corresponding (lattice) path integrals, and then to
extract the slope of the ratio $R^{SUM}$. We call this technique
{\it summation method}. \\

Clearly, according to 
eq. (\ref{eq_MN_sigma}), one can determine $\langle N|\bar{q}q|N\rangle$
also by a direct lattice calculation of $M_N(m_q)$ at several 
quark masses $m_q$. However, this would require to derive $M_N(m_q)$
numerically with respect to $m_q$,
which in practice leads to
a significant loss of statistical accuracy. We will come back to this
point later on. Note that this method does not discriminate between
connected and disconnected insertions, and it cannot be used to calculate
the $y$ parameter, eq. (\ref{eq_ypar_def}). Applied in the quenched 
approximation, it yields only the connected part of $\sigma_{\pi N}$.   

One could argue, that the requirement of the nucleon being in its
ground state might not be strictly fulfilled for the 3-point correlator
$G_{NqN}$. Although $x_0$ can be made large,
there are always short (time) distance contributions 
from the correlation of the scalar insertion $(\bar{q}q)(y_0)$ with the
proton, since $G_{NqN}$ contains the sum over $y_0$. To check on the
influence of such contributions we sketch here briefly a third method   
to compute  $\langle N|\bar{q}q|N\rangle$. This method, which we call
{\it plateau method}, uses the correlation function
\begin{equation}
C_{NqN}(x_0,y_0) = 
\sum_{\vec{x}}\sum_{\vec{y}}
\langle N^{\dagger}(\vec{0},0) \bar{q}(\vec{y},y_0)q(\vec{y},y_0)
N(\vec{x},x_0) \rangle \;,
\label{eq_scalar_insertion_plateau}
\end{equation}
in which we do not sum over $y_0$.
The connection to the matrix element $\langle N|\bar{q}q|N\rangle$ 
can be most easily obtained in the (Euclidian) continuum\footnote{The
corresponding calculation on the lattice would have to be performed
in the somewhat cumbersome transfer matrix
formalism\cite{lattice_textbooks}. The result is however identical if we
assume an infinitely extended lattice in time.}.
The prescriptions for shifts  of operators $\cal O$ in space and time
are given by
\begin{equation}
{\cal O}(\vec{0},x_0) = 
e^{i{\cal\vec{P}}\vec{x}}{\cal O}(\vec{x},x_0)e^{-i{\cal\vec{P}}\vec{x}}
\quad \mbox{and} \quad
{\cal O}(\vec{x},0) = 
e^{-{\cal H}x_0}{\cal O}(\vec{x},x_0)e^{{\cal H}x_0}\;.
\label{eq_shift_operators}
\end{equation}
$\cal P$ and $\cal H$ denote the momentum and Hamilton
operators. Inserting complete sets of energy eigenstates $|n\rangle$
\begin{equation}
\sum_n |n\rangle \langle n | = 1
\label{eq_completeness}
\end{equation}
to the right and to the left of $\bar{q}(\vec{y},y_0)q(\vec{y},y_0)$
in $C_{NqN}$, shifting all operators to $(\vec{0},0)$, and 
carrying out the spatial sums one finally gets for the 
(field theoretically connected) part
\begin{eqnarray}
\lefteqn{C_{NqN}(x_0,y_0) - 
G_{NN} \langle \sum_{\vec{y}}
\bar{q}q(\vec{y},y_0)\rangle}\hspace{3.0cm}
 \nonumber \\ 
& & = \sum_{n,n'} \langle 0|N^{\dagger}|n \rangle
\langle n |\bar{u}u |n'\rangle \langle n'|N|0 \rangle
e^{-E_n y_0} e^{-E_{n'}(x_0 - y_0)} \;.
\label{eq_scalar_insertion_energy}  
\end{eqnarray}
The ground state $|n\rangle=|n'\rangle=|N\rangle$ is isolated
only if both time separations, $(x_0-y_0)$ and $y_0$, become large. 

Note that eq. (\ref{eq_scalar_insertion_energy}) is independent
of $y_0$ in the large (time) distance limit. In this region, the ratio
\begin{equation}
R^{PLA}(x_0,y_0) = \frac{C_{NqN}}{G_{NN}} - \langle \sum_{\vec{y}}
\bar{q}q(\vec{y},y_0)\rangle
 = \langle N|\bar{q}q|N\rangle
\label{eq_plateau_ratio}
\end{equation}
is just the matrix element of interest. Thus, instead of looking at a 
linear slope, one has to identify a plateau region in $(x_0,y_0)$.

The {\it plateau method} is clearly superior to the {\it summation
  method} with respect to a reliable ground state identification.
However, the ratio $R^{PLA}$ might be more noisy than
 ${R}^{SUM}$, as one refrains from accumulating statistics by summing
over $y_0$. We will see below, that statistical fluctuations are indeed
a major problem in the determination of disconnected
contributions. In this case, even the {\it summation method} might
fail, since contributions from the region $y_0 \gg x_0$ occur, which
do not add to the signal but largely enhance the noise. \\
  
The  {\it plateau accumulation method} ({\it PAM}) 
tries to avoid the drawbacks of the methods described above.
It is defined by
\begin{equation}
R^{PAM}(x_0,\Delta x_0,\Delta x'_0)
 = \sum_{y_0=\Delta x_0}^{x_0-\Delta x'_0} {R}^{PLA}(x_0,y_0)\;,
\label{eq_mplateau_meth_def}
\end{equation}  
with $1 \leq \Delta x_0,\Delta x'_0 \leq x_0$. 
The asymptotic time dependence is given by
\begin{equation}
R^{PAM}(x_0,\Delta x_0,\Delta x'_0) = A +
\langle N |\bar{q}q|N \rangle (x_0 -\Delta x'_0 - \Delta x_0) \;.
\label{eq_mplateau_meth_asympt}
\end{equation}
With {\it PAM} the summation is restricted to the region in $y_0$ where
ground state behavior of the 3-point correlator can be
expected. Clearly, to ensure ground state behavior, one has to 
vary $\Delta x_0$, $\Delta x'_0$, and to compare the results.  

In the next paragraph we will briefly explain how
the correlations functions $G_{NN}$, $G_{NqN}$ and $C_{NqN}$ 
can be  determined on the lattice.
     
\subsection{What has to be computed \label{sec_what_has}} 

Assume that we would have been able to create a statistically
significant set of gauge configurations [U] by some fancy 
Monte Carlo algorithm. Then, to evaluate the path integral 
according to eq. (\ref{eq_path_lattice_det}),
one `simply' determines the function $F(U,M^{-1}(U))$ for each
of the correlation functions $G_{NN}$, $G_{NqN}$ and $C_{NqN}$,
calculates its values on each gauge configuration, and finally averages
over all configurations. The purpose of this paragraph is to construct
$F$ for the above correlations functions, and to explain
how it can be calculated numerically on each gauge configuration. \\
   
The key quantity, from which all our correlation functions can be
constructed, is the propagator $\Delta_q$ of a quark $q$, 
 created at $(\vec{x},x_0)$ and annihilated at $(\vec{y},y_0)$ 
\begin{equation}
\Delta_q(\vec{x},x_0;\vec{y},y_0) = \bar{q}(\vec{x},x_0)q(\vec{y},y_0)\;.  
\label{eq_qprop_def}
\end{equation}
Once we know the corresponding function $F_{\Delta_q}$, it is just a
matter of combinatorics to determine $F_{G_{NN}}$, $F_{G_{NqN}}$ and
$F_{C_{NqN}}$.

$F_{\Delta_q}$ can be most easily obtained in the generating
functional formalism\cite{lattice_textbooks,analytic_path}. One finds
\begin{equation}
F_{\Delta_q}(\vec{x},x_0;\vec{y},y_0) 
= M^{-1}(\vec{y},y_0;\vec{x},x_0) \;,
\end{equation} 
where $M$ is the fermion matrix, c.f. eq. (\ref{eq_fermion_matrix}).
If we insert the nucleon operator, eq. (\ref{eq_proton_op}), into
eq's. (\ref{eq_proton_prop}), (\ref{eq_scalar_insertion_prop}), and 
(\ref{eq_scalar_insertion_plateau}) 
and contract $\bar{q}$ and $q$ in all possible ways, we can 
express $F_{G_{NN}}$, $F_{G_{NqN}}$ and $F_{C_{NqN}}$ in terms of
$M^{-1}$. Note  that $\bar{q}$ and $q$
are anti-commuting Grassmann variables
\begin{equation}
\{\bar{q}(x),\bar{q}(y)\} = \{q(x),q(y)\} =
\{q(x),\bar{q}(y)\} = 0 \;. 
\label{eq_grassmann_commutators}
\end{equation}
 We do not give here the explicit expressions for all these functions
as those are quite lengthy. We merely want to discuss their
structure in order to reveal how the different parts can be computed.

The propagator of a nucleon that travels from the source
$(\vec{0},0)$ to the sink $(\vec{x},x_0)$
contains the product of three quark propagators
\begin{eqnarray}
G_{NN}^{\gamma,\gamma'}(0;x) &=& 
\langle F_{G_{NN}}^{\gamma,\gamma'}(0;x)\rangle
\label{eq_proton_prop_exp}
\\
&=& \langle (M^{-1}_u)^{\gamma,\gamma'}(x;0)
\left[ C\gamma_5 M^{-1}_d(x;0) C \gamma_5\right]^{\alpha,\alpha'}
(M^{-1}_u)^{\alpha,\alpha'}(x;0) \nonumber \\
& & +
(M^{-1}_u)^{\gamma,\alpha'}(x;0)
\left[ C\gamma_5 M^{-1}_d(x;0) C \gamma_5\right]^{\alpha,\alpha'}
(M^{-1}_u)^{\alpha,\gamma'}(x;0) \rangle \;. \nonumber 
\end{eqnarray}
Here we have suppressed the color  structure on the r.h.s..
With the source point  $(\vec{0},0)$ being fixed, the computational problem
in exploiting this equation is just 
to determine one row\footnote{In fact, as each fermion has 3 color
and 4 Dirac degrees of freedom, one has two compute 12 rows of
$M^{-1}$.} of the inverse of the fermion matrix, i.e. to
solve
\begin{equation}
M_{x;x'} F_{\Delta_q}(0;x') = \delta_{x,0} \;.
\label{eq_inversion_simple}
\end{equation}
Although $M$ is a huge matrix of size 
$(N_s^3 \times N_t \times 3 \times 4 \times 2)^2$, the solution
of this  equation is not a severe problem\footnote{This is true for
typical lattice sizes and quark masses used in current simulations.}
if one uses
fast iterative inverters \cite{CG,MR,BICG_stab}.\\  

We now turn to the calculation of $G_{NqN}$. The application of the
contraction procedure reveals two important differences to 
eq. (\ref{eq_proton_prop_exp}). First of all, there is an additional term 
$F_{G_{NN}}(0;x) \sum_y Tr(M^{-1}(y;y))$. This term can be identified
immediately with the disconnected contribution to $G_{NqN}$, since
$M^{-1}(y,y)$ describes just the quark propagation from 
$y \rightarrow y$. Thus,
\begin{equation}
G_{NqN}^{disc} = \langle F_{G_{NN}}(0;x) \sum_y Tr(M^{-1}(y;y) \rangle
\,.
\label{eq_NqN_disc}
\end{equation}
In our graphical language, $G_{NqN}^{disc}$ is represented by 
fig. \ref{fig_schema_con_dis}(b). The computational expense to
calculate this contribution is huge. We will come back to this point
below.

The second difference concerns the connected part of $G_{NqN}$.
Since we have inserted the operator $\sum_y (\bar{q}q)(y)$, one of
the nucleon valence quarks, started at the source $0$,
will always meet the `inserted'
quark at $y$, where the (scalar) interaction is located, 
and then travel to its destination at $x$. Thus, the propagator of 
this quark is modified
\begin{equation}
M^{-1}_q(x,0) \rightarrow 
\tilde{M}^{-1}_q(x,0) = \sum_y M^{-1}_q(x,y)M^{-1}_q(y,0) \,.
\label{eq_insertion_prop}  
\end{equation}
Graphically, such a modified quark propagator is represented by   
the upper quark line of fig.\ref{fig_schema_con_dis}(a).

Therefore, $G_{NqN}$ can be written in the form of
eq. (\ref{eq_proton_prop_exp}), but with one quark propagator
replaced by a modified propagator.
Choosing the notation
$G_{NN} = {\cal F}(M_u^{-1},M_d^{-1},M_u^{-1})$ one gets in the case
of an $u$-quark insertion
\begin{equation}
G_{NqN}^{con}(u) = {\cal F}(\tilde{M}_u^{-1},M_d^{-1},M_u^{-1}) +
{\cal F}(M_u^{-1},M_d^{-1},\tilde{M}_u^{-1}) \;, 
\label{eq_uins_sum}
\end{equation}
and for the $d$-quark insertion
\begin{equation}
G_{NqN}^{con}(d) = {\cal F}(M_u^{-1},\tilde{M}_d^{-1},M_u^{-1})\;.
\label{eq_dins_sum}
\end{equation}
Clearly, there is no such contribution for the $s$-quark.
 
The numerical computation of $\tilde{M}^{-1}$ can be done in a 
two step procedure \cite{maiani_ratio}. One first determines 
$M^{-1}(x,0)$ according to eq. (\ref{eq_inversion_simple}). Then,
because of
\begin{equation}
M(x,x')\tilde{M}^{-1}(x',0) = M(x,x') M^{-1}(x',y)M^{-1}(y,0) = M^{-1}(x,0)\:, 
\label{eq_insertion_identity}
\end{equation}  
one replaces the r.h.s. of eq. (\ref{eq_inversion_simple}) by
$M^{-1}(x,0)$ and solves this equation with the standard method.
Thus, the computational effort to determine $\tilde{M}^{-1}(x,0)$ 
is just twice the effort for the quark propagator.  \\
 
We didn't comment yet on the structure of $C_{NqN}$ in terms of quark
propagators. As the difference to $G_{NqN}$ is only that the sum
over $y_0$ is missing, we could write down the connected and disconnected
parts of $C_{NqN}$ exactly in the form of 
eq's. (\ref{eq_uins_sum}), (\ref{eq_dins_sum}), and (\ref{eq_NqN_disc}), albeit
omitting the sum over $y_0$. We could also use the two step
procedure outlined above for the connected part. However, in doing so 
one would have to apply the second step of this procedure for each
time slice separately, increasing the numerical effort linearly with
the number of time slices. 
In order to avoid this increase one can use alternatively the `insertion'
procedure invented by Martinelli and Sachrajda \cite{new_ins}. This
method allows for an arbitrary variation of $y_0$ without additional
inversions. 

We have seen already, that $\tilde{M}^{-1}$ can be easily obtained
from the solution of eq. (\ref{eq_inversion_simple}) if the
the source, i.e. it's r.h.s., is suitably modified. This concept can be
extended to even more involved effective propagators. Martinelli
and Sachrajda use the main part of the proton propagator as a source.
Schematically, one has       
\begin{equation}
\epsfxsize=11cm
\epsfbox{martinelli_insertion.eps}\nonumber
\label{eq_martinelli_insertion}
\end{equation}
The r.h.s. of this equation is given by a combination of quark propagators.
Note that the timeslice $x_0$ of the nucleon sink is fixed. 
$C_{NqN}$ is obtained from the product of the solution to 
eq. (\ref{eq_martinelli_insertion}), $\tilde{C}(y,0)$, and $M^{-1}(y;0)$,
for arbitrary $y_0$. The expected ground state behavior, 
eq. (\ref{eq_plateau_ratio}), can be identified as follows: One first
chooses a large value of $x_0$. Then one varies $y_0$ between $0$ and
$x_0$ to obtain the plateau region. \\
  
The disconnected parts of $G_{NqN}$
and $C_{NqN}$ have quite simple structures. All we need to know is
the nucleon propagator and the `quark loop', $\sum_{y}Tr M^{-1}(y,y)$.  
However, if one would use eq. (\ref{eq_inversion_simple}) to determine
the trace, the computational effort would be $N_s^3\times N_t$ times
the cost for the calculation of a single quark propagator, a
prohibitively large amount of computer time.

The key to reduce this amount drastically can again be found in a
suitable modification of the source term of eq. (\ref{eq_inversion_simple}).
Suppose we would insert the source $I$, with $I_y=1$, to the
r.h.s. . Then, the solution $\tilde{M}^{-1}$ to this equation is
given by
\begin{equation}
\tilde{M}^{-1} = \sum_y M^{-1}(x,y)I_y = \sum_y M^{-1}(x,y) \;. 
\label{eq_walls_M}
\end{equation} 
Summation over $x$ yields
\begin{equation}
\sum_x \sum_y M^{-1}(x,y) = \sum_y Tr M^{-1}(y,y) + 
\sum_{x,y;x \neq y} M^{-1}(x,y) \;.
\label{eq_walls_sumM}
\end{equation}
As the second term on the r.h.s. is not gauge invariant, one expects
it to cancel out in the process of averaging over all gauge
configurations. Note however, that this cancellation is mingled with
the `normal' statistical fluctuations of $\sum_y Tr M^{-1}(y;y)$. 
Since one is in practice always limited to a finite sample of gauge
configurations, a systematic bias due to a non-cancellation of
the second term could be easily overlooked. 
Furthermore, one would expect an increase of the statistical
uncertainty, since the noise of this term adds to the noise
of $\sum_y Tr M^{-1}(y,y)$.   
 
Thus, it is highly desirable to modify this 
`volume source' method\cite{japan_volume_source} such that it allows
to remove $\sum_{x,y;x \neq y} M^{-1}(x,y)$ on each single gauge
configuration. This is achieved by the `stochastic estimator'
methods\cite{bitar_gauss,liu_z2}, which we will now explain briefly. 

The major difference between the `volume source' method and the 
stochastic estimator technique is that  the volume source
vector $I$ is replaced by a random vector $\eta$, which
has the properties
\begin{equation}
\langle \eta_i\rangle = 0 \quad \mbox{and} \quad
\langle \eta_i \eta_j \rangle = \delta_{i,j} \;.
\label{eq_stoch_vec_def}
\end{equation}
Note that the brackets denote the average over the distribution
of random numbers on each gauge configuration.
For each choice $s$
of a random source $\eta^s$ one gets 
\begin{equation}
\tilde{M}^{-1}\eta^s = \sum_y M^{-1}(x,y)\eta^s_y  
\label{eq_stochest_M}
\end{equation} 
as a solution to the modified eq. (\ref{eq_inversion_simple}).
The average over $N_{E}$ random sources then yields
\begin{equation}
\frac{1}{N_{E}}\sum_{s=1}^{N_{E}}
\sum_{x,y} \eta^s_x M^{-1}(x,y)\eta^s_y 
= \frac{1}{N_{E}}\sum_{s=1}^{N_{E}}\left[
 \sum_y \eta^s_y M^{-1}(y,y) \eta^s_y + 
\sum_{x,y;x \neq y} \eta^s_x M^{-1}(x,y) \eta^s_y \right] \;.
\label{eq_stochest_sumM}
\end{equation}
For large $N_{E}$ the second term on the r.h.s. cancels according
to eq. (\ref{eq_stoch_vec_def}), and the first term reduces to the
desired result $Tr M^{-1}$ on each single gauge configuration.
The onset of asymptotics and the statistical accuracy of $Tr M^{-1}$
can be monitored within the course of sampling estimates.
We emphasize that both these 
properties depend on the choice of the random number distribution.
It has been shown in ref's.\cite{sesam_disc,liu_z2_sup} that a (complex)
$Z_2$ distribution\cite{liu_z2} is superior to a Gaussian 
distribution\cite{bitar_gauss}. Unfortunately it turns out that 
even then the number of estimates $N_E$ has to be chosen of
$O(300)$ for current lattice sizes. 
 
We mention finally that the stochastic estimator technique allows to compute
single matrix elements of $M^{-1}$. For example, using $Z_2$ 
noise\footnote{For $Z_2$ noise, the property 
of the average $\langle \eta_i \eta_i\rangle = 1$ holds for each
estimate, i.e. $\eta_i \eta_i =1$.}, the combination
\begin{equation}
\frac{1}{N_{E}}\sum_{s=1}^{N_{E}}
\sum_{y} \eta^s_{z_0} M^{-1}(x_0,y)\eta^s_y 
= \frac{1}{N_{E}}\sum_{s=1}^{N_{E}}\left[
  M^{-1}(x_0,z_0)  + 
\sum_{y \neq z_0} \eta^s_{z_0} M^{-1}(x,y) \eta^s_y \right] \;,
\label{eq_stochest_singleM}
\end{equation}
yields $M^{-1}(x_0,z_0)$ in the limit $N_E \rightarrow \infty$.

\subsection{Lattice Results \label{sec_sigma_results}}

\subsubsection{Quenched QCD}

The numerical determination of disconnected contributions by application
of the methods outlined above requires both, high statistics of
gauge configurations and, at least if one uses stochastic estimator
techniques, a multiple of compute power (O(300)) compared to standard
propagator calculations. Early, exploratory lattice 
simulations \cite{maiani_ratio,sigma_wuppertal_old,new_ins}  
were therefore restricted to the calculation of the connected part of
$\sigma_{\pi N}$ in the quenched approximation. As a result, an 
astonishingly small value, $\sigma_{\pi N}^{con} \simeq 15$MeV, was found
in these calculations, suggesting that the disconnected contribution
to the full pion-nucleon-sigma term should be approximately twice this
number.

\begin{figure}
\begin{center}
\vskip -0.9 cm
\noindent\parbox{10.0cm}{
\parbox{10.0cm}{
\epsfysize=9.0cm
\epsfbox{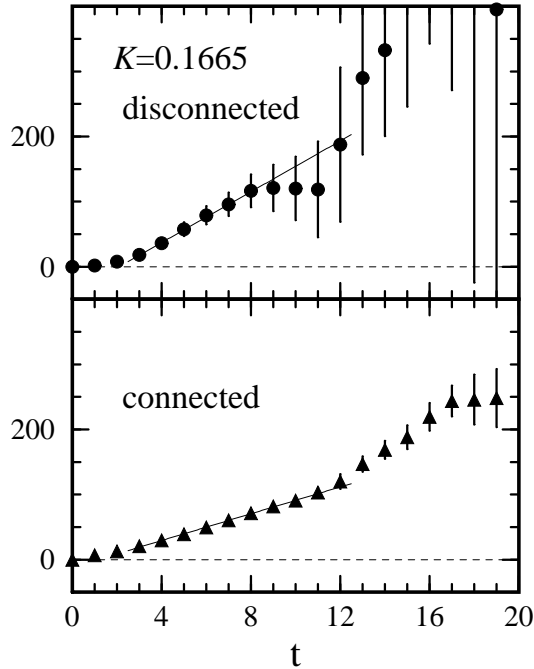}
}
}
\end{center}   
\caption{\label{fig_japan_sigma_raw}{\it Ref. \cite{japan_nsigma}:
The ratio $R^{SUM}(t)$
 for the connected and disconnected contributions (u+d insertion) 
to the scalar  density amplitude of a proton  
$\langle P|\bar{u}u+\bar{d}d|P\rangle$, at a quark mass corresponding
to $m_{\pi}/m_{\rho}=0.604$. 
 }} 
\end{figure}

Due to the advances in computer speed  and algorithms, it is nowadays
possible to investigate connected and disconnected parts in some
detail. A high statistics simulation (300-400 gauge configurations,
depending on the quark mass) has been performed by 
Fukugita et al.\cite{japan_nsigma} at a lattice cutoff of
$a^{-1}\simeq 1.5$GeV. They used a slightly modified wall
source technique for the disconnected contributions and analyzed
all signals with the summation method, eq. (\ref{eq_scalar_ratio}).

To illustrate the statistical quality of their data we show in 
fig. \ref{fig_japan_sigma_raw} the ratio $R^{SUM}(t)$ both for the
connected and the disconnected contributions to $\sigma_{\pi N}$.
Note that, although the errors to the disconnected data are large, the
expected linear rise of the signal can nevertheless be identified
in both cases.

A similar analysis, although with a significantly smaller statistical
sample (24 gauge configurations for the connected and 
50 for the disconnected part)
has been done by Dong et al.\cite{liu_nsigma} at a larger cutoff,
$a^{-1}\simeq 2$GeV. They have applied the
stochastic estimator technique, eq. (\ref{eq_stochest_sumM}), 
with (complex) $Z_2$ noise and 300 estimates per configuration.

\begin{figure}
\begin{center}
\vskip -4.0cm
\epsfxsize=8cm
\epsfbox{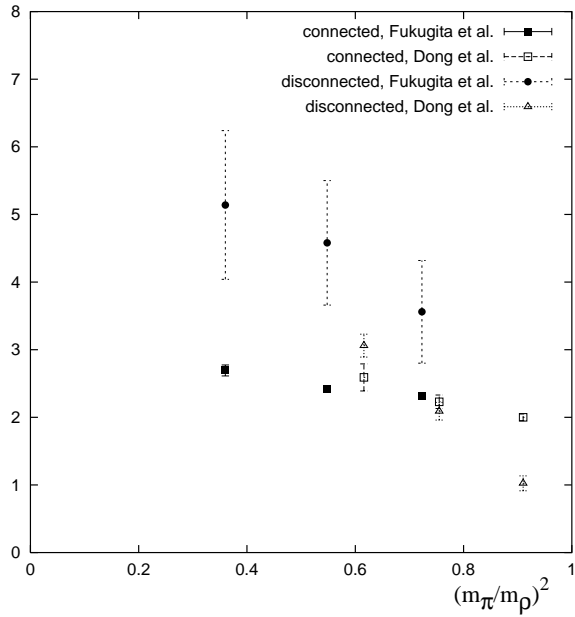}
\end{center}   
\caption{\label{fig_sigma_con_dis_quenched}{\it Connected and disconnected 
contributions to the scalar density amplitude of a proton,
$\langle P|\bar{u}u+\bar{d}d|P\rangle$. The data has been taken from
Fukugita et al.\cite{japan_nsigma} and from Dong et al.\cite{liu_nsigma}. 
To compare the results of different simulations we have renormalized
the amplitudes consistently with the tadpole improved renormalization
factor $Z_s$\cite{sw_tadpole_imp}. 
 }} 
\end{figure}

In fig.\ref{fig_sigma_con_dis_quenched} we
show the results of both simulations as a function of 
$(m_{\pi}/m_{\rho})^2$. 

The data for the connected insertion is consistent within statistical
errors, whereas the estimates of ref.\cite{liu_nsigma} for the
disconnected  contribution appear to be significantly lower than those
of \cite{japan_nsigma}. This might indicate a finite cutoff effect,
but in view of the very limited sample of gauge configurations used in  
ref.\cite{liu_nsigma} it is likely that the statistical errors
have been underestimated substantially. 
 
Obviously, the mass dependence of the connected  and the disconnected
part is quite different.
The disconnected part grows significantly
with decreasing quark mass whereas the connected part is basically
unaffected by the variation of mass.

An extrapolation (see below) to the light quark mass,
$(m_{\pi}/m_{\rho})^2_{phys}=0.033 $, consolidates the presumption
from the results of the early quenched calculations. One finds
\begin{equation}
R^{quen}_{dc} = 
\frac{\langle P|\bar{u}u + \bar{d}d|P\rangle_{disc}}
{\langle P|\bar{u}u + \bar{d}d|P\rangle_{con}}
= \left\{ 
\begin{array}{cc}
2.23(52) & \mbox{ref. \cite{japan_nsigma}} \\
1.8(1)   & \mbox{ref. \cite{liu_nsigma}} 
\end{array} \right. 
\;,
\label{eq_con_disc_ratio_quenched}
\end{equation}
and 
\begin{equation}
\sigma_{\pi N}^{quenched} 
= \left\{ 
\begin{array}{cc}
43(7) \mbox{MeV} & \mbox{ref. \cite{japan_nsigma}} \\
50(3) \mbox{MeV}  & \mbox{ref. \cite{liu_nsigma}} 
\end{array} \right. 
.
\label{eq_sigma_quenched}
\end{equation} 
Within errors the quenched findings of $\sigma_{\pi N}$ are
consistent with the estimates extracted from experiment. The
disconnected contribution appears to be much larger than the
connected one. Thus, according to eq. (\ref{eq_y_alternative}), one
needs a strong breaking of flavor symmetry for the disconnected part
to achieve a low value of $y$. \\  

Note that $\sigma_{\pi N}$ is a renormalization group
invariant quantity. Thus, the lattice result, multiplied by the
lattice scale yields directly the physical result.  

The matrix element and the quark mass individually are connected to
their respective physical values by renormalization factors 
$Z_s = 1/Z_m$, which depend on the bare coupling $g_0$ and the lattice
quark mass  
\begin{equation}
\langle P|\bar{u}u + \bar{d}d|P\rangle^{phys} =
Z_s \langle P|\bar{u}u + \bar{d}d|P\rangle^{latt} \quad , \quad
m_q^{phys} = Z_m m_q^{latt} \;.
\label{eq_phys_latt_ren}
\end{equation}
We will discuss the methods to determine these factors in the next
chapter. Here, we mention however, that the lattice result
for the light quark mass
used to obtain $\sigma_{\pi N}$ corresponds to a physical mass 
$m_l(2\mbox{GeV})_{\bar{MS}} \simeq 5\mbox{MeV}$.    \\

We did not comment yet on the ansatz used for the extrapolation of the
matrix elements to the light quark mass. Since both simulations have
only three data points at their disposal, and in view of the fact that
the statistical errors to the disconnected parts are large, one
would not expect to be able to resolve reliably contributions
from higher orders in the quark mass.
Correspondingly, ref. \cite{japan_nsigma} uses a
linear ansatz. 

Inspired by  chiral perturbation theory, 
ref.\cite{liu_nsigma} extrapolates with a non-linear, non-analytic
ansatz of the form
$C + D\sqrt{m_q}$. This, of course, leads to an enhancement of
the results at the light quark mass. However, even in high precision
calculations of the hadron spectrum \cite{cp_pacs_spectrum}, 
with $O(800)$ gauge configurations and quark masses corresponding to 
$(m_{\pi}/m_{\rho})^2 = 0.16$, one does not find numerical evidence
for the presence of such non-analytic contributions.\\

We have seen above, that the ratio of disconnected to connected
contributions to $\sigma_{\pi N}$ is astonishingly large in the
quenched approximation. Assuming an approximate flavor symmetry of
the disconnected parts, one would conclude that the $y$ parameter,
defined in eq. (\ref{eq_ypar_def}), takes the value $y=2/3$. 
This is much higher than the phenomenological expectation,
$y \simeq 0.2-0.4$, and it would indicate that the strange quark 
contribution to the nucleon mass,  $m_s \langle N|\bar{s}s|N \rangle$
is as large as $\sim 400$ MeV \cite{gasser_nsigma}.

To determine $y$ and to check the assumption of flavor symmetry,
one has to calculate the (purely disconnected) strange density
matrix element of the proton. In the
quenched approximation, this is done in two steps.

First, one adjusts the 
lattice quark mass such that the physical strange meson spectrum 
($K,K^*,\Phi$) is reproduced\footnote{The value for the strange quark
mass depends ($O(10\%)$)on the physical quantity
($m_K$ or $m_{K^*}, m_{\Phi}$)
used to define it\cite{yoshie_ref}. This however does not influence the
results discussed below.}. This determines the strange quark mass.
In the second step one extrapolates the matrix element as a function
of the valence quark mass of the proton to the light quark mass, while
keeping the loop quark fixed at the strange mass. This then yields the
lattice value of $\langle N|\bar{s}s|N \rangle$.

Using  this procedure, ref. \cite{japan_nsigma} finds 
\begin{equation}
y = 0.66(15)\;,
\label{eq_japan_y}
\end{equation}  
which is almost exactly the value one expects from the assumption of
flavor symmetry of the disconnected contributions. The result appears
plausible, since the quenched QCD vacuum is not sensitive to quark 
mass and quark flavor.

Note that $y$ depends on the ratio
$Z_s(m_s)/Z_s(m_l)$. Since the change of $Z_s$ with the lattice quark
mass is expected to be small in the light and strange quark region,
this ratio should be close to 1. Using
the tadpole improved perturbative renormalization scheme 
\cite{sw_tadpole_imp}, which has been applied in ref. \cite{japan_nsigma}, 
one finds $Z_s(m_s)/Z_s(m_l)\simeq 1.06$. \\

It has been suggested by Laga\"{e} and Liu \cite{liu_z_factors} that the
lattice quark mass dependent part of the renormalization factor 
would be quite different for connected and  disconnected amplitudes.    
Use of their result would induce a strong symmetry breaking in the
disconnected sector.   
The authors of ref. \cite{liu_nsigma} have applied this modified
renormalization procedure to their raw data.
On top of this they have used the
non-analytic ansatz for the chiral extrapolation discussed above. 
Both manipulations
lower the value of $y$, and they find $y=0.36(3)$.

We do however not agree  with the suggestion of Laga\"{e} and Liu,
at least in the case of scalar quark loops,
for the following reason: Since the combination
$m_q \langle P|\bar{q}q|P\rangle$ is a renormalization group invariant
quantity, both parts of the amplitude, connected and disconnected,
will be renormalized by the same factor $Z_S = 1/Z_m$, where $Z_m$
renormalizes the quark mass. Choosing different
factors, $Z_S^{con} \neq Z_S^{disc}$ would violate the renormalization
group invariance.
  
In summary, quenched lattice simulations find  
$\sigma_{\pi N}$ to be consistent with the phenomenological
expectation. It turns out that the ratio of disconnected to connected
contributions is close to 2 and that the flavor symmetry of the 
disconnected part in nearly unbroken. This leads to a large value
of the $y$ parameter. It is therefore of utmost interest to study
these aspects in full QCD lattice calculations. 
  
\subsubsection{Full QCD \label{sec_sigma_full}}

We have seen above, 
c.f. fig. \ref{fig_sigma_con_dis_quenched}, that quenched simulations
with a statistics of $O(300)$ configurations find statistical errors
on disconnected contributions of $\sim 30\%$.
This is bad news if one plans to determine such
contributions in full QCD, for two reasons. 
First, the creation of
gauge configurations in full QCD enhances the computational demands  
by a factor of $O(1000)$. Thus, one needs a `high end' supercomputer
to generate statistical samples with comparable size.
Secondly, a comparison of quenched and full
QCD spectrum results at equal statistics revealed 
\cite{sesam_light_spectrum}, that the
statistical fluctuations on full QCD configurations might be much
larger than in quenched QCD.
Early, exploratory full QCD simulations 
\cite{mtc_full_sigma,gupta_full_sigma} have therefore not been able
to extract possible unquenching effects on $\sigma_{\pi N}$ and $y$.\\

Recently, the SESAM collaboration has generated 4 samples of 200 
statistically independent gauge configurations with $n_f=2$ mass 
degenerate Wilson quarks, on  a $n_s^3 \times n_t = 16^3 \times  32$
lattice, and a lattice cutoff $a^{-1}_{\rho} \simeq 2.3$GeV.
The samples correspond to different sea quark masses, with
$m_{\pi}/m_{\rho} = 0.833(3),0.809(15),0.758(11)$
and 0.686(11)\cite{sesam_light_spectrum}. 

The analysis\cite{sesam_nsigma,sesam_disc} of these
configurations indeed showed that the standard
procedures, i.e. {\it summation method} combined with the wall source
or the stochastic estimator technique -- fail in producing a
reasonable signal for the disconnected contribution to $\sigma_{\pi N}$.
The situation can be significantly improved however, if one uses 
{\it PAM}, together with a stochastic estimator technique with
(complex) $Z_2$ noise. 

\begin{figure}[htb]
\begin{center}
\vskip -3.0cm
\noindent\parbox{15.0cm}{
\parbox{7.0cm}{\epsfxsize=7.0cm\epsfbox{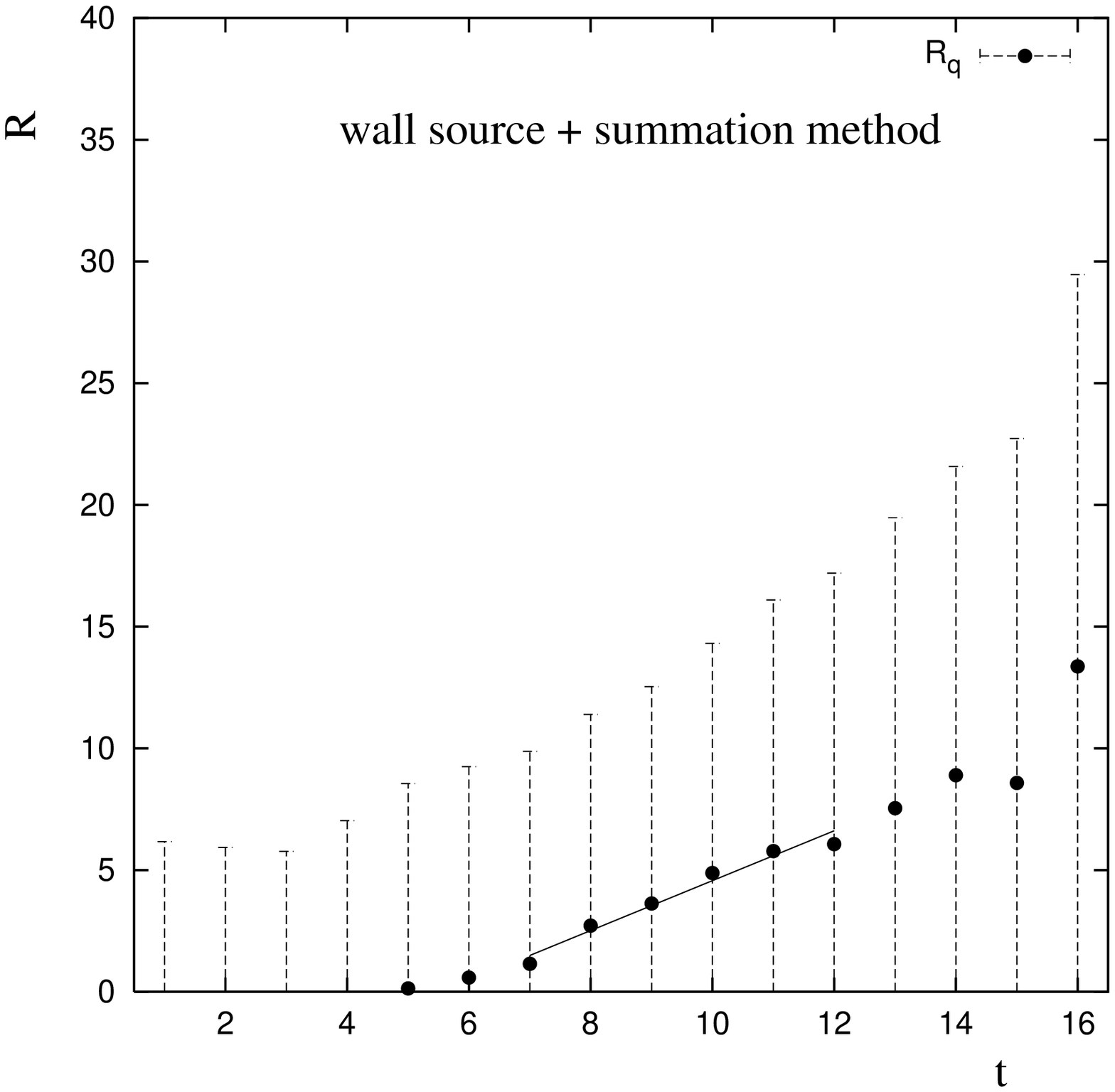}}
\parbox{7.0cm}{\epsfxsize=7.0cm\epsfbox{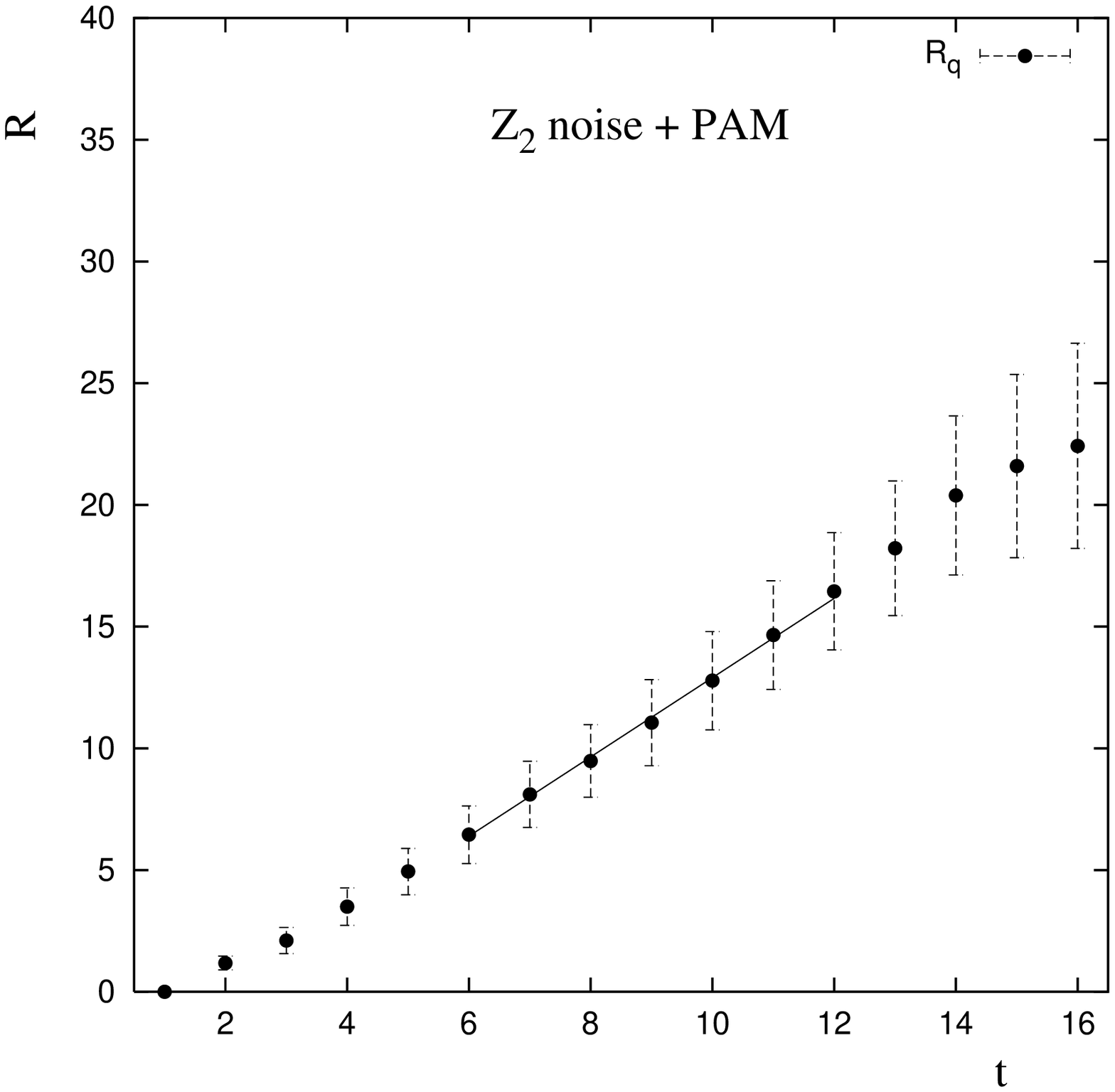}}
}
\caption{\label{fig_disc_comp_wall_best} {\it Disconnected
contribution to $\sigma_{\pi N }$. Comparison of wall
    source + summation method with $Z_{2}$ stochastic estimator +
    PAM at a sea quark mass corresponding to $m_{\pi}/m_{\rho} = 0.833$. 
The solid lines indicate the region of ground state behavior of the signal.}}
\end{center}
\end{figure}

\begin{figure}[htb]
\begin{center}
\vskip -3.0cm
\noindent\parbox{15.0cm}{
\parbox{7.0cm}{\epsfxsize=7.0cm\epsfbox{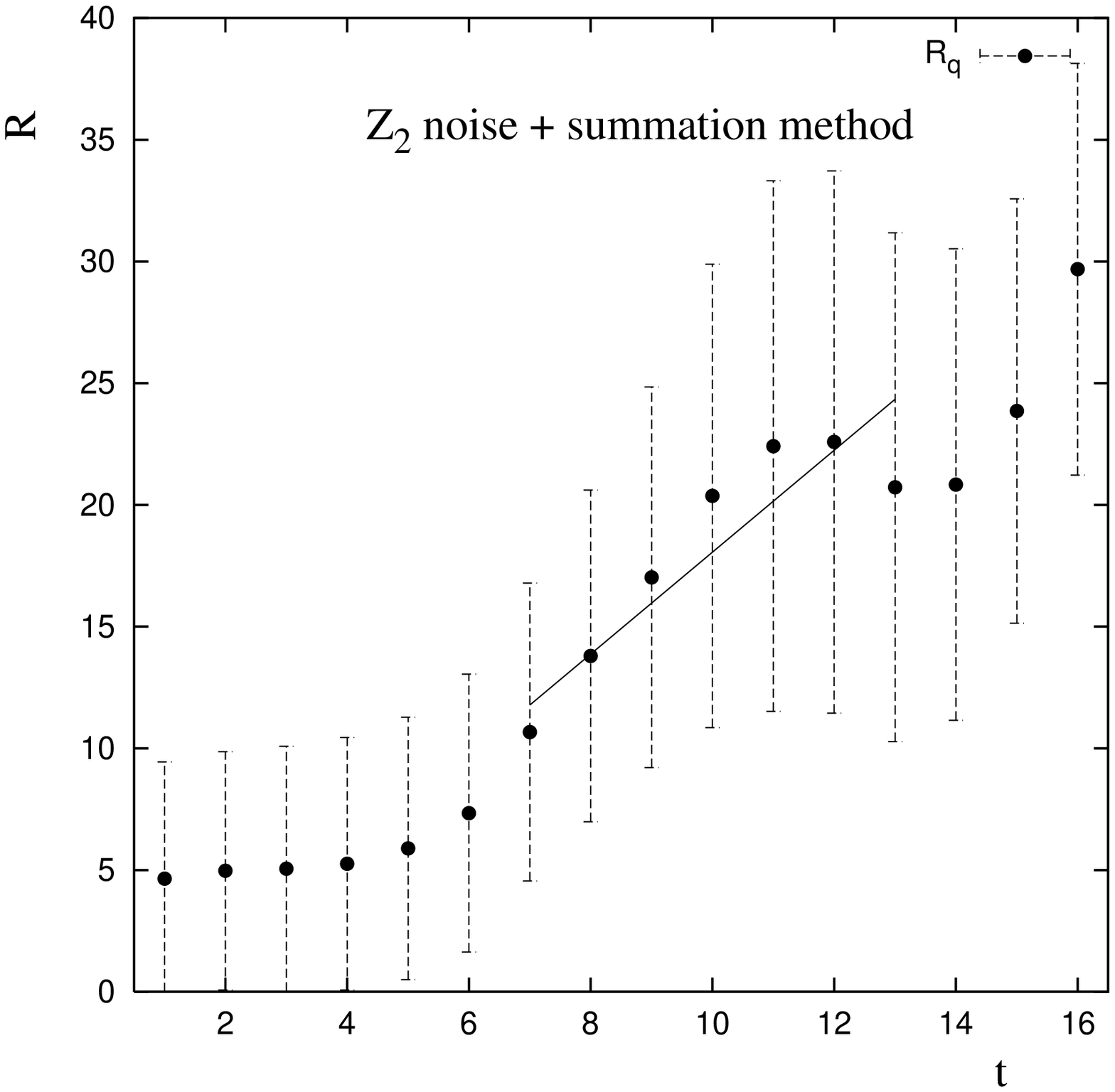}}
\parbox{7.0cm}{\epsfxsize=7.0cm\epsfbox{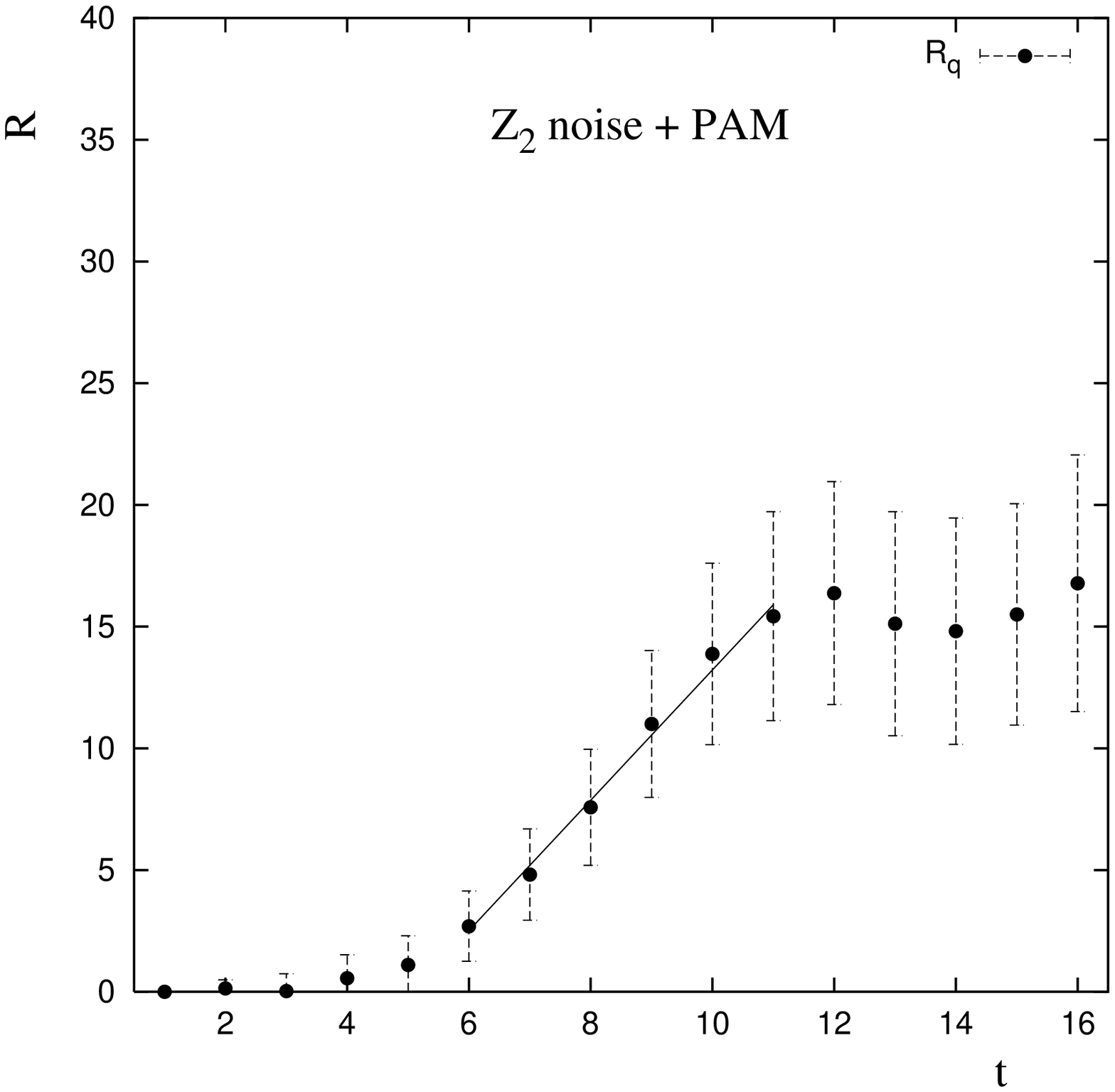}}
}
\caption{\label{fig_disc_comp_z2_best} {\it Disconnected contribution
to $\sigma_{\pi N}$. Summation method with PAM at a sea quark
mass corresponding
 to $m_{\pi}/m_{\rho} = 0.686$. In both methods, the stochastic
 estimator technique with complex $Z_{2}$ noise has been used.
 The solid lines indicate the region of ground state behavior of the signal.}}
\end{center}
\end{figure}

We illustrate this result in figs. \ref{fig_disc_comp_wall_best} and
\ref{fig_disc_comp_z2_best}
, where we compare the signals for the ratio
$R^{PAM}$, created using the {\it $Z_2$ stochastic estimator} method
with those for $R^{SUM}$, generated with the {\it wall source}
and, alternatively, with {\it $Z_2$ stochastic estimator} technique. 

Obviously, the former combination
($R^{PAM}$+  $Z_2$ stochastic estimator) is superior with respect to the
statistical quality of the signal. However, the errors are even then  
of $O(30\%)$.

\begin{figure}[htb]
\begin{center}
{\epsfxsize=10.0cm\epsfbox{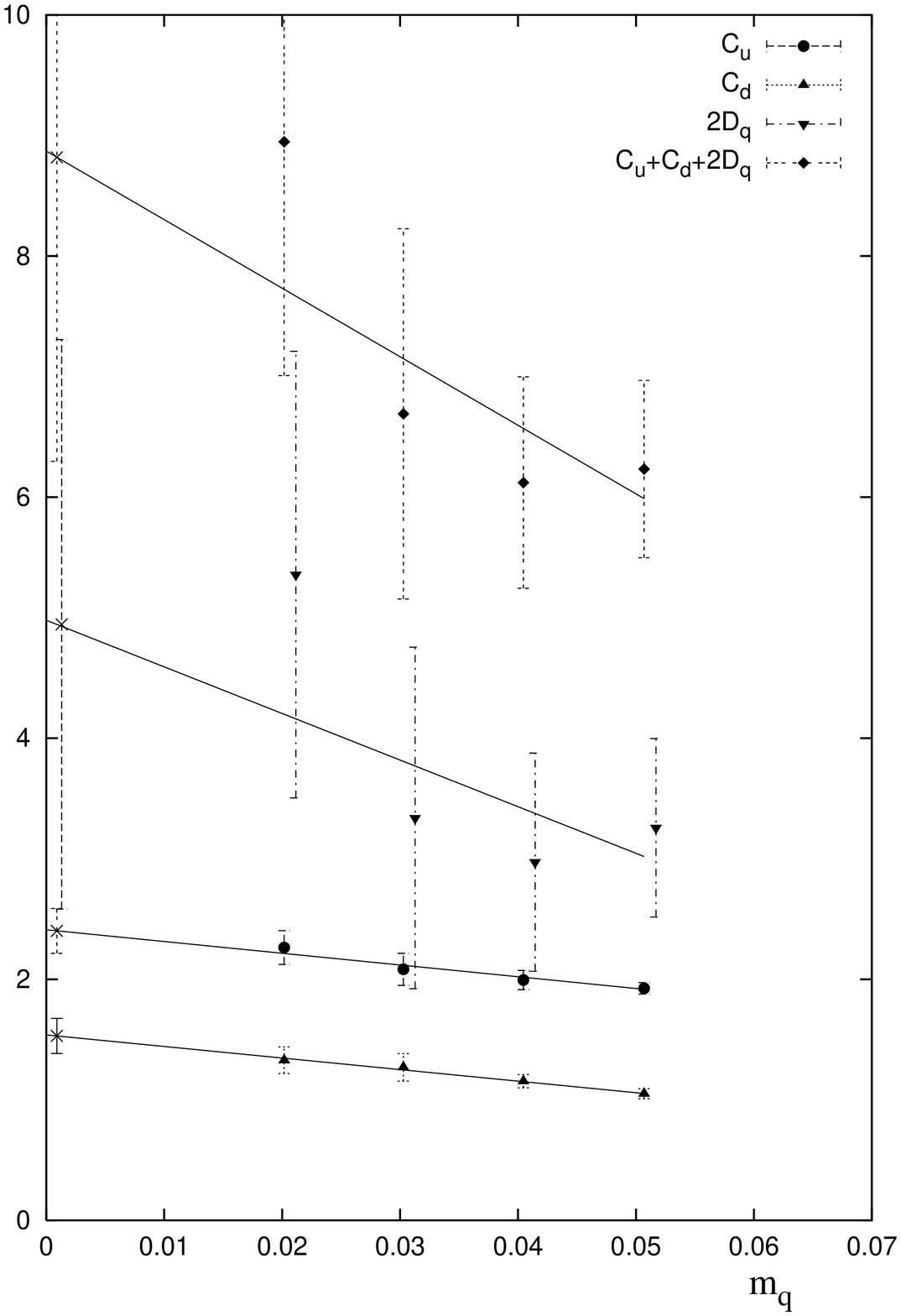}}
\caption{\label{fig_extrap_ins} {\it Full QCD: Linear extrapolation
 of connected and disconnected amplitudes to the light quark mass, from
 ref.\cite{sesam_nsigma}.  Bursts indicate the
results of the extrapolation. $C_q$ denotes the connected
, and $D_q$ the disconnected contributions to $\sigma_{\pi N}$.
}}
\end{center}
\end{figure}

Using the {\it plateau method} for the connected contributions and
{\it PAM} + stochastic $Z_2$ noise for the disconnected part, the
authors of ref. \cite{sesam_nsigma} have determined the scalar
amplitudes of the proton at the 4 SESAM quark masses. The results are
shown in fig. \ref{fig_extrap_ins}, together with the linear
extrapolations to the light quark mass. 

At first sight the full QCD results look quite similar to those of
the quenched analysis, c.f. fig. \ref{fig_sigma_con_dis_quenched}. To
see this in some more detail we compare in
tab. \ref{tab_renorm_numbers} the $n_f=2$ results at the light quark
mass with the findings of of ref. \cite{japan_nsigma}. 

\begin{table}[ht]
\begin{center}
\begin{tabular}{|c|c|c|c|c|c|}
\hline 
  &         &         &          &          &           \\
     & $Z_S(C_u+C_d)$ & $Z_S(2D_q)$    & sum    & $Z_S^{-1}m_l
           a^{-1}$[MeV] & $a^{-1}$[GeV] \\
   &        &         &          &          &           \\ \hline
 ref.\cite{japan_nsigma}   & 2.62(6)  & 5.8(1.4)  & 8.6(1.4) & 5.0(2)
           & 1.45(2)
  \\
  quenched    &       &        &         &          &  \\ \hline  
 ref.\cite{sesam_nsigma}                 & 3.11(25) & 3.9(1.8)   &
           7.0(2.0) & 2.7(2)\cite{sesam_quarkmass}  & 2.30(6)  \\
  $n_f=2$     &      &        &         &          &  \\ \hline  
\end{tabular}
\caption{\label{tab_renorm_numbers}{\it Results at the light quark
mass. For comparison all data has been renormalized by the tadpole
improved perturbative renormalization factors $Z_S$, $Z_M$ 
\cite{sw_tadpole_imp}.
}}
\end{center}
\end{table}
 
It turns out that, although the amplitudes are consistent within
errors, the relative weight of disconnected and connected amplitudes
is quite different. One finds
\begin{equation}
R^{full}_{dc} = 1.26(57) \;,
\end{equation}   
compared to $R^{quen}_{dc} = 2.23(52)$. This might give a first
indication  for the influence of sea quarks on scalar density
nucleon matrix elements\footnote{Some caution in the interpretation
of this difference is in order since the results have been obtained
at unequal lattice cutoffs.}.
As we mentioned above, the ratio $R_{dc}$ is one of the two 
quantities which fix the value of the $y$-parameter.\\

A consistency check to the sum of amplitudes can be done in full QCD
by use of the Feynman-Hellmann theorem, eq. (\ref{eq_MN_sigma}).
From a quadratic fit to the proton mass data, the authors 
of ref. \cite{sesam_nsigma} find 
\begin{equation}
Z_S \frac{\partial M_N}{\partial m_q}(m_l) = 9.2(3.9)\;.
\label{eq_fh_amp}
\end{equation} 
This is in agreement  with the result of the direct determination. 
Note, that the statistical error in eq. (\ref{eq_fh_amp})
is larger. \\

From tab. \ref{tab_renorm_numbers} one extracts a rather small value
of the pion-nucleon-sigma term
\begin{equation}
\sigma_{\pi N} = 18(5)\mbox{MeV}\;.
\end{equation}
Clearly, this is almost exclusively due to the weight of the light
quark, which is lower by nearly a factor 2 in the $n_f=2$ case
compared to the quenched result. We emphasize, that in both cases
the standard lattice definition\footnote{The standard definition
of the bare quark mass is given in terms of hopping parameters $\kappa$
by $m_q a = 1/2(1/\kappa - 1/\kappa_c)$. The critical parameter
$\kappa_c$ is determined such that the pion mass vanishes for this
choice of $\kappa$.} of quark masses\cite{gupta_lat97} has
been used. \\

It has been argued by R.~Gupta \cite{gupta_lat97} that the
perturbative renormalization would be inappropriate in full QCD,
and that non-perturbative contributions due to the presence of sea
quarks would enhance $Z_m$ by approximately a factor of 2. 
This however does not affect the result for $\sigma_{\pi N}$, as this
is a renormalization group invariant quantity.    

Another explanation could come from the lattice definition of the
quark mass. Due to finite cutoff effects, different definitions of
physical quantities can lead to different answers on the lattice.
Of course, the results have to agree in the continuum limit.

It is known from quenched simulations, that the standard
definition of the lattice quark mass, which is extracted from the
discretized Hamilton density, and the PCAC definition, which uses the
axial Ward identity \cite{gupta_lat97}, do not agree at finite
lattice cutoff. At $a^{-1}\simeq 2$GeV, the (perturbatively
renormalized) quark masses differ by about $30\%$. A scaling analysis
revealed \cite{cp_pacs_burkhalter} that the cutoff dependence of the
quark mass in the standard definition is much weaker than in the
PCAC definition, thus favoring the former. 

The situation could be different in full QCD simulations. Indeed, the
(preliminary) results of the CP-PACS
collaboration\cite{cp_pacs_burkhalter}  indicate a stronger
cutoff dependence of the quark mass in the standard definition. For a
definite conclusion on that issue however more data at different
cutoffs are needed.\\

We now turn to the determination of the $y$ parameter in full QCD.
In principle, since one needs a strange quark loop contribution in this
context,
a fully consistent QCD lattice calculation would require both, light and
strange quarks to be present in the quark sea. Such a $n_f \ge 3$ simulation 
with non-degenerate quarks will surely be one of the next large
projects in lattice QCD, but is has not been performed yet. 

Within the framework of QCD simulations with degenerate light sea
quarks, one has to rely on a {\it semi-quenched} analysis, where the
strange quark has no counterpart in the sea of quarks. This at least
accounts for the light quark mass dependence of the vacuum and is
therefore more appropriate than a pure quenched analysis of $y$.

The {\it semi-quenched} method is technically very similar to the
quenched procedure outlined above. The physical difference is that
the mass of the light valence quark is always identified with the sea
quark mass.   
 
Using this procedure, the SESAM collaboration finds\cite{sesam_nsigma}
\begin{equation}
Z_S \langle P|2 \bar{s}s| P\rangle = 4.1(1.5) \;,
\end{equation}   
which is in accordance with the result from the light quark region,
given in tab.\ref{tab_renorm_numbers}. This indicates, that the
flavor symmetry of the disconnected contributions to the scalar
proton amplitude is still intact, even in the presence of light sea 
quarks.

Thus, only $R_{dc}$ serves to reduce the value of the $y$ parameter.
Correspondingly, the result
\begin{equation}
y^{n_f=2} = 0.59(13)
\end{equation}
differs only slightly from the quenched value. 

\subsection{Discussion}
 
Let us shortly comprise the physics information that evolves from 
lattice QCD simulations discussed above. Taking the results at face
value, the following picture arises:
In quenched QCD, $\sigma_{\pi N}$ is consistent with the
`experimental' value. However, the major contribution ($2/3$)
comes from the vacuum polarization (disconnected) part of the
amplitude. As those parts turn out to be flavor invariant, 
the $y$ parameter picks up a rather large value ($2/3$), in sharp
contrast to the phenomenological expectation.

In full QCD, with two degenerate sea quarks, the connected part is
slightly enhanced whereas the disconnected amplitude  
is decreased. The sum of both contributions is still close to
the quenched result. Because the  
light quark mass is found to be quite small, the full QCD lattice
result differs from the `experimental' value by about a factor of
two. The flavor symmetry of the disconnected parts seems not to
be hampered by the presence of light sea quarks. Thus, the $y$
parameter decreases only slightly.\\
 
Let us now discuss what could be done to consolidate this picture.
First of all, the statistical errors to the disconnected contributions
are too large. Although the stochastic estimator technique with
complex $Z_2$ noise in combination with {\it PAM} 
reduces these errors considerably, one is still confronted with
uncertainties of $O(30\%)$, on gauge configuration samples of $O(200)$.   

In the case of quenched QCD it should be no major problem to reduce
these errors by a factor of 10, simply by exploiting the 
computational power of currently available parallel supercomputers.

For full QCD
calculations the situation looks less promising. One can expect
however, that the signal to noise ratio will improve on
larger lattices, due to self averaging effects.

Once one has succeeded in reducing the statistical errors to a few
percent, a scaling analysis of $\sigma_{\pi N}$ and $y$ can be
done. This will remove possible cutoff effects and lead to reliable
continuum values for $\sigma_{\pi N}$. Since the $y$ parameter depends
on (a ratio of) renormalization factors it will be advantageous to use
non-perturbative renormalization 
techniques \cite{luescher_nonpert,giusti_nonpert} instead of the
standard (tadpole improved) perturbative methods. This might weaken
the cutoff dependence of $y$ considerably.

The last step in a full QCD determination of $y$ will be to include
sea quarks of different masses into the generation of vacuum
configurations. A lattice simulation with two degenerate light and two
degenerate strange quarks should be feasible even with current
algorithmic and computational facilities, as it avoids
the sign problem of the fermion determinant \cite{gottlieb}. 
\newpage

\section{The Flavor Singlet Axial Coupling of the Proton}

About a decade ago the European Muon 
Collaboration (EMC) \cite{EMC_exp} announced
measurements of the first moment of the polarized proton structure
function $g_1^p$, which implied an unexpectedly low value for the
flavor singlet axial coupling of the proton $G_A^1 \simeq 0$, 
defined by
\begin{equation}
s_{\mu} G_A^1 = \langle p,s | A_{\mu}^0(q)|p,s\rangle_{q^2=0}\;. 
\label{eq_ga1_def}
\end{equation}
$s_{\mu}$ is the proton spin vector, and  $A_{\mu}^0$ is the
flavor singlet axial current
\begin{equation}
A_{\mu}^0 = {\cal{Z}}_A^{S} \sum_{q=u,d,s} \bar{q}\gamma_{\mu}\gamma_5 q \;.
\label{eq_a0_def}
\end{equation}
${\cal{Z}}_A^{S}$ denotes the renormalization
constant of the singlet current in the continuum.\\

In the simple parton model, $G_A^1$ can be interpreted as the fraction
of the proton spin carried by the quarks. Thus, on first sight, the
EMC result indicated that the quarks would not contribute
substantially to the proton spin: The confusion which followed from
this naive interpretation was called the `proton spin crisis'.

On top of this, the EMC result stands in contradiction to the
Ellis-Jaffe sum rule \cite{ellis_jaffe_sr}, which can be derived in
the OZI (Zweig rule) limit of QCD. This indicated already that the
proton spin problem might be deeply connected to the quantum field 
theoretical properties of QCD.\\

Since then, in a tremendous effort, both on the experimental
\cite{SMC_exp,SLAC_exp1,SLAC_exp2,SMC_exp_new,SLAC_exp_new} and on the
theoretical side \cite{review_ga}, the EMC result
has been refined and a clear
qualitative understanding of the phenomenon could be achieved. 

The latest analysis of experimental results \cite{SMC_exp_new},
including all proton, 
deuteron and neutron data, yields\footnote{This is at a
  renormalization scale of $Q_0^2=5\mbox{GeV}^2$.}
\begin{equation}
G_A^1 = 0.29\pm 0.06 \;,
\label{eq_ga1_exp}
\end{equation}    
which is significantly below the Ellis-Jaffe prediction of
$(G_A^1)^{EJ} = 0.58$.

From the theoretical investigations it appears that it is the influence
of the $U(1)$ axial anomaly, which
leads to  the small value of $G_A^1$. This immediately removes
the `proton spin crisis', as an interpretation in terms of the simple
parton model is no longer valid. We mentioned in the
introduction that the axial anomaly is closely connected to the
topological properties of the QCD vacuum. Thus, the anomalous value
of $G_A^1$ is one of the rare opportunities for a quantitative study
of topological aspects.\\ 
 
The lattice approach provides (at least) two different methods for the
calculation of $G_A^1$.

The first one, which we will call the
{\it direct method}, is to evaluate the r.h.s. of eq. (\ref{eq_ga1_def}). 
Since this is a flavor singlet matrix element, one expects
contributions from  both the connected and the disconnected parts,
depicted in fig. \ref{fig_schema_con_dis}. If it turns out, that
the smallness of $G_A^1$ originates from sizeable (negative) 
disconnected contributions, one could immediately clarify why and
to what extent the Ellis-Jaffe sum rule fails:
Disconnected diagrams are excluded in the OZI limit of QCD.
    
The second method, which we will denote as the {\it topological method},
focuses on the topological aspects  -- charge and susceptibility -- of QCD.    
Here, one calculates the correlation of the proton propagator and
the topological charge density. A success of this approach with respect to
a small value of $G_A^1$ would encourage for a detailed, quantitative
investigation of the connection between the axial anomaly, QCD
topology and instantons.

Both methods will be explained below in detail.

\subsection{Preparation}

\begin{figure}
\begin{center}
\epsfxsize=12cm
\epsfbox{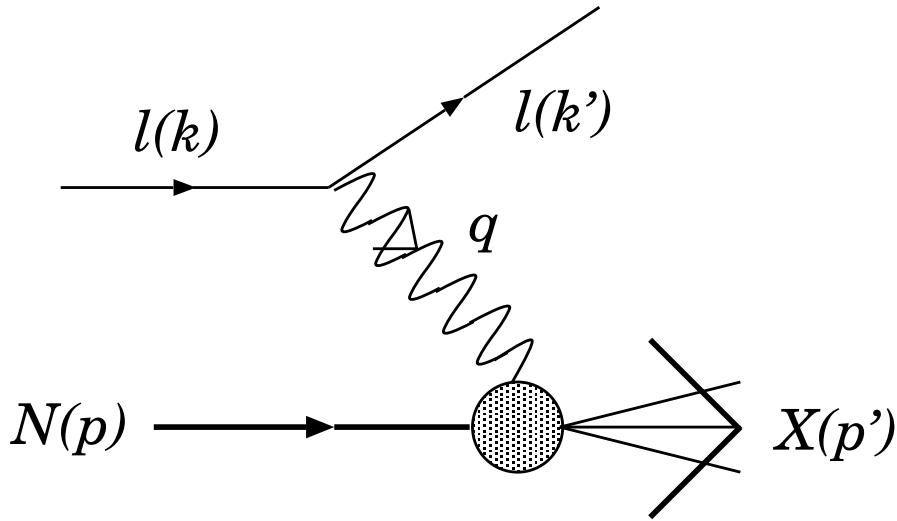}
\caption{\label{fig_dis_scatt}{\it Inclusive polarized DIS scattering.
 }} 
\end{center}
\end{figure}

Let us briefly recall how $G_A^1$ is connected to
the polarized deep inelastic scattering (DIS) of leptons
and protons, shown schematically in fig.\ref{fig_dis_scatt}.

The differential cross section for lepton proton DIS 
in the laboratory frame is given by
\begin{equation}
\frac{d^2 \sigma}{d\Omega dE'} = \frac{1}{2M}\frac{\alpha^2}{q^4}
\frac{E}{E'} L_{\mu \nu} W^{\mu \nu}\;,
\label{eq_dis_crossection}
\end{equation}  
where $M$ is the proton mass, $E$ and $E'$ are the energies of the
incoming and outgoing lepton, and $L_{\mu \nu}$ and $W_{\mu \nu}$ denote
the leptonic and hadronic tensors, respectively.
$W_{\mu \nu}$ can be parameterized by two spin independent structure
functions, $F_1$ and $F_2$, and by two spin sensitive functions, 
$g^p_1$ and $g^p_2$
\newpage
\begin{eqnarray}
W_{\mu \nu} &=& F_1 \left(-g_{\mu \nu} + \frac{q_\mu q_\nu}{q^2} \right)
            + F_2 \left( p_\mu - \frac{pq}{q^2}q_\mu\right)
                  \left( p_\nu - \frac{pq}{q^2}q_\nu\right)/(p q)
             \label{eq_dis_hadronic_tensor}\\
            &+& i \frac{M}{pq}\epsilon_{\mu\nu\rho\sigma}q^{\rho}
                \left\{ s^{\sigma}g^p_1 + 
                        \left[ s^{\sigma} -
            \frac{sq}{pq}p^{\sigma}\right]g^p_2 
           \right\} \;. \nonumber
\end{eqnarray} 
$s$ denotes the polarization vector of the proton. 

The polarized structure functions $g^p_1$ and $g^p_2$ can be extracted
experimentally from the measurement of the asymmetries
\begin{equation}
A_{\parallel} = 
\frac{d\sigma^{\uparrow\downarrow} - d\sigma^{\uparrow\uparrow}}
     {d\sigma^{\uparrow\downarrow} + d\sigma^{\uparrow\uparrow}}
     \;,\quad
A_{\perp} =
\frac{d\sigma^{\downarrow\rightarrow} - d\sigma^{\uparrow\rightarrow}}
     {d\sigma^{\downarrow\rightarrow} + d\sigma^{\uparrow\rightarrow}}
     \;, 
\label{eq_dis_asymmetries}
\end{equation}
where the arrows indicate the polarizations of the lepton (first arrow) 
and the proton (second arrow): up and down arrows stand for
longitudinal polarizations, and the right arrow denotes a proton spin
direction transverse to the lepton momentum and towards the direction
of the scattered lepton.

On the theoretical side, $g^p_1$ is given by the proton matrix element
of two electromagnetic currents carrying large space like momentum.
Its first moment
\begin{equation}
\Gamma_1^p(Q^2) \equiv \int_0^1 dx g^p_1(x,Q^2) \;,\quad
Q^2 = -q^2\;, \quad x=\frac{Q^2}{2pq}
\label{eq_1st_moment_g1}
\end{equation} 
can be derived using the operator product expansion (OPE)\cite{wilson_ope}.
Including terms up to twist two one obtains
\begin{equation}
J^\rho(q)J^\sigma(-q)  \stackrel{Q^2 \rightarrow \infty}{\sim}
2 \epsilon^{\rho \sigma \nu \mu} \frac{q_\nu}{Q^2}
\left[ 
C_1^{NS}(\alpha_s)\left( A_\mu^3 + \frac{1}{\sqrt{3}}A_\mu^8\right)
 + \frac{2}{3}C_1^{S}(\alpha_s)A_\mu^0
\right] \;,
\end{equation}  
and correspondingly
\begin{equation}
\Gamma_1^p(Q^2) = \frac{1}{12}C_1^{NS}\left( G_A^3 +
  \frac{1}{3}G_A^8\right) + \frac{1}{9}C_1^{S}G_A^1 \;.
\end{equation} 
The Wilson coefficients $C_1^{NS}$ and $C_1^{S}$ are both known to
$O(\alpha_s^3)$ \cite{ga_wilson_coeff}. The non-singlet form factors 
$G_A^3$, $G_A^8$ and the singlet coupling $G_A^1$ are given by the forward
proton matrix elements of the (renormalized) axial currents
\begin{eqnarray}
\langle p,s|A_\mu^3| p,s\rangle &=& s_\mu \frac{1}{2}G_A^3 \\
\langle p,s|A_\mu^8| p,s\rangle &=& s_\mu \frac{1}{2\sqrt{3}}G_A^8 
\label{eq_ga8}\\
\langle p,s|A_\mu^0| p,s\rangle &=& s_\mu G_A^1 \;, \label{eq_ga1}
\end{eqnarray}
with
\begin{eqnarray}
A_\mu^{3} &=& {\cal{Z}}_A^{NS}
\frac{1}{2}(\bar{u}\gamma_{\mu}\gamma_5 u -\bar{d}\gamma_{\mu}\gamma_5 d)
 \nonumber \\
A_\mu^{8} &=& {\cal{Z}}_A^{NS}
\frac{1}{2\sqrt{3}}(\bar{u}\gamma_{\mu}\gamma_5 u + 
 \bar{d}\gamma_{\mu}\gamma_5 d -
2\bar{s}\gamma_{\mu}\gamma_5 s)  \\
A_\mu^0 &=& {\cal{Z}}_A^{S}
\sum_{q=u,d,s}\bar{q}\gamma_{\mu}\gamma_5 q \;.\nonumber
\end{eqnarray}
Note that we have introduced different renormalization factors
for the non-singlet (${\cal{Z}}_A^{NS}$) and the singlet 
(${\cal{Z}}_A^{S}$) axial currents.
This is necessary, since the conservation of $A^{3,8}_\mu$ is only
violated by the quark mass, whereas the singlet current is not
conserved in the limit
of vanishing quark mass, due to the presence of the axial anomaly.
  
The non-singlet couplings can be extracted from neutron and hyperon
$\beta$ decays. One finds\footnote{$G_A^8$ has been determined from hyperon
decays under the assumption of $SU(3)$ flavor symmetry. Thus, the
uncertainty on this quantity might be underestimated.}
\cite{ga3_exp,ga8_exp} 
\begin{equation}
G_A^3 = 1.2670 \pm 0.0035\; ,\quad G_A^8 = 0.579 \pm 0.025 \;.
\label{eq_exp_ga3_ga8}
\end{equation}
Thus, the flavor singlet coupling $G_A^1$ is determined by the
measurement of $\Gamma^p_1$.\\

In the OZI limit of QCD, where (strange) quark loops do not contribute,
eq's. (\ref{eq_ga8}) and (\ref{eq_ga1}) imply $G_A^1 = G_A^8$, in sharp
contrast to the experimental result and to the fact that the $Q^2$ 
dependence of $G_A^1$ is different from that of $G_A^8$.
The QCD parton model can
qualitatively repair this failure by the assumption of a
large gluonic contribution to $g_1^p$. It turns out that singlet quark
and gluon distributions mix under
renormalization\cite{parton_review}. In a particular renormalization
scheme, the Adler-Bardeen scheme, it is possible to shift the 
anomalous $Q^2$ dependence completely to the gluonic part, leaving
intact the naive interpretation of the singlet quark distributions as
the fraction of the proton spin carried by the quarks.    
However, the QCD parton model is unable to make any quantitative
prediction for $G_A^1$.\\

In this situation, a non-perturbative method, which has
direct access to the singlet proton amplitude, c.f. eq. (\ref{eq_ga1}), 
is badly needed. A lattice computation of this amplitude can reveal
whether, and to what extend, QZI violating contributions determine
the value of $G_A^1$. In the language of fig.\ref{fig_schema_con_dis},
the latter correspond to the disconnected parts of the singlet proton
amplitude, whereas the connected
contributions should still obey the Ellis-Jaffe sum rule.

Before we explain how such a {\it direct} calculation can be
performed on the lattice we will discuss shortly the second approach to
$G_A^1$ which exploits the close connection between flavor singlet
axial current, axial anomaly and topological charge.

The axial vector Ward identity
\begin{equation}
\partial_\mu A^0_\mu = 2i \sum_{i=1,n_f} m_{q_i} \bar{q}_i\gamma_5 q_i + 
2 n_f \frac{\alpha_s}{4\pi} tr G_{\mu\nu}\tilde{G}^{\mu\nu}\;,  
\label{eq_axvec_ward}
\end{equation} 
with $n_f$ being the number of quark flavors and $G$($\tilde{G}$)
the (dual) tensor of gluon field strength, states that the axial vector
flavor singlet current is not conserved in the chiral limit due to
existence of an anomalous pure vacuum contribution  
$tr G_{\mu\nu}\tilde{G}^{\mu\nu}$, called the axial anomaly. 

The anomaly is deeply
connected to the topological structure of the
vacuum\cite{thooft_topol}. In particular, the topological charge
$Q$ (topological index, Pontryagin index), which labels
the different topological sectors of the vacuum, can be expressed by
\begin{equation}
Q = -\frac{1}{16\pi^2} \int d^4x\, tr G_{\mu\nu}\tilde{G}^{\mu\nu} \;.
\label{eq_topol_charge_def}
\end{equation}  
Thus, the axial anomaly is just the topological (charge) density
${\cal Q}(x) \equiv tr G_{\mu\nu}\tilde{G}^{\mu\nu}$ of the vacuum.

We can now reformulate eq. (\ref{eq_ga1}) in terms of ${\cal Q}$.
In the chiral limit ($m_q=0$), a Fourier transformation of 
eq. (\ref{eq_axvec_ward}) yields
\begin{equation}
i q_\mu \langle p',s|A^0_{\mu}(q)| p,s  \rangle = 2 n_f
\frac{\alpha_s}{4\pi}\langle p',s|{\cal Q}(q)| p,s  \rangle \;,
\label{eq_ga1_topol_predef}
\end{equation}
where $p$,$p'$ are the 4-momenta of the incoming and outgoing proton,
and $\vec{q} = \vec{p} - \vec{p}'$. 
If we choose the proton polarization $s$ to be (anti)parallel to the
direction of the momentum transfer $q$,
\begin{equation}
\frac{\vec{s}}{|\vec{s}|} = \pm \frac{\vec{q}}{|\vec{q}|}\;,
\end{equation}
we obtain in the rest frame of the proton at zero momentum transfer
\begin{equation}
G_A^1 = s_{\mu}\langle p,s|A_\mu^0| p,s\rangle =
\lim_{\vec{q}\rightarrow \vec{0} }
2 n_f \frac{\alpha_s}{4\pi}
\frac{-i|\vec{s}|^2}{\vec{q} \cdot \vec{s}} \langle p',s |{\cal Q} |p,s \rangle\;.
\label{eq_ga1_topol_def}
\end{equation}
Thus, one arrives at a topological expression for the singlet axial
coupling. We call the lattice procedure to determine the disconnected 
matrix element on the r.h.s. the {\it topological method}. Details 
will be given below. Let us mention however, that  this method has a
serious shortcoming in the quenched approximation\cite{gupta_ga1}.  
Because of the fact that the $\eta'$ meson becomes massless in the
chiral limit of quenched QCD, the desired matrix element 
$\langle p',s |{\cal Q} |p,s \rangle$ diverges in this limit.
Therefore, such calculations can be done only in full QCD.

An alternative, less fundamental, expression for $G_A^1$ in
terms of topological quantities can be
found using the composite operator propagator-vertex (CVP) 
method\cite{ga_anomaly2,ga_anomaly3,shore_cpv}. Here one uses the OPE
to decompose the singlet part of $\Gamma_1^p$, 
eq. (\ref{eq_1st_moment_g1}), into composite operator propagators
$\langle 0|T {\cal O}_i {\cal O}_j| 0\rangle$ at zero momentum 
and vertex functions 
$\Gamma$
\begin{equation}
(\Gamma_1^p)_{sing} = \frac{1}{9}\frac{2}{2M} 2 n_f C_1^s(\alpha_s)
\left[ \langle 0| T {\cal Q}{\cal Q}| 0\rangle \Gamma_{Qpp} +
  \langle 0|T {\cal Q} \Phi_{5}| 0 \rangle \Gamma_{\Phi_{5}pp} \right]\;,
\label{eq_cpv_g1p}
\end{equation}
with $\Phi_{5} = \sum_{i=1,n_f} \bar{q}_i \gamma_5 q_i$.
The topological susceptibility 
\begin{equation}
\chi(k^2) = \int dx e^{ikx}\,i \langle 0|T {\cal Q}(x){\cal Q}(0)|0\rangle
\label{eq_topsus_def}
\end{equation}
vanishes in the chiral limit of full QCD at zero momentum transfer.
The remaining propagator on the r.h.s. of eq. (\ref{eq_cpv_g1p}) is given
by the square root of the first moment $\chi'$ of the topological 
susceptibility\cite{ga_anomaly2}. Thus, if the vertex function 
$\Gamma_{\Phi_{5}pp}$ and the first moment of $\chi$ would be known,
one could calculate the singlet axial vector coupling. Unfortunately,
the vertex function cannot be calculated directly from first
principles, making the CPV approach somewhat less rigorous than 
the {\it topological method}. However, it can be estimated
e.g. in the QZI limit of QCD. The determination of $\chi'$ is a
typical task for non-perturbative methods like lattice QCD or QCD
spectral sum rules (QSSOR). The authors of ref.\cite{ga_anomaly3} have
used the latter. They found\footnote{A very preliminary result of
an exploratory full QCD lattice simulation with Kogut-Susskind
fermions has been reported by the authors of ref. \cite{chip_digiacomo}:  
$\sqrt{\chi'(0)}=16(3)$MeV .}
$\sqrt{\chi'(0)}/_{Q_0^2=10 {GeV}^2} = 23.2\pm 2.4$ MeV
and, correspondingly $G_A^1 =0.35\pm 0.05$, consistent
with the result of the latest experimental determination, quoted
in eq. (\ref{eq_ga1_exp}). It will be interesting to see, whether this
result can be consolidated by a full QCD lattice computation.\\

Finally, let us comment on the apparent incompatibility in the
physical interpretation of the two approaches.
From a success of the {\it direct method} one would conclude that
the correlation of vacuum quark loops with the proton is responsible
for the small value of $G_A^1$, whereas one expects in 
the {\it topological method} that topological excitations of the
vacuum decrease the flavor singlet axial coupling.
We want to emphasize that these are
not different mechanisms: Topologically non trivial vacuum
configurations influence of course the correlation of quark loops with
the proton, since it is exclusively  mediated by the vacuum. 
From that point of view, one would consider the (disconnected) quark
loop contributions to $G_A^1$ as being induced by the topological
properties of the vacuum.
  
\subsection{Direct Method\label{sec_direct_meth}}

According to eq. (\ref{eq_ga1}) we need to compute the matrix element
$\langle p,s |\bar{q}\gamma_{\mu}\gamma_5|p,s\rangle$. The techniques
which are employed for this task are quite similar to
those introduced in section \ref{sub_oper}. Again, one exploits ratios
of 3- and 2-point correlation functions at large time distances,
namely
\begin{eqnarray}
\lefteqn{R^{SUM}_A(x_0,\mu) = \frac{G_{NqN}^A(x_0,\mu)}{G_{NN}(x_0)}
- \langle \sum_{\vec{y},y_0} (\bar{q}\gamma_{\mu}\gamma_5
q)(\vec{y},y_0) \rangle} \hspace{1.5cm} \nonumber \\
& & \stackrel{x_0 \rightarrow \infty}{\longrightarrow} \quad 
A + \langle N |\bar{q}\gamma_{\mu}\gamma_5 q|N\rangle\,x_0 \;,
\label{eq_ra_sum}
\end{eqnarray}
\begin{eqnarray}
\lefteqn{R^{PLA}_A(x_0,y_0,\mu) =
\frac{C_{NqN}^A(x_0,y_0,\mu)}{G_{NN}(x_0)}
- \langle \sum_{\vec{y}} (\bar{q}\gamma_{\mu}\gamma_5
q)(\vec{y},y_0) \rangle} \hspace{1.5cm} \nonumber \\
& & \quad \stackrel{x_0 \rightarrow \infty}
{\longrightarrow}
\langle N |\bar{q}\gamma_{\mu}\gamma_5 q|N\rangle\ \;, 0 \ll y_0 \ll x_0 \;, 
\label{eq_ra_pla} 
\end{eqnarray} 
and
\begin{equation}
R^{PAM}_A(x_0,\Delta x_0,\Delta x'_0,\mu)
 = \sum_{y_0=\Delta x_0}^{x_0-\Delta x'_0} {R}^{PLA}_A(x_0,y_0,\mu)\;.
\label{eq_ra_pam}
\end{equation} 
The axial 3-point functions $G_{NqN}^A$ and $C_{NqN}^A$ are
defined as
\begin{eqnarray}
C_{NqN}^A(x_0,y_0,\mu) &=& 
\sum_{\vec{x}}\sum_{\vec{y}}
\langle N^{\dagger}(\vec{0},0) 
\left(\bar{q}\gamma_{\mu}\gamma_5 q\right)(\vec{y},y_0)
N(\vec{x},x_0) \rangle \quad, \quad \\
G_{NqN}^A(x_0,\mu) &=& \sum_{y_0} C_{NqN}^A(x_0,y_0,\mu) \;.
\end{eqnarray}

One might suspect that the application of the {\it summation method}
is not justified, since a term 
$ \sim \bar{q}\gamma_{\mu}\gamma_5 q$ is not present in the 
QCD action $S_{QCD}$,
and thus the Feynman-Hellmann theorem cannot be used directly
to derive eq. (\ref{eq_ra_sum}). However, if one  starts from a slightly
modified action
\begin{equation}
S_{QCD}(\lambda) = S_{QCD} + \lambda \sum_y (\bar{q}\Gamma q)(y)
\quad, \quad \Gamma =\gamma_{\mu},\gamma_5,\gamma_{\mu}\gamma_5,...
\end{equation}
and repeats the steps which lead to eq. (\ref{eq_scalar_ratio}) one
arrives at
\begin{equation}
R^{SUM}_A(x_0,\lambda) = 
\frac{G_{NqN}^A(x_0,\lambda)}{G_{NN}(x_0,\lambda)} 
- \langle  \sum_{y}(\bar{q}\Gamma q)(y) \rangle
= A(\lambda) + 
\frac{\partial M_N(\lambda)}{\partial \lambda}x_0\;.
\end{equation}
The Feynman-Hellmann theorem, see eq. (\ref{eq_feynman_hellmann}), yields
\begin{equation}
\frac{\partial M_N(\lambda)}{\partial \lambda} =
\langle N(\lambda)|\bar{q}\Gamma q| N(\lambda)\rangle \;.
\end{equation}
Finally, in the limit $\lambda \rightarrow 0$, one obtains 
eq. (\ref{eq_ra_sum}).\\
 
The numerical methods used to calculate the connected contributions
to $R^{SUM}_A$ and $R^{PLA}_A$ are identical with those described in
section \ref{sec_what_has}. \\

To determine the disconnected contributions, one has to evaluate
\begin{equation}
(C_{NqN}^A)^{disc} = \langle F_{G_{NN}}(0,x)
\sum_{\vec{y}} 
Tr(\gamma_{\mu}\gamma_5 M^{-1}(\vec{y},y_0;\vec{y},y_0))\rangle
\end{equation}
and
\begin{equation}
(G_{NqN}^A)^{disc} = \sum_{y_0} (C_{NqN}^A)^{disc}\;.
\label{eq_disc_ga}
\end{equation} 
In principle, the standard estimator techniques, described in 
section \ref{sec_what_has}, can be applied. However, these methods
become inefficient if one computes off-diagonal elements of
$M^{-1}$ in spin space, which occur for example in the combinations
$Tr(\gamma_{1,2}\gamma_5 M^{-1})$. In these cases, large diagonal
contributions add to the 'noise term' on the r.h.s. of 
eq. (\ref{eq_stochest_singleM}), and one would have to enhance the number
of estimates or gauge configurations drastically to suppress these terms.

Two alternative ways to cope with this
problem have been proposed in ref's.\cite{sesam_spin_explizit,sesam_nsigma} and 
ref.\cite{sesam_substr_diag}. The first method simply circumvents
the problem by
splitting  the stochastic inversion, eq. (\ref{eq_stochest_M}), into 
four pieces, each of them corresponding to one spin
component. It turns out that, with an equal computational effort, this
{\it spin explicit method} reduces the statistical error on the
estimate of $Tr(\gamma_{1,2}\gamma_5 M^{-1})$ by about 40 - 50$\%$.

The second technique reduces the fluctuations from large diagonal
contributions by subtracting the estimate for the on-diagonals
\begin{equation}
E[M^{-1}(x_0,x_0)] = \frac{1}{N_E} \sum_{s=1}^{N_E}\sum_y
\eta^s_{x_0} M^{-1}(x_0,y)\eta^s_y
\end{equation}
 from the estimate $E[M^{-1}(x_0,z_0)]$ for the off-diagonals, given
by eq. (\ref{eq_stochest_singleM}). This yields an improved estimate
\begin{equation}
E[M^{-1}(x_0,z_0)]^{impr} = E[M^{-1}(x_0,z_0)] - E[M^{-1}(x_0,x_0)]
\frac{1}{N_E}\sum_{s=1}^{N_E} \eta^s_{z_0} \eta^s_{x_0}\;.
\end{equation}
We emphasize that this {\it diagonal improved method}
works equally well as the  {\it spin explicit method}.  

\subsubsection{Renormalization}

The axial current $A_{\mu}$ is not conserved in the continuum due to
the presence of finite quark masses and to the axial anomaly. Thus it
gets renormalized by quantum fluctuations. The effects of the two
chiral symmetry breaking mechanisms on the renormalization of $A_{\mu}$
are however quite different. 

The quark masses induce only a soft
symmetry breaking, i.e. the dimension of the relevant operator is
less than 4, and the renormalization factor ${\cal{Z}}_A^{NS}$ 
is cutoff independent and finite. In particular, one obtains 
${\cal{Z}}_A^{NS} = 1$ in the limit of vanishing quark mass. Non-singlet
axial currents  are renormalized by ${\cal{Z}}_A^{NS}$ alone, as the
axial anomaly is absent in this case.

The symmetry breaking caused by the axial anomaly is not soft
(dimension=4), and $A_{\mu}$ picks up an additional renormalization
which diverges logarithmically if the cutoff is 
removed\cite{renorm_anomalie}. This leads
to a non-zero anomalous dimension which induces a 
renormalization point dependence on ${\cal{Z}}_A^S$\cite{renorm_anomaly}.    
Fortunately, the renormalization point dependence of $G_A^1$ is weak.
Increasing the renormalization scale $\mu^2$ from 10GeV$^2$ to
infinity decreases the value of $G_A^1$ by about 10$\%$ \cite{SMC_exp_new}.
 
The situation is more complicated in lattice QCD. Lattice
discretizations of the fermion action, which avoid the fermion
doubling problem, introduce additional chiral symmetry breaking terms.
It has been shown by Karsten and Smit\cite{fermion_doubling2} for
the Wilson type of discretization that these terms are essential in
reproducing the correct form of the axial anomaly in the continuum limit. 
However, the renormalization of the (axial) currents becomes
non-trivial. Although the lattice value, $Z_A^{NS}$, 
is still finite, the condition
$Z_A^{NS} = 1$ at $m_q=0$ is no longer valid.

Much progress  has been
achieved in the  determination of the lattice values of
$Z_A^{NS}$,  which depend
on the lattice coupling and  quark mass. We will 
comment on the different methods below. 
  
Not much is known about the lattice renormalization factor $Z_A^S$ 
of the singlet axial current. In first order lattice perturbation
theory, the logarithmically divergent contribution from axial anomaly
is absent, and $Z_A^S = Z_A^{NS}$\cite{fermion_doubling2}.  
Unfortunately, a two loop or even a non-perturbative determination
of $Z_A^{NS}$ has
not been performed yet. In view of the large statistical uncertainties
in the disconnected parts  of  flavor singlet axial current matrix
elements (see below) it appears to be justified to work with the assumption
$Z_A^S = Z_A^{NS}$. However, one should keep in mind that a possible 
improvement of the statistical quality would have to go hand in hand
with  a more precise determination of $Z_A^{S}$. 

The early estimates of $Z_A^{NS}$ were based on first order lattice
perturbation theory calculations, where the bare lattice coupling
was taken as the expansion parameter\cite{fermion_doubling2,groot_renorm}.  
It was demonstrated however through several examples \cite{sw_tadpole_imp}
that lattice perturbation theory in the bare coupling $g_0$ is ill
behaved, and that large higher order corrections can introduce 
considerable uncertainties on the renormalized matrix elements. 

The convergence of the perturbative expansion can be improved,
if one uses a renormalized coupling rather than $g_0$ , defined by a proper 
renormalization condition \cite{parisi,sw_tadpole_imp}.  
Defining an 
improved coupling $\alpha_V$ by the static potential $V$
\begin{equation}
V(q) = - \frac{C_f 4\pi\alpha_V(q)}{q^2} \quad, \quad C_f = \frac{4}{3}\,,
\end{equation}
Lepage and Mackenzie \cite{sw_tadpole_imp} have worked out
an improved perturbative expansion on the lattice, which in many cases
exhibits a faster convergence than the expansion in $g_0$. Moreover
they demonstrated that the improvement is mainly due to the inclusion
of tadpole-like diagrams, which, if ignored, induce a slow convergence
of the expansion. Most of the recent lattice simulations have used
this 'tadpole improved' perturbation theory in first order 
to renormalize their results. We emphasize however that even with this method
higher order contributions can be sizable. Thus, a scaling analysis
of the renormalized results is still mandatory. 

To overcome this problem, non-perturbative renormalization techniques
have been developed, which avoid the use of
lattice perturbation 
theory\cite{martinelli_wardid_ren1,martinelli_wardid_ren2,martinelli_ri_ren,luescher_schroed_ren}.
For the determination of $Z_A^{NS}$ two different procedures, the Ward
identity method (WI)\cite{martinelli_wardid_ren1} and the regularization
scheme independent method (RI)\cite{martinelli_ri_ren} are equally
appropriate. The WI method exploits the fact that the Ward identities
on the lattice have the same structure as in the continuum. This
defines a set of non-linear equations for the renormalization factors,
which can be solved numerically on a given lattice.

Clearly, the WI
method is restricted to those operators where Ward identities do
exist for. The RI method is more general in that respect and can be
applied in principle to arbitrary composite operators. In this scheme the
renormalization condition for an operator ${\cal{O}}$ is
 imposed on the corresponding amputated Green function, computed  
between off shell quark states of large virtuality. This defines a 
renormalized operator $\cal{O}^R$, and thus a renormalization constant
$Z_O$, which is independent of the regularization
scheme\cite{martinelli_independ}. A disadvantage of the RI method is
that one has to work in a fixed gauge. This makes it necessary
to study the influence of lattice Gribov copies
on the results \cite{gribov_lattice}.

Both non-perturbative methods imply a numerical lattice determination
of several correlation functions. Thus, uncertainties due to finite
size effects, statistical accuracy and, in the case of the WI method,
ground state identification may still be present.\\ 

One may ask of course to what extent the value of $Z_A^{NS}$
 depends on the method used for its determination. To illustrate this
point we show\footnote{This table has
been originally compiled by the authors of 
ref.\cite{rapuano_renorm}.} in tab.\ref{tab_za_values} the
results for $Z_A^{NS}$
from the different methods, calculated with the Wilson action in the
 quenched approximation and at several lattice couplings
 $\beta=6/g_0^2$.  
\begin{table}[ht]
\begin{center}
\begin{tabular}{|c|c|c|c|c|c|}
\hline 
     &              &       &      &         &    \\
$\beta$ & $a^{-1}$[GeV] \cite{rapuano_renorm} & SPT & TIPT & 
WI \cite{martinelli_wardid_ren2} &  RI \cite{rapuano_renorm} \\
     &              &       &      &         &    \\
\hline  
6.0  &     2.258(50) & 0.87 & 0.78 & 0.85(7) & 0.81(1)  \\
6.2  &    2.993(94) & 0.87 & 0.79  &         & 0.81(1) \\
6.4  &    4.149(161) & 0.87 & 0.80 &         & 0.82(2) \\ 
\hline   
\end{tabular}
\caption{\label{tab_za_values}
{\it
Results
for $Z_A^{NS}$ in the chiral limit at a renormalization
scale $\mu = a^{-1}$ in the
standard Wilson discretization. Denotation: SPT: first
order standard (bare) perturbation theory; TIPT: first order tadpole
improved perturbation theory; WI: non-perturbative Ward identity
method; RI: non-perturbative regularization scheme independent method.
}}
\end{center}
\end{table}

Obviously, the effect of variation of methods
is not dramatic ($5-10\%$), compared to the large statistical
 uncertainties on the disconnected part of the axial flavor singlet
 current. Given that, together with the fact that remormalization
effects on $G_A^1$ are small in the continuum, one can hope, for lack
of better wisdom,  that
 the one loop result for the
 renormalization of this current, $Z_A^{S} = Z_A^{NS}$, is a 
currently tolerable approximation.

\subsection{Topological Method \label{sec_topol_meth}}

To determine $G_A^1$ from eq. (\ref{eq_ga1_topol_def}) by a 
lattice calculation one needs to
define a discretized lattice operator for the topological charge density 
${\cal Q}^L$ which agrees with the continuum definition of  ${\cal Q}$
in the limit $a \rightarrow 0$, e.g.
\begin{equation}
{\cal Q}^L(x) \stackrel{a \rightarrow 0}{\longrightarrow} a^4 {\cal Q}(x) +
O(a^6) \;.
\label{eq_q_naive_limit}
\end{equation}
A straightforward way to define  ${\cal Q}^L$ starts from the
discretized version of the field strength tensor $G_{\mu\nu}^L$,
\begin{eqnarray}
G_{\mu\nu}^L(x) &=& \frac{1}{8}
 \left[ 
P_{\mu,\nu}(x) + P_{-\mu,-\nu}(x) + P_{\nu,-\mu}(x) + P_{-\nu,\mu}(x)
\right.
\nonumber \\
&& \left. 
    -\; P_{\nu,\mu}(x) - P_{-\nu,-\mu}(x) - P_{-\mu,\nu}(x) -
P_{\mu,-\nu}(x)
\right] \;,
\end{eqnarray}
where $P_{\mu,\nu}(x)$ is the product of gauge parallel transporters
$U(x)$ around an elementary plaquette
\begin{equation}
P_{\mu,\nu}(x) =
U_{\mu}(x)U_{\nu}(x+a{\hat{\mu}})U_{-\mu}(x+a{\hat{\mu}}+a{\hat{\nu}})
U_{-\nu}(x+a{\hat{\nu}})\;. 
\end{equation}
From that it follows for the topological charge density\cite{fabricius}
\begin{equation}
{\cal{Q}}^L(x) = \frac{1}{64} \sum_{\mu\nu\rho\sigma}^{\pm 4}
\tilde{\epsilon}_{\mu\nu\rho\sigma} Tr \left(
  P_{\mu,\nu}(x)P_{\rho\sigma}(x) \right)\quad,
\label{eq_q_latt_def}
\end{equation} 
where $\tilde{\epsilon}_{\mu\nu\rho\sigma}$ is the Levi-Civita tensor
for $\mu,\nu,\rho,\sigma > 0$, and 
$\tilde{\epsilon}_{\mu\nu\rho\sigma} =
-\tilde{\epsilon}_{-\mu\nu\rho\sigma}$, etc. .\\

To relate the lattice definition of the topological charge density to
the respective continuum quantity one needs to renormalize ${\cal{Q}}^L$.
In principle, this is a highly non-trivial task, since $\cal{Q}$
mixes with $\partial_\mu A_\mu$ and with the pseudo scalar current
$\sum_{i=1}^{n_f} m_i \bar{q}_i \gamma_5 q_i$ under 
renormalization\cite{ga_anomaly2}. It has been shown however in first
order lattice perturbation theory\cite{q_renorm_latt} that the mixing
is quite small ($\le 5\%$) for bare lattice couplings 
$g_0 \le 1$. Thus one can concentrate on the determination of one
multiplicative renormalization constant $Z_Q$,
\begin{equation}
{\cal{Q}} \simeq Z_Q {\cal{Q}}^L \;. 
\end{equation}
Unfortunately, it turns out that the one loop correction to $Z_Q$ is
large\cite{q_renorm_latt,q_renorm_latt1}, making a non-perturbative determination highly desirable. Such
a non-perturbative renormalization method has been developed by the
authors of ref's. \cite{q_renorm_nonpert1,q_renorm_nonpert2}. Roughly
speaking it consists of preparing a classical gauge configuration
of given topological charge and then to add gauge fluctuations by a
Monte Carlo process. A comparison of the integer valued `classical' 
topological charge with the one extracted from the `fluctuating' gauge
configurations allows for the determination of $Z_Q$.

On the other hand one can try to improve the lattice operator
$\cal{Q}^L$, eq. (\ref{eq_q_latt_def}), such that the higher order
corrections to $Z_Q$ are made small or might even vanish. The link
smearing technique \cite{q_latt_impr}, which supplies a non-local
version of eq. (\ref{eq_q_latt_def}), reduces the higher order
corrections substantially. However, the `ultima ratio' with respect to
renormalization properties is the so called geometric definition of
$\cal{Q}^L$, given by M.~L\"uscher\cite{luescher_q_def}. Here,
$\cal{Q}^L$ is constructed such that the corresponding topological
charge takes only integer values even in the case of `fluctuating'
gauge configurations. It follows that $Z_Q=1$, i.e. higher order
corrections are absent.

A comparison of the different definitions of topological charge on
the lattice has been given by the authors of ref.\cite{comp_q_latt} in
the quenched approximation. It turns out that, using the 
non-perturbative renormalization procedure described above, all
definitions lead to consistent results. However, the geometrical
definition and the definition of eq. (\ref{eq_q_latt_def}) suffer from
large statistical noise. Thus, for practical simulations, one should 
use\footnote{We did not discuss the `fermionic' definition of
the topological charge density  
\begin{equation}
{\cal{Q}}^L = m_q \langle Tr(\gamma_5 M^{-1}) \rangle \;, 
\label{eq_q_ferm}
\end{equation}
where $m_q$ is the quark mass and $M^{-1}$ the quark propagator, 
which has been proposed by Smit and Vink \cite{q_smit_vink}, inspired 
by the Atiyah-Singer index theorem. Although stochastic estimator
techniques enable for an accurate  calculation \cite{viehoff_latt98} 
of the r.h.s. of eq.(\ref{eq_q_ferm}), the renormalization properties
of have not been studied yet in detail.}
the link smearing technique \cite{q_latt_impr} to calculate $\cal{Q}^L$.\\

Having succeeded in defining suitable lattice operators  $\cal{Q}^L$
there remains still a potentially severe problem for full QCD lattice
applications\footnote{Quenched QCD does not suffer from this problem.}
to extract reliable results from numerical simulations.  
It has been obtained in ref. \cite{boyd_q_staggered} for the
staggered formulation of lattice fermions that the standard
updating algorithm, the Hybrid Monte Carlo, tends to loose its
ability to move between different sectors of topological charge if the
lattice quark mass is chosen at or below  $ma \simeq 0.01$, which
corresponds to $m_{\pi}/m_{\rho} \simeq 0.57$. Thus, in this region
of quark mass, the
results might be biased by a non-Gaussian distribution of gauge
configurations with respect to topology. The authors of
ref.\cite{sesam_q_wilson} have performed a similar investigation using
dynamical Wilson fermions with quark masses down to a value
corresponding to $m_{\pi}/m_{\rho} \simeq 0.69$, finding no
significant decrease of the Hybrid Monte Carlo's mobility. At
$m_{\pi}/m_{\rho} \simeq 0.56$ ref.\cite{sesam_q_wilson} obtains still
transitions between different topological sectors, but with a reduced
frequency. This indicates that either the number of Hybrid Monte Carlo
updates has to be increased at small quark 
mass\footnote{Ref.\cite{sesam_q_wilson} estimates an increase by a
  factor of four when decreasing the quark mass from $m_{\pi}/m_{\rho}
  \simeq 0.69$ to $m_{\pi}/m_{\rho} \simeq 0.56$.} or one has to find
an alternative updating algorithm which does not suffer from this kind
of immobility.\\

Assuming that we have successfully created a statistically
representative sample of vacuum configurations we are now prepared
to evaluate  the r.h.s. of eq. (\ref{eq_ga1_topol_def}) on the lattice.   
Obviously, two steps are necessary for this task: First one has to
determine the matrix element $\langle p',s|{\cal{Q}^L}(q)|p,s \rangle$
for several momenta $p,p'$ and secondly one has to extrapolate the
results to the point $q= p-p' = 0$. Concerning the second step we 
emphasize that it is advantageous to work in the brick-wall system 
\begin{equation}
|\vec{p}| = |\vec{p'}| \quad,\quad q_0 = 0\;.
\label{eq_momentum_cond}
\end{equation}   
Within this setting, the (unwanted) matrix element
$q_0\langle p',s|A_0^0|p,s\rangle$ is absent in
eq. (\ref{eq_ga1_topol_predef}) even for $\vec{q} \neq 0$.
Thus, there is no need to remove it by the extrapolation
$\vec{q} \rightarrow 0$ in eq. (\ref{eq_ga1_topol_def}).

To extract the matrix element $\langle p',s|{\cal{Q}^L}(q)|p,s \rangle$     
one starts from  the 3-point correlation function
\begin{equation}
A(\vec{y},\tau,\vec{x},t) = 
\langle P^{\dagger}(\vec{0},0)T_s
{\cal{Q}}(\vec{y},\tau)P(\vec{x},t)\rangle \;,
\end{equation}     
where $P$,$P^{\dagger}$ are interpolating proton operators and $T_s$ 
is the spin projection operator. The expectation value $A$ can be
determined on the lattice by
calculating the correlation of the (spin projected) proton propagator
with the topological charge density.

In Fourier space one obtains for given final momentum of the proton
$\vec{p}$ and momentum 
transfer\footnote{Note that the initial proton
momentum $\vec{p}'$ is implicitly fixed by
$\vec{p}' = \vec{p} - \vec{q}$.}
$q$ 
\begin{equation}
\tilde{A}(\vec{q},\tau,\vec{p},t) = \sum_{\vec{x},\vec{y}} 
A(\vec{y},\tau,\vec{x},t) e^{i\vec{y}\vec{q}} e^{i\vec{x}\vec{p}} \;.
\end{equation} 
At large time distances $t - \tau$ and $\tau$ the interpolating proton
operator projects only on the proton ground state. This yields  
\begin{eqnarray}
\lefteqn{\tilde{A}(\vec{q},\tau,\vec{p},t) 
\stackrel{t-\tau,\tau \rightarrow \infty}{\longrightarrow }} \\
&& \langle 0|P^{\dagger}|P(p',s)\rangle \langle P(p,s)|P|0 \rangle
\langle P(p',s)|{\cal Q}|P(p,s)\rangle \,\times\,
e^{-E(p)(t-\tau)}e^{-E(p')\tau}\;.
\nonumber
\end{eqnarray}
The proton amplitudes $\langle 0|P^{\dagger}|P(p',s)\rangle$ and
$\langle P(p,s)|P|0 \rangle$ are determined from the asymptotic 
time dependence of
the proton propagator $\Delta_P(p,t)$, see
eq. (\ref{eq_proton_prop_time_dep}).
Using the condition 
$|\vec{p}| = |\vec{p'}|$ (see above) one finally gets
\begin{equation}
\langle P(p',s)|{\cal Q}|P(p,s)\rangle = 
\lim_{t-\tau,\tau \rightarrow \infty} 
\frac{\tilde{A}(\vec{q},\tau,\vec{p},t)}{\Delta_P(p,t)}\;. 
\end{equation}   

From a computational point of view one might think that the
topological method is superior to the direct method since it
circumvents the very costly and statistically noisy calculation of
disconnected (fermionic) contributions with stochastic
estimator methods.
However it has some serious disadvantages which might be quite
difficult to overcome. First of all, eq. (\ref{eq_ga1_topol_predef}),
which  connects the axial current matrix element with the topological
charge density, is valid only in the chiral limit.
Apart from the fact that this makes an extrapolation in $m_q$ 
necessary\footnote{The fermion matrix approaches a zero mode at 
$m_q=0$. Thus, the quark propagator diverges at this point. A direct
lattice simulation at $m_q=0$ is therefore excluded.}
one has to neglect symmetry breaking effects
according to $m_{u,d} \neq 0$. Secondly one has to calculate
$\tilde{A}$ and $\Delta_P$ at several momenta $\vec{p},\vec{q}$.
The signals from momentum nonzero correlation functions are however very
noisy on the lattice and one needs high statistics and well suited
guesses for the wavefunctions to extract the required ground state
information. On top of this the extrapolation in $q$ might have to
span a quite long range. The smallest nonzero momentum on a lattice 
with periodic boundary conditions is given by
\begin{equation}
p_{min}= \frac{2\pi}{N_s} \times a^{-1} \;.
\end{equation}
With typical lattice parameters $N_s \simeq 20$ and 
$a^{-1} \simeq 2$GeV one gets the rather large value
$p_{min} \simeq 600$MeV.

Last nor least, as we mentioned above, the application of the
topological method is not justified in the quenched 
approximation \cite{gupta_ga1}. In full QCD one has to fight with
a possible stiffness of the updating algorithm at small sea quark
masses with respect to
tunneling between different topological sectors\footnote{Of course
this problem affects also the direct method.}.
 
Exploratory lattice studies have been performed in 
quenched\cite{gupta_ga1,digiacomo_ga_quen} and
in full \cite{mtc_ga_full,digiacomo_ga_full} QCD to check on the
feasibility of this method. It turned out that the statistical quality
of the momentum nonzero correlation functions and, in full QCD, the
stiffness of the updating algorithm with staggered fermions 
are the major hindrances in
extracting reliable results. Thus, being optimistic, one can hope that
the topological method might become useful with high statistics full
QCD simulations using Wilson-like fermions. Such simulations are
being performed currently by the CP-PACS\cite{cp_pacs_burkhalter}
and T$\chi$L\cite{tchil_collab} collaborations.

\subsection{Lattice Results}

\subsubsection{Quenched QCD}

The first steps towards a lattice determination of $G_A^1$ have been
done several years ago by calculating the connected contributions
to the matrix elements $\langle P|\bar{u}u|P\rangle$ and
 $\langle P|\bar{d}d|P\rangle$
\cite{wup_ga_old,liu_ga_con,gupta_full_sigma}.
Although the statistical accuracy of
these simulations has been rather low (less than 40 gauge
configurations have been used), the results have been quite
encouraging as they yielded estimates for the non-singlet axial
couplings $G_A^3$ and $G_A^8$ in the ball-park of the
experimental findings, eq. (\ref{eq_exp_ga3_ga8}). 

Recently, a more precise estimate of $G_A^3$ in the quenched
approximation has been calculated by the QCDSF collaboration
\cite{schierholz_ga}. They used the $O(a)$ improved 
Sheikoleslami-Wohlert action with a non-perturbatively
 determined \cite{sw_luescher_imp} `clover' coefficient $c_{sw}$, and
non-perturbative estimates for the axial vector
renormalization constant $Z_A^{NS}$. 
\begin{figure}[htb]
\begin{center}
{\epsfxsize=10.0cm\epsfbox{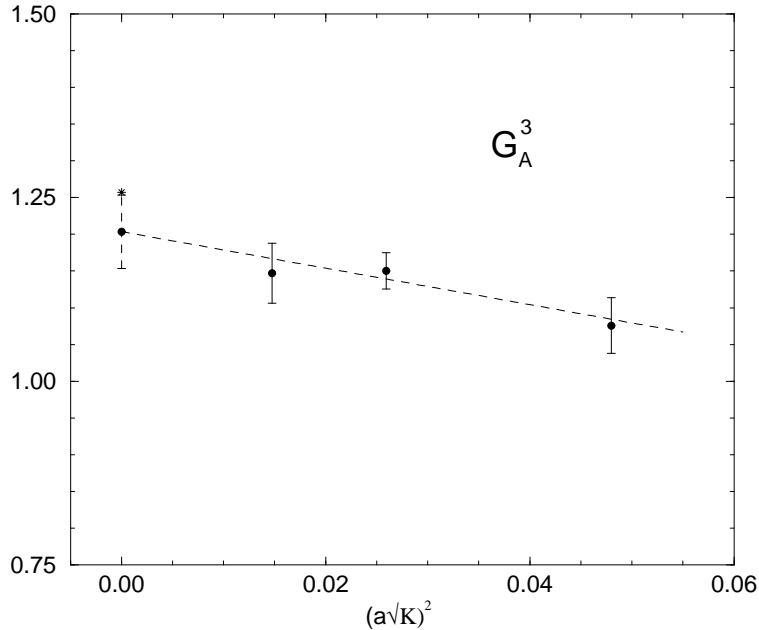}}
\caption{\label{fig_schierholz_ga} {\it 
QCDSF collaboration \cite{schierholz_ga}: Continuum extrapolation
of $G_A^3(a^2)$. The star indicates the position of the experimental
result.
}}
\end{center}
\end{figure}
The simulation has been performed at 3 different values of the
bare coupling, $\beta=6.0,6.2,6.4$, which correspond to lattice
cutoffs in the range $a^{-1} \simeq 2 - 3.5$Gev. The size of their
statistical samples varies from $O(500)$ gauge configurations at 
$\beta=6.0$ to $O(100)$ at $\beta=6.4$. Fig. \ref{fig_schierholz_ga} 
displays the results for $G_A^3$ as a function of the (square of)
the lattice spacing $a$ in units of the string tension $\sqrt{K}$.
It turns out that the cutoff effects in this range of the bare
coupling are small ($O(10\%)$). A linear extrapolation\footnote{Using an
$O(a)$ improved action one expects the cutoff effects to appear at
$O(a^2)$. Thus, a linear extrapolation of $G_A^3(a^2)$ is justified.}
to the continuum limit, $a=0$, yields a value for $G_A^3$
which is slightly lower, but,
within errors, still consistent with the experimental result. Thus,
one can expect from quenched lattice simulations to find a value of
$G_A^3$, which is close to experiment. Turning this conclusion around,
the deviation of a lattice estimate of $G_A^3$ at a finite cutoff
$a^{-1}$ from the experimental result 
can serve as an indicator for systematic effects
in the lattice calculations of the flavor non-singlet couplings.
For want of something more precise, this deviation may be used also as an
(lower limit) estimate for cutoff effects present in lattice
determinations of $G_A^1$.\\  
   
In
quenched QCD, two such simulations have been performed up to now with
the standard Wilson action, one
by Fukugita et al. \cite{japan_ga} and one by Dong et al. \cite{liu_ga}.
The former has been done with accurate statistics ($O(300)$ 
gauge configurations), but at a somewhat large coupling, $\beta=5.7$,
which corresponds to $a^{-1} \simeq 1.5$GeV. The latter has operated 
closer to the continuum, at $\beta=6.0$, but with a very low
statistics (24 configurations), which might lead to a substantial 
under-estimation of the errors.

Using $G_A^3$ as an indicator for the size of cutoff effects,
one finds in both
cases that the uncertainty of their lattice estimates for $G_A^1$ 
is dominated by the large statistical noise found in the disconnected
contributions. Before we discuss the results of 
ref's. \cite{japan_ga,liu_ga} in somewhat more detail we will
illustrate briefly why one indeed expects considerable statistical
fluctuations in those parts. 
  
\begin{figure}[htb]
\begin{center}
{\epsfxsize=14.0cm\epsfbox{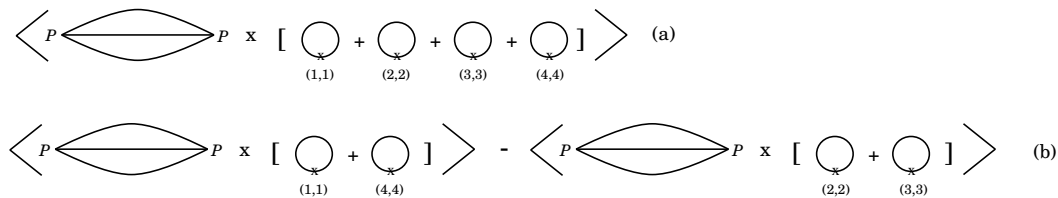}}
\caption{\label{fig_schema_bad_signal} {\it 
(a) Disconnected contribution to $\sigma_{\pi N}$; (b) Disconnected
contribution to $\langle P |\bar{q}\gamma_3\gamma_5 q|P\rangle$. 
}}
\end{center}
\end{figure}
We have seen in chapter \ref{sec_sigma_results} that the disconnected
contribution to $\sigma_{\pi N}$ is much more noisy than the connected
one. This is because
both, the noise from the estimator method and from the inherent quantum
fluctuations add up on top of the disconnected signal.
From a numerical point of
view the situation is even more disadvantageous for the disconnected
contributions to $G_A^1$. This is illustrated in 
fig. \ref{fig_schema_bad_signal}. According to eq. (\ref{eq_NqN_disc})   
the disconnected part of $\sigma_{\pi N}$ is given by the correlator
of the nucleon propagator and the sum over quark loops with Dirac
components ($\mu,\mu$), $\mu = 1,2,3,4$. Since the contributions are
of similar size for each of the Dirac components, the summation helps
to stabilize the signal. The disconnected part of $G_A^1$,
 see eq. (\ref{eq_disc_ga}), however is determined by the
difference of two such noisy correlators. \\   
\begin{figure}[htb]
\begin{center}
{\epsfxsize=9.0cm\epsfbox{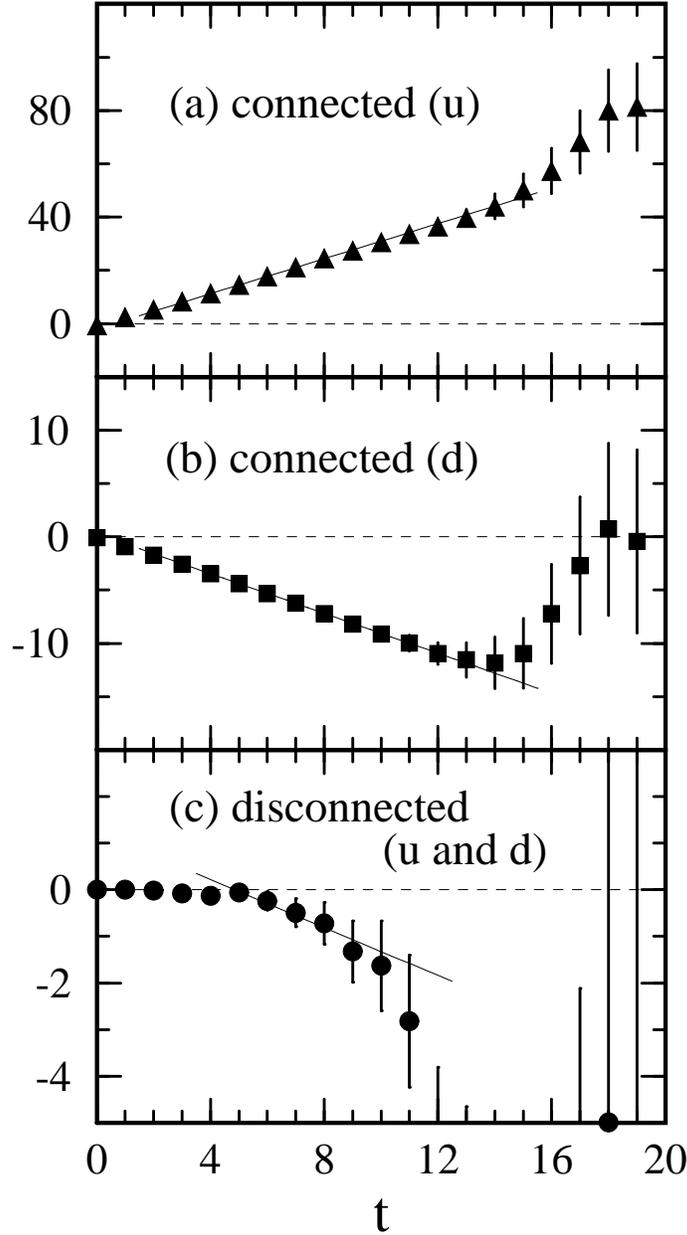}}
\caption{\label{fig_japan_ga_con_dis} {\it 
 Connected u-quark (a), connected d-quark (b) and disconnected
contributions $G_A^1$ as found by ref. \cite{japan_ga} at a quark mass
corresponding to $m_{\pi}/m_{\rho} \simeq 0.745$. The disconnected
part has been estimated with the (modified) volume source technique.
The solid lines indicate the results of the fits according to
eq. (\ref{eq_ra_sum}).}}
\end{center}
\end{figure}
In fig. \ref{fig_japan_ga_con_dis} we display the signals for
connected and disconnected contributions to $G_A^1$ as found by the
authors of ref. \cite{japan_ga}, at a quark mass corresponding to
$m_{\pi}/m_{\rho} \simeq 0.745$. The connected parts exhibit a 
clear linear slope, see eq. (\ref{eq_ra_sum}), whereas the
disconnected contribution is very noisy and disappears at $t=12$. 
\begin{figure}[htb]
\begin{center}
{\epsfxsize=9.0cm\epsfbox{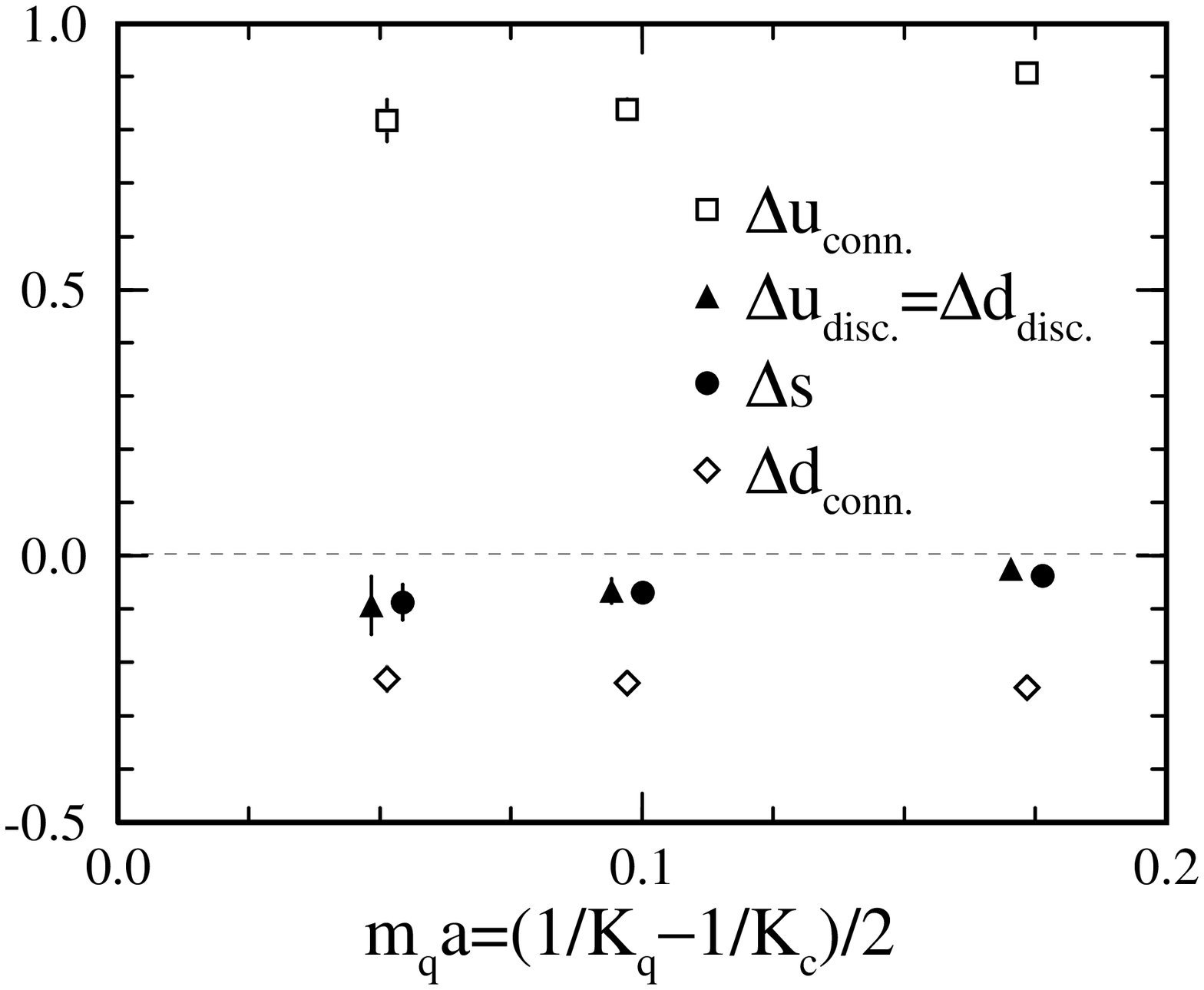}}
\caption{\label{fig_japan_ga_mq} {\it 
Ref. \cite{japan_ga}: Connected and disconnected amplitudes
$\Delta q = \langle P|\bar{q}\gamma_3 \gamma_5 q| P\rangle$ as
a function of the bare quark mass in lattice units. The data has been
multiplied by the 1st order tadpole improved renormalization constant
$Z_A^{S}=Z_A^{NS}$  
.}}
\end{center}
\end{figure}

Using the results of linear fits to the signals at several quark masses one
can perform an extrapolation to the chiral limit. The quark mass
dependence of connected and disconnected contributions is shown in
fig. \ref{fig_japan_ga_mq}. Note that the amplitudes have been multiplied
by the 1st order tadpole improved renormalization constant 
$Z_A^{S}=Z_A^{NS}$. To obtain the strange quark contribution to
the singlet coupling, the `loop' quark mass has been fixed to the
strange quark mass\footnote{The lattice value of the strange quark
mass $m_s$ is determined by the condition 
$m_K(m_s,m_l)/m_{\rho}(m_l) = 0.64$, where $m_l$ is the light quark
mass.}, while extrapolating in the light (valence) quark mass of the proton.
   
We emphasize that the disconnected parts are of opposite sign 
to the connected ones, thus lowering the value
of $G_A^1$. Obviously, they agree
within errors over the entire mass range. This indicates  flavor 
symmetry of the disconnected amplitudes. The mass dependence of all the
amplitudes is weak, and a linear extrapolation seems well justified.  

At the chiral limit, the authors of ref. \cite{japan_ga} obtain
\begin{equation}
G_A^3 = \Delta u - \Delta d = 0.638(54) + 0.347(46) = 0.985(25)
\label{eq_japan_res_ga3}
\end{equation}
and
\begin{equation}
G_A^1 = \Delta u + \Delta d + \Delta s = 0.638(54) - 0.347(46) -
0.109(30) = 0.18(10)
\label{eq_japan_res_ga1}\;,
\end{equation} 
where $\Delta q = \langle P|\bar{q} \gamma_3 \gamma_5 q| P \rangle$.     
Note that $G_A^1$ is significantly lower than the OZI estimate
$G_A^1 \simeq G_A^8 = 0.579(25)$, even within a statistical
uncertainty of about $50\%$. 

Unfortunately, the systematical errors
are not well under control. As mentioned above, the comparison of
the lattice estimate of $G_A^3$  with the experimental result can be
taken to give at 
least a rough impression on the size of systematical effects. From the
inspection of eq. (\ref{eq_japan_res_ga3}) and eq. (\ref{eq_exp_ga3_ga8}) 
one estimates that those are of the order of $30\%$. 

A similar behavior of connected and disconnected amplitudes has
been obtained by the authors of ref. \cite{liu_ga}. They find estimates
$G_A^1=0.25(12)$, $G_A^3=1.20(10)$, and $G_A^8=0.61(13)$, which are
much closer to the experimental results. Since they have worked at a
larger cutoff, $a^{-1} \simeq 2$GeV compared to  $a^{-1} \simeq
1.5$GeV in ref. \cite{japan_ga}, the systematical errors might be indeed
reduced. However, as mentioned above, the statistical sample
which underlies this calculation is rather small, i.e. $1/10$ of
the sample used in  ref. \cite{japan_ga}. Thus, some caution in
interpreting these findings is in order. \\
   
Nevertheless, the results found in quenched lattice calculations of
$G_A^1$ are encouraging, as they yield  quantitative support for
the assumption that the small value of $G_A^1$ can be explained by
negative contributions from quantum fluctuations.   
 
It will be interesting to see whether this remains valid in full QCD.  
 
\subsubsection{Full QCD}

Recently, the SESAM collaboration has performed a full QCD lattice
calculation of $G_A^1$ \cite{sesam_ga}. The analysis is  based on 4 sets of
200 statistically independent gauge configurations, which have been
generated in the Wilson discretization scheme with $n_f=2$ dynamical
fermions at $\beta=5.6$. The cutoff $a_{\rho}^{-1} =2.3$GeV
corresponds approximately to that of a quenched calculation at
$\beta=6.0$ \cite{sesam_light_spectrum}.
\begin{figure}[htb]
\begin{center}
\vskip -5.0cm
{\epsfxsize=11.0cm\epsfbox{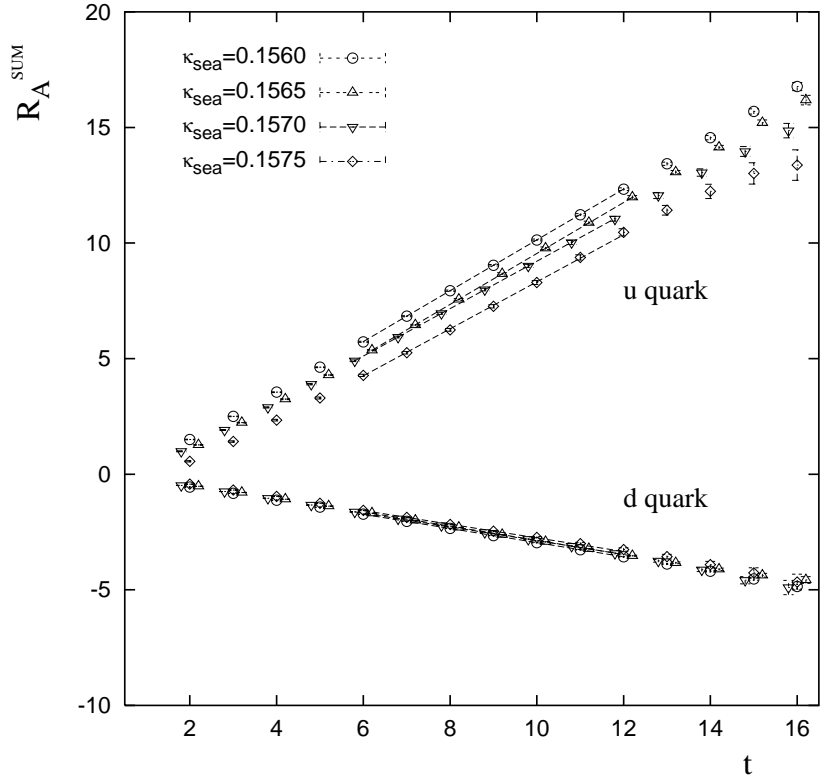}}
\caption{\label{fig_sesam_connected_ga} {\it Full QCD, SESAM 
collaboration \cite{sesam_ga}: Connected contributions to $G_A^1$
at different quark masses. The slope of the `u quark' signals
corresponds to the (spin averaged) amplitude 
$s_{\mu} \langle P |\bar{u}\gamma_mu \gamma_5 u|P \rangle$, that
of the  `d quark' signals to
$s_{\mu} \langle P |\bar{d}\gamma_mu \gamma_5 d|P \rangle$. The fits
(range and value) are indicated by dashed lines. 
}}
\end{center}
\end{figure}

SESAM has applied the summation method, see eq. (\ref{eq_ra_sum}), and the
standard insertion technique, c.f. section \ref{sec_what_has},
to calculate the connected contributions. We illustrate the quality 
of the corresponding signals in fig. \ref{fig_sesam_connected_ga}.
Obviously the expected linear rise of $R^{SUM}_A(t)$ is manifest for
all four quark masses. 
\begin{figure}[htb]
\begin{center}
\vskip -3.0cm
\noindent\parbox{15.0cm}{
\parbox{7.0cm}{\epsfxsize=7.0cm\epsfbox{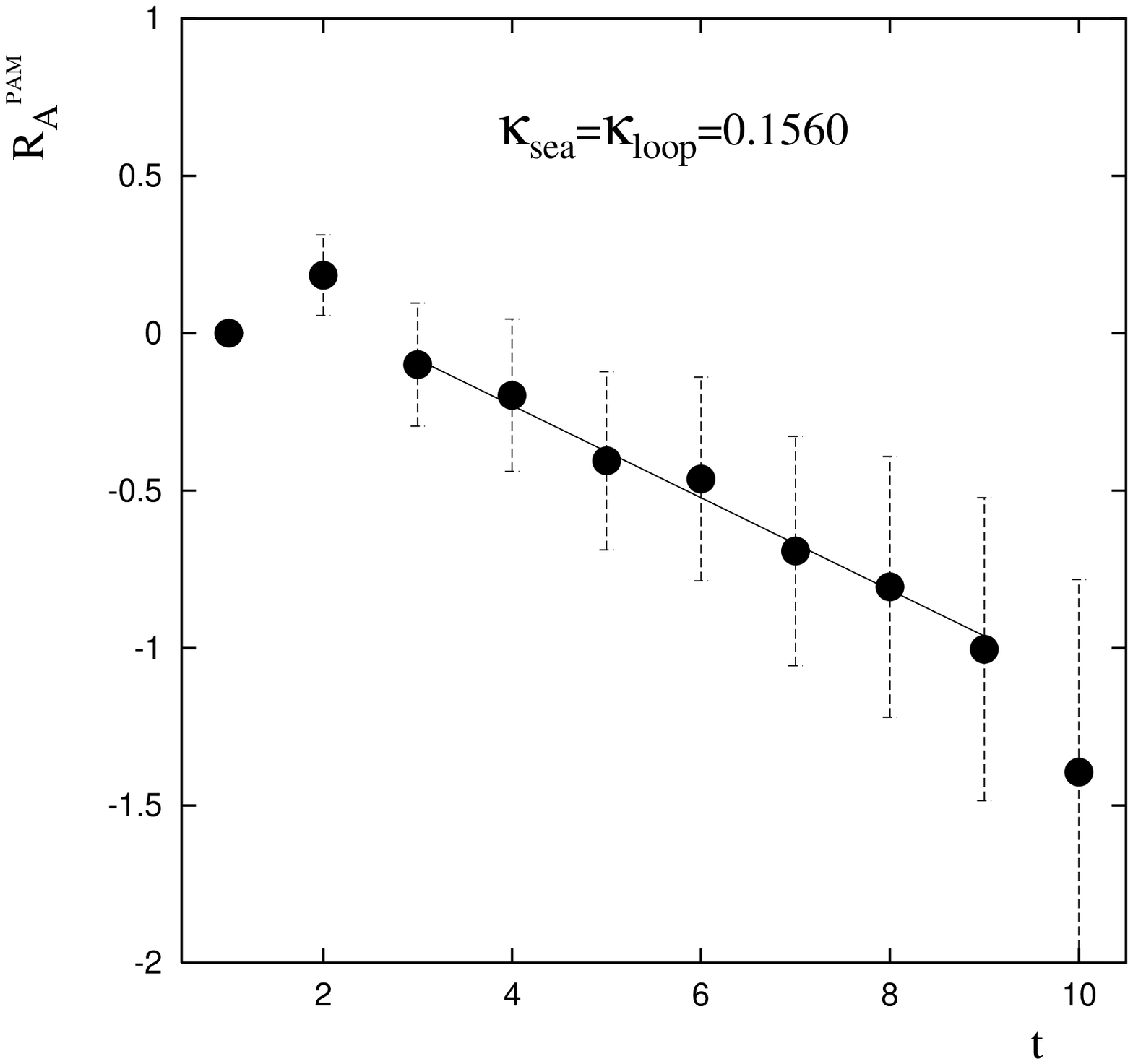}}
\parbox{7.0cm}{\epsfxsize=7.0cm\epsfbox{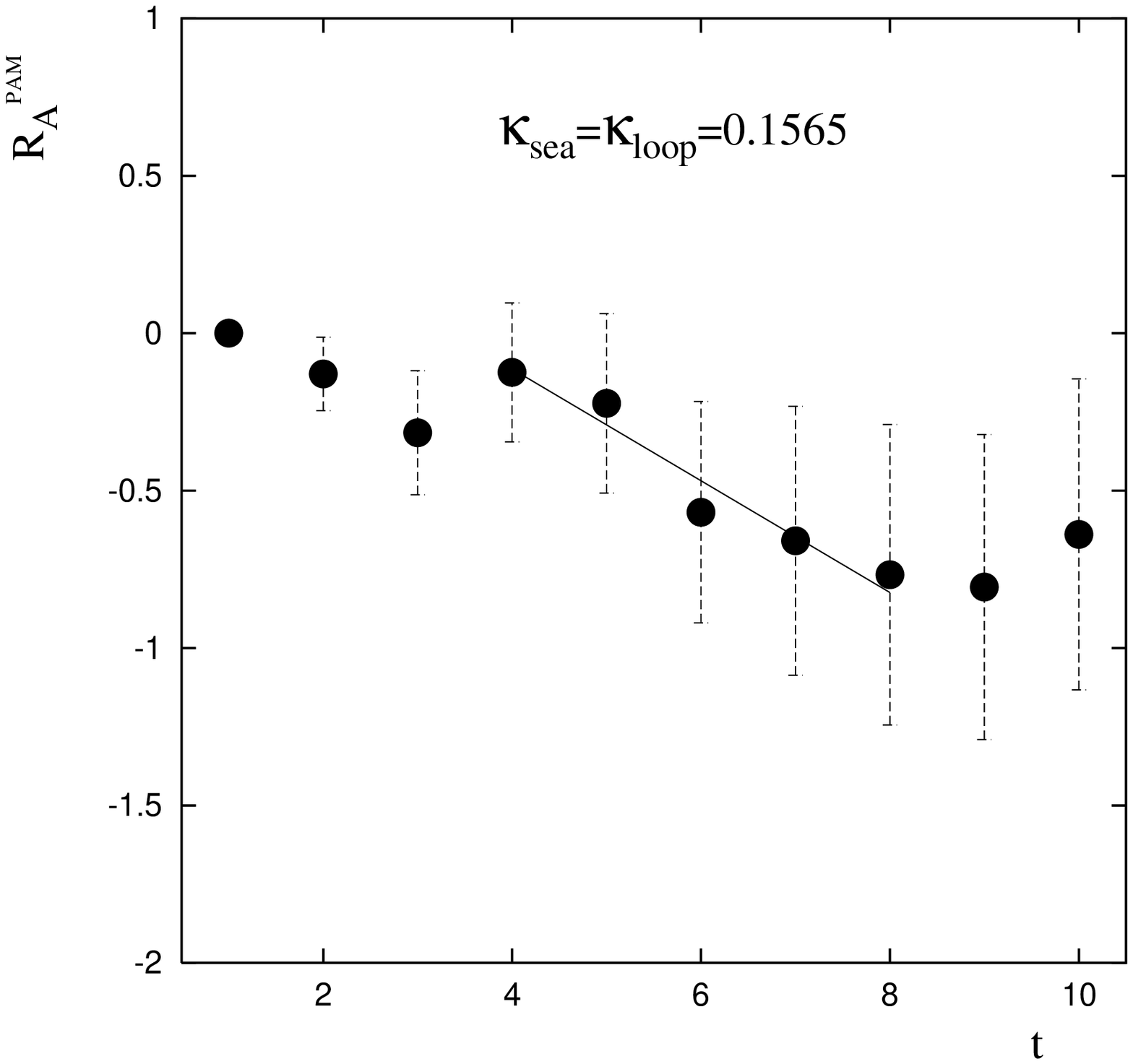}}
\\ \linebreak
\vskip -4.0cm
\parbox{7.0cm}{\epsfxsize=7.0cm\epsfbox{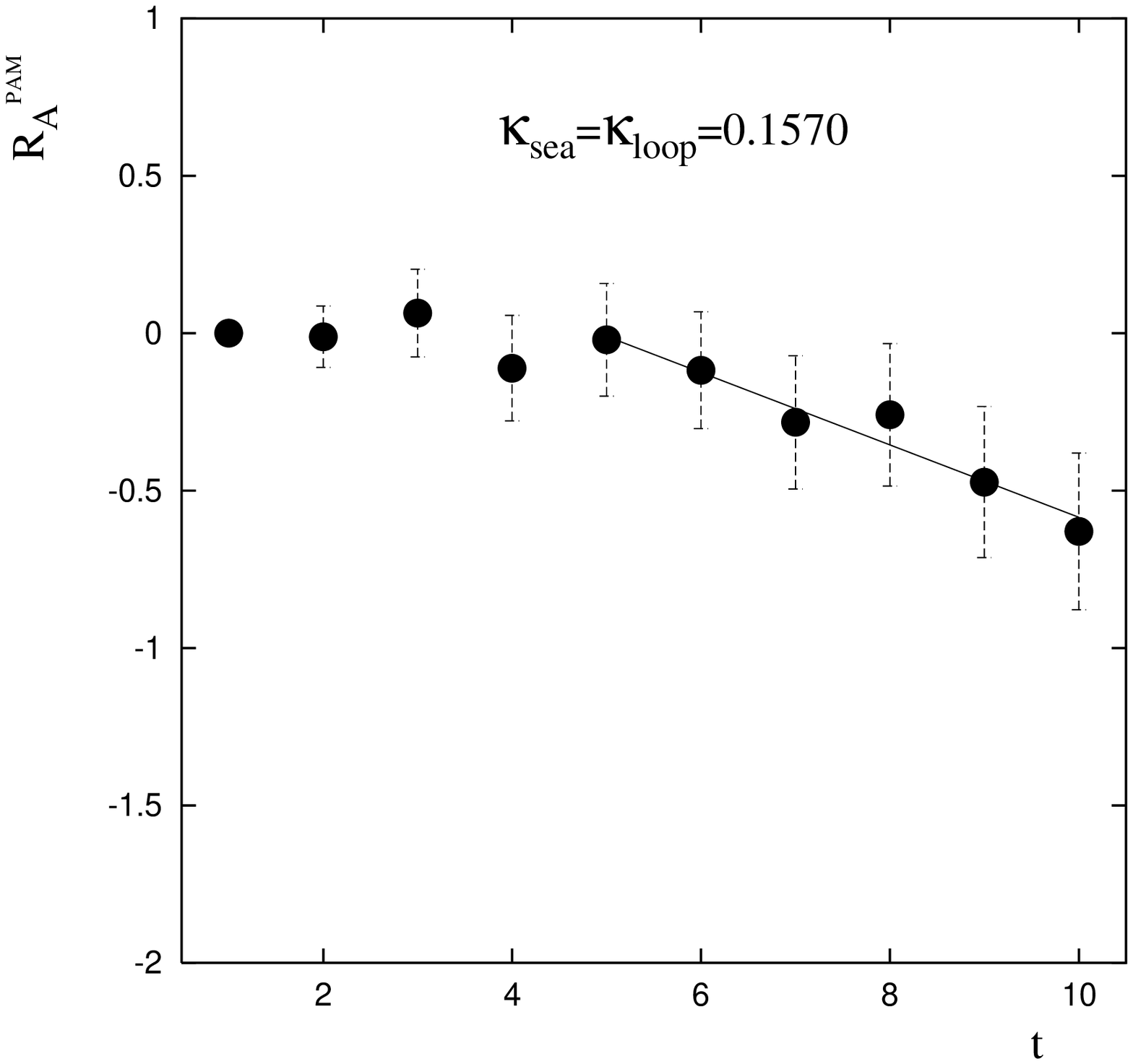}} 
\parbox{7.0cm}{\epsfxsize=7.0cm\epsfbox{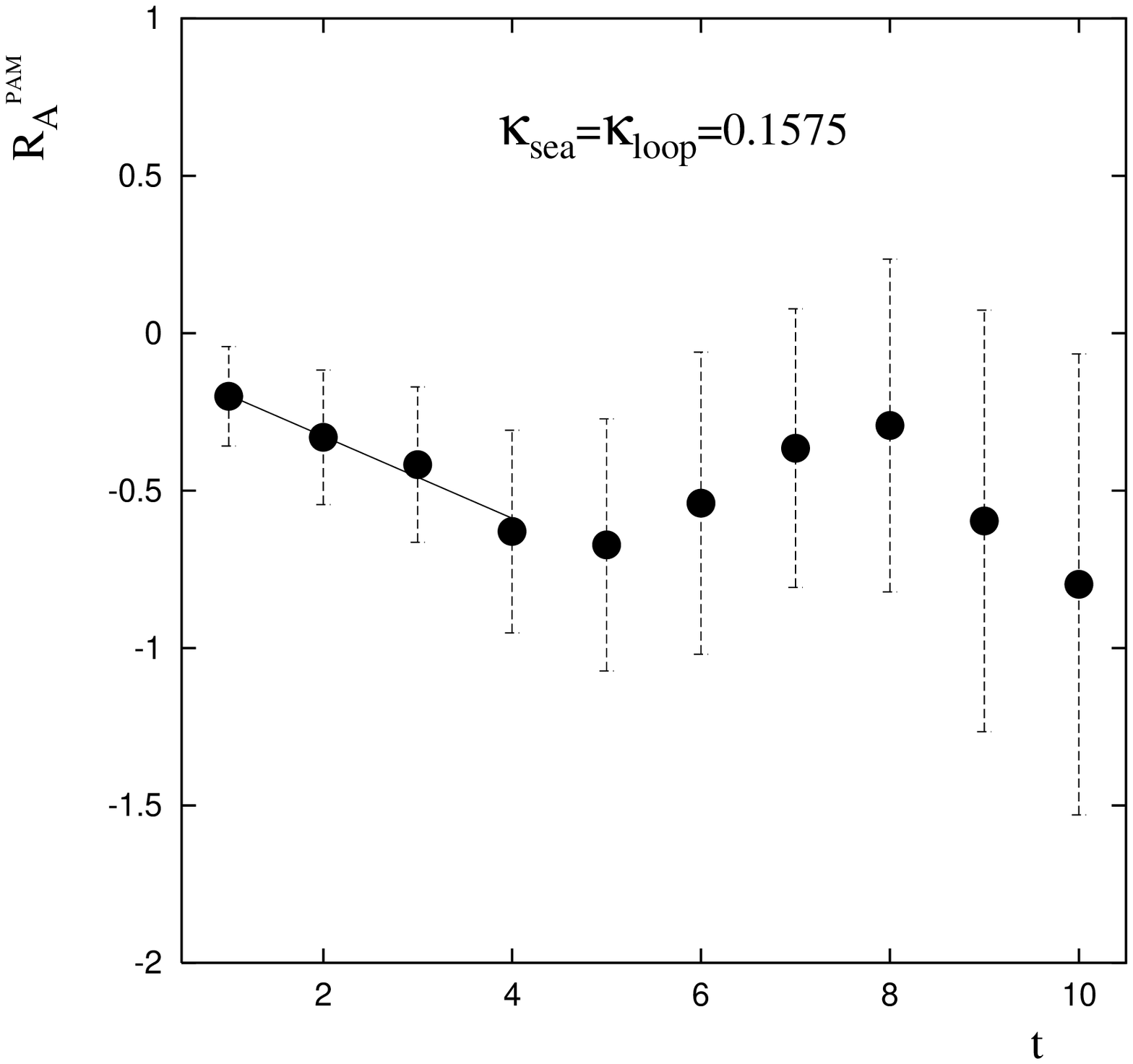}}\\
}
\caption{\label{fig_sesam_ga_disc} {\it Full QCD, SESAM collaboration 
\cite{sesam_ga}: The ratio $R_A^{PAM}(t)$ with
$\Delta {t_0}=\Delta t=1$ for the disconnected amplitudes 
$D_q =\langle P(\kappa_{sea})|
(\bar{q}\gamma_3\gamma_5 q)(\kappa_{sea})| P(\kappa_{sea})\rangle$ at four
sea quark masses. The fits (range and value) are indicated by solid lines.}}
\end{center}
\end{figure}

The disconnected amplitudes have been determined with the 
PAM method, eq. (\ref{eq_ra_pam}). To calculate the numerator of
$R_A^{PAM}$ SESAM has used the spin explicit stochastic estimator
method, see section \ref{sec_direct_meth}, with complex $Z_2$
noise and 100 estimates per spin component. This procedure
has been proven to be much more advantageous than the standard summation
method in calculations of the disconnected contributions to 
$\sigma_{\pi N}$ \cite{sesam_nsigma}. The signals for the axial vector
insertions are shown in fig. \ref{fig_sesam_ga_disc}. Although a
linear behavior can be identified, the
statistical quality of the data is low. This is similar to the
situation found in the quenched calculations with (un-improved) volume 
source and summation methods, see fig. \ref{fig_japan_ga_con_dis}.    
The apparent non-improvement of PAM and spin explicit stochastic
estimator technique might be due to larger statistical fluctuations
in full QCD compared to quenched QCD. As mentioned above, such a
behavior has been recognized
in the analysis of the light spectrum, where, at equal sample size and
lattice setup, the statistical errors in full QCD exceed those of
quenched QCD by more than a factor of 2 \cite{sesam_light_spectrum}. 
 
\begin{figure}[htb]
\begin{center}
\vskip -4cm
\epsfxsize=10.0cm
\centerline{\epsfbox{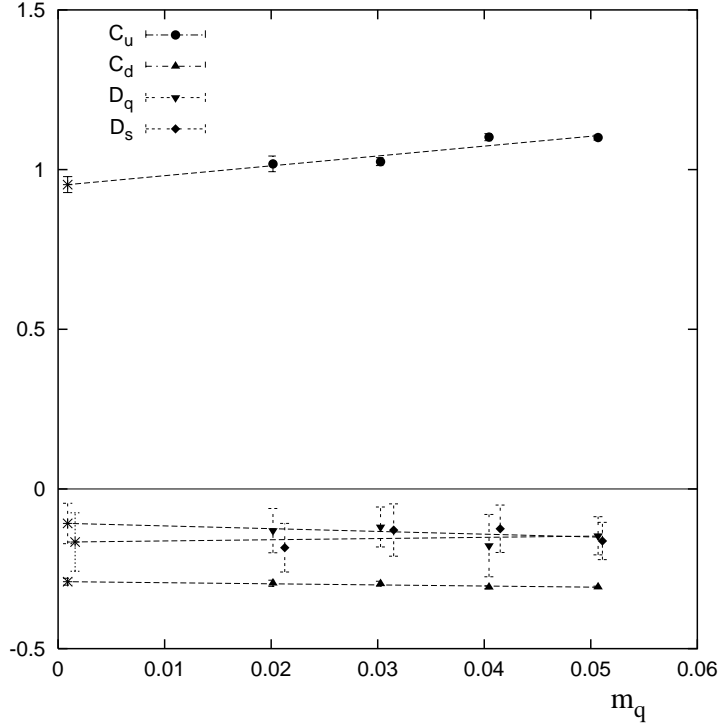}}
\caption{\label{fig_sesam_ga_mq} {\it Full QCD, ref. \cite{sesam_ga}:
Extrapolation of the
unrenormalized lattice amplitudes to the light quark mass. 
The fits are indicated by dashed lines, the results of the fits
by bursts. The amplitudes are defined as follows:
$C_u =s_{\mu}\langle P(m_q)|(\bar{u}\gamma_{\mu}\gamma_5 u)(m_q)|
P(m_q)\rangle_{con}$  (spin averaged);
$C_d =s_{\mu}\langle P(m_q)|(\bar{d}\gamma_{\mu}\gamma_5 d)(m_q)|
P(m_q)\rangle_{con}$  (spin averaged);
$D_q =\langle P(m_q)|(\bar{q}\gamma_{3}\gamma_5 q)(m_q)| P(m_q)\rangle_{disc}$;
$D_s =\langle P(m_q)|(\bar{s}\gamma_{3}\gamma_5 s)(m_s)| P(m_q)\rangle_{disc}$
.}}
\end{center}
\end{figure}
From the analysis of $R_A^{SUM}$ and $R_A^{PAM}$ at several quark
masses one extracts the mass dependence of connected and disconnected
contributions and extrapolates to the light quark mass. The data and
the extrapolations  are compiled 
in fig. \ref{fig_sesam_ga_mq}. Obviously, the full QCD results are not
very different from the quenched estimates of ref. \cite{japan_ga},
which are shown in fig. \ref{fig_japan_ga_mq}. In both cases one finds
a weak mass dependence of the amplitudes. The disconnected insertions
are less than zero, and their flavor symmetry appears to be maintained.    
 
At the light quark mass, the full QCD estimates for the axial vector coupling
constants read
\begin{equation}
G_A^1 = Z_A^{S}\left[C_u + C_d + 2D_q + D_s \right] = 0.20(12) 
\label{eq_sesam_ga1_res}
\end{equation}
and
\begin{equation}
G_A^3 = Z_A^{NS}\left[C_u - C_d \right] = 0.907(20) \;,\;
G_A^8 = Z_A^{NS}\left[C_u + C_d \right] = 0.484(18) 
\label{eq_sesam_ga3_ga8_res}\;,
\end{equation}
where the 1st order tadpole improved renormalization 
\begin{equation}
Z_A^{S} = Z_A^{NS} = \frac{1}{2\kappa}
(1 - \frac{3 \kappa}{4 \kappa_c})
(1 - 0.31\alpha_{\overline{MS}}({ 1 \over a}))
\end{equation}
has been applied\footnote{SESAM has used
$\alpha_{\overline{MS}}({ 1 \over a}) = 0.215$,
$\kappa=\kappa_l=0.15846$ for the light quark insertions
($C_u$,$C_d$,$D_q$), and $\kappa=\kappa_s=0.15608$ for $D_s$.}.   
As in the quenched case at $\beta=5.7$, the estimate of the triplet
coupling differs from the experimental
result by about $30\%$. This points to the presence of sizeable
systematic uncertainties due to insufficient renormalization and
finite cutoff effects. However, the major source of uncertainty in $G_A^1$
is given by the substantial amount of statistical fluctuations.  
 
Within these uncertainties one finds no significant difference
between quenched and full QCD results. This might indicate that the
disconnected contributions to the flavor singlet axial coupling of the
proton are dominated by the gluonic properties of the vacuum.
 
\subsection{Summary}

Within statistical and systematical uncertainties of about $50\%$, 
lattice simulations consolidate the assumption that the influence
of QCD vacuum properties leads to a small value of $G_A^1$.
With the direct method, the latter are reflected in the disconnected
contributions to the flavor singlet axial vector amplitude of the
proton. Since quenched and full QCD calculations yield similar
results, one might suspect that this phenomenon is dominated by
gluons.

Use of the topological method would allow to investigate the influence
of the axial anomaly on the value of $G_A^1$ in more detail. 
In particular one could study explicitely it`s dependence
on the topological sectors of the QCD vacuum.  
However,
as we have pointed out above, this approach suffers from serious
numerical problems, and one has not succeeded yet to  extract
reliable estimates for $G_A^1$ by application of  this technique.\\

Several strategies to improve this situation come in mind.
With the direct method, the accuracy limiting factors are the
statistical noise of the disconnected amplitudes and the neglect of
higher order or non-perturbative contributions to $Z_A^S$. The former
can be reduced by use of larger lattices, where self averaging effects
will help, higher statistics and adapted estimator techniques.
To improve on the latter one needs, at least, a 2nd order calculation
of $Z_A^{S}$ in the framework of lattice perturbation theory.
Even then, as can be seen from the findings of the QCDSF
collaboration, one might have to fight against cutoff effects. Thus, 
a scaling analysis of $G_A^1$ appears to be unavoidable.

To apply the topological method successfully, one needs to generate
(physically) large vacuum configurations at  small cutoff. This
would enable for a calculation of the correlation of the proton
with the topological charge density 
at small quark masses and momentum transfer. Fortunately, a
non-perturbative renormalization procedure for the topological charge
is available. 
  
With current quenched or ($n_f=2$) full QCD simulations, the strange
quark contribution to $G_A^1$ might not be estimated
correctly . In a 
realistic setup one would have to include a strange sea quark. 
From the numerical point of view a lattice simulation with 2 (degenerate)
light quarks and 2 (degenerate) strange quarks appears feasible.

\newpage 
\section{The Mass of the $\eta'$ Meson and the $U(1)$ Problem}

\subsection{The Problem}

In a classical field theory, symmetries of the Lagrangian must be
reflected by symmetries of the physical phenomena which such a theory
is supposed to describe. In a quantum field theory, this is not
necessarily true, since quantum fluctuations of the vacuum can spoil 
the classical symmetries. However, one then expects to obtain
remnants of this symmetry breaking in the physical
observables, e.g. the occurrence of (massless) Goldstone bosons
in case of a spontaneously broken global symmetry and of massive
gauge bosons and Higgs particles if a (local) gauge symmetry is violated. 

The QCD Lagrangian exhibits a 
$U(3)\times U(3) = SU(3)\times SU_{A}(3)\times U(1) \times U_{A}(1)$
global flavor symmetry \cite{glashow_u3_u3} in the chiral limit,
$m_u=m_d=m_s=0$. Most of the implications of this large
symmetry on the physics of strong interaction has been understood
already in the early days of current algebra and the quark model.  

The $SU(3)$ flavor symmetry is reflected in the
multiplet structure of the particle spectrum[\cite{eightfold_way},
whereas the 
$SU_{A}(3)$ axial flavor symmetry is supposed to be broken
spontaneously. The corresponding 
Goldstone bosons can be identified with the 8 light pseudo scalar
mesons ($\pi^{0,\pm}$,$K^{0}$,$\bar{K}^{0}$,$K^{\pm}$,$\eta$), and the
fact that those have non vanishing and non degenerate masses can be 
explained by an additional 
explicit breaking of the $SU(3)$ and $SU_{A}(3)$ symmetries due
to $m_s \neq m_d \neq m_u \neq 0$. \\

The $U(1)$ flavor symmetry predicts baryon number conservation, which  
is well realized in nature. Likewise, one would expect an
`{\it axial} baryon number conservation', i.e. a parity doubling of baryon
states, from the $U_{A}(1)$ symmetry. Since the particle spectrum does
not exhibit such a parity doubling, one would suppose that this
symmetry is broken spontaneously. However, there is no light
pseudo scalar meson which could be identified with the corresponding
Goldstone boson. The $\eta'$ meson has the correct quantum numbers, but
it is simply too heavy. This is the so called $U(1)$ problem of QCD:
The $U_{A}(1)$ symmetry of the QCD Lagrangian seems to be broken by
quantum fluctuations, but apparently there is no remnant of this
symmetry breaking visible in the particle spectrum.  

This problem is nearly as old as QCD itself. After 
Glashow\cite{glashow_u3_u3} had noted in 1968 the cogency
of a pseudo scalar meson with a mass of the order of the
pion mass, it was shown by Weinberg\cite{weinberg_m_eta}, using
conventional soft pion methods, that this mass should indeed be smaller than
$\sqrt{3}\times m_{\pi}$.

Since then a natural solution to the $U(1)$ problem has been
achieved\footnote{For reviews on the $U(1)$ problem see
\cite{coleman_u1,wilczek_u1,thooft_rev}.} by the study
of the connection between the axial anomaly, topological excitations
of  the vacuum and instantons\cite{thooft_eta,instantons_others}. 

The physical picture which emerges from this work is (roughly) as follows.
The conservation of the $U_A(1)$ symmetry current $A_{\mu}^0$, 
eq. (\ref{eq_a0_def}), is formally
violated by the presence of the axial anomaly
$2 n_f \frac{\alpha_s}{4\pi} tr G_{\mu\nu}\tilde{G}^{\mu\nu}$, c.f.
eq. (\ref{eq_axvec_ward}). One might wonder however, how such a
contribution can make a physical effect at all, if one assumes {\it random}
quantum fluctuations of the gauge fields. In this case, a possible
contribution from the axial anomaly should just average away.

A possible solution to this 
problem came from G.'t Hooft \cite{thooft_eta}, who showed that
non-trivial topological
excitations of the vacuum, instantons, can produce a non vanishing
contribution to physical observables.
  
In this scenario, the
$U_A(1)$ symmetry breaking would not be spontaneous and thus
one would not expect to find a corresponding pseudo scalar 
Goldstone boson\footnote{It has been argued frequently, 
see e.g. ref.\cite{christos_u1}, that, even in the presence of the
axial anomaly, it is still possible to define
a conserved flavor singlet axial vector current. Thus, the $U(1)$ 
symmetry would be broken spontaneously and there should still exist
a corresponding Goldstone boson. Since this conserved current
is not gauge invariant, the Goldstone boson could be
realized as a Kogut-Susskind
ghost\cite{kogut_pole} in the unphysical, gauge variant sector of QCD.
The physical interpretation of such an unphysical construction however
is by far not clear and requires at least additional assumptions about
the nature of the vacuum\cite{thooft_rev}.}. Remnants of this
new kind of symmetry breaking would
however still be observable in nature. For example 
the $\eta'$ meson would acquire it's large mass from instanton
contributions. If it were possible to switch off the topological vacuum
excitations, the $\eta'$ meson should become light and would behave
like an `ordinary ' Goldstone particle.\\  

Indeed, such a behavior has been obtained in the framework of a
$1/N_c$ expansion, $N_c$ being the number of colors, by 
Witten \cite{witten_eta} and 
Veneziano\cite{veneziano_eta}. They showed that in the limit
$N_c \rightarrow \infty$, where the axial anomaly is absent, the
masses of $\eta'$ and $\pi$ become degenerate. In first order $1/N_c$
this degeneracy is lifted and, in an $SU(3)$ flavor symmetric world,
one obtains
\begin{equation}
m_{\eta'}^2 - m_{\pi}^2 = m_0^2 = \frac{2 n_f \chi}{f_{\pi}^2}\;.
\label{eq_witten_veneziano}
\end{equation} 
Here $n_f$ denotes the number of flavors and $\chi$ is the
topological susceptibility defined in eq. (\ref{eq_topsus_def}), at
$k^2=0$. Taking  into account an explicit $SU(3)$ flavor symmetry
breaking and neglecting the effects of a (small) $\eta-\eta'$ mixing,
one would expect
\begin{equation}
m_0^2 \simeq m_{\eta'}^2 - (4m_K^2 + 3m_{\pi}^2 + m_{\eta}^2)/8
\simeq (867\mbox{MeV})^2\;,\; \chi \simeq (182\mbox{MeV})^4\;.
\label{eq_m0_phys}
\end{equation}   

In summary, there is a lot of (theoretical) evidence, that the
topological properties of the QCD vacuum are indeed likely to solve
the $U(1)$ problem.  
The question remains however, whether this qualitative evidence
holds also quantitatively.  
To answer this question one would first
have to show that the large $\eta'$ mass, $m_{\eta'}=958$MeV,  is
really a property of QCD. Secondly, one would like to see that
the topological excitations in full QCD are such that they 
yield an additional contribution to $m_{\eta'}$ of size $m_0\simeq 867$MeV.
Last not least one would have to study the instanton content of QCD
in detail to learn about the mechanism of how the $\eta'$ acquires it's
mass. \\

Being a non-perturbative method, lattice gauge theory is perfectly suited to
study quark-antiquark bound states like octet and singlet mesons.
It has been argued in ref's. \cite{wilczek_u1,thooft_rev} that 
the contributions of topological vacuum excitations could be
overlooked by standard perturbative expansions. Clearly, with lattice QCD  
this flaw is absent. Within the course of a Monte Carlo
process, QCD vacuum configurations are explicitly 
generated\footnote{One has to ensure of course that the Monte Carlo
algorithm tunnels sufficiently often between different topological
sectors \cite{sesam_q_wilson}}. Thus,
one has direct access to topological properties of each single
configuration. This enables, in principle, for a detailed study of the
connection between instantons, topological excitations and the
$\eta'$ mass. 
    
\subsection{Lattice Technique} 

The mass of a (full) QCD bound state can be extracted from the asymptotic
behavior of the momentum zero projection of it's propagator.

Let us consider, for example, a one flavor pseudo scalar meson
with an interpolating operator $P=\bar{q}\gamma_5 q$. The time
dependence of it's (momentum zero) propagator is 
given by\footnote{This is
true for an infinitely extended lattice in time. For a finite lattice
with time extend $T$ and (anti)periodic boundary conditions the
asymptotic form is given by 
$\lim_{t \rightarrow \infty}G(0,t) = |\langle 0|P^{\dagger}|n_0 \rangle|^2
\left[e^{-m_{n_0}t} + e^{-m_{n_0}(T-t)}\right]$. For the sake of
clarity we will however restrict the discussion to infinitely extended
lattices.} 
\begin{equation}
G_P(0,t) = \sum_{\vec{x}}\langle P^{\dagger}(\vec{0},0) P(t,\vec{x})\rangle =
\sum_n |\langle 0|P^{\dagger}|n \rangle|^2 e^{-m_n t}\;
\stackrel{t\rightarrow \infty}{\longrightarrow}\;
|\langle 0|P^{\dagger}|n_0 \rangle|^2 e^{-m_{n_0} t}\;,
\label{eq_mesprop_tdep}
\end{equation}  
where $n_0$ is the lowest eigenstate of the QCD
Hamiltonian with quantum numbers $J^{PC}=0^{-+}$.

Alternatively one can express $G_P(0,t)$ in terms of quark operators
\begin{equation}
G_P(0,t) = 
\sum_{\vec{x}}\langle (q\gamma_5 \bar{q})(\vec{0},0)
                     (\bar{q}\gamma_5 q)(\vec{x},t)\rangle\;,
\end{equation}
or, performing all possible contractions of quark operators, in terms
of quark propagators $\Delta_q$, defined in eq.(\ref{eq_qprop_def})
\begin{eqnarray}
\lefteqn{G_P(0,t) = }&& \label{eq_mesprop_qprop} \\
&& \sum_{\vec{x}} \left\{
\langle Tr(\Delta_q \Delta_q^{\dagger})(0,t)\rangle -
\langle Tr(\Delta_q\gamma_5)(0,0) Tr(\Delta_q\gamma_5)(t,t) \rangle
\right\} \equiv C_P(0,t) - D_P(0,t)\;. \nonumber
\end{eqnarray}     
Thus, a one flavor meson propagator has always two
contributions: The first term on the r.h.s. of
eq.(\ref{eq_mesprop_qprop}), $C_P$,  is called
`connected', since the meson states at time 0 and time $t$ are
explicitly connected by quark lines. 
Correspondingly, the second term, $D_P$,
is called `disconnected'. Note that this term violates the OZI rule.
It is highly sensitive to the details of the vacuum structure since
the correlation between the quark loops $Tr(\Delta_q\gamma_5)$ at 
time 0 and $t$ is mediated by quantum effects (multiple gluons and
internal quark loops). Both contributions are schematically depicted
in fig.\ref{fig_schema_eta}.\\

The physical light mesons are not made of one quark flavor. They are
rather 2-flavor triplet (pions), 3-flavor octet (Kaons,$\eta$) or 3-flavor
singlet ($\eta'$) combinations of quarks. Just from combinatorics
one expects the
disconnected contributions to cancel each other largely for the non singlet 
combinations\footnote{Up to a presumably negligible flavor symmetry
breaking, they cancel exactly.}, and to contribute additively
to the singlet state.

From this point of view, one can identify the connected contribution
with the normal `Goldstone part' of the pseudo scalar mesons and the
disconnected contribution with the anomalous `topological part'.\\

We could now proceed and calculate $G_P$ according to
eq.(\ref{eq_mesprop_qprop}) on a representative sample of gauge
configurations. Then one would extract $m_{\eta'}$ from a fit to its
asymptotic form, eq.(\ref{eq_mesprop_tdep}). However this might not be
advantageous for several reasons. First, one would like to
concentrate on the (anomalous) mass contribution from the
disconnected part. Secondly, one should try to keep statistical
errors, which arise mostly from the large fluctuations of he
disconnected part, 
as small
as possible. For both these reasons one favors to consider the
ratio 
\begin{equation}
R_{DC}(t) = \frac{n_f D_P(0,t)}{C_P(0,t)} \;,
\end{equation} 
rather than to $G_P$ itself. The factor $n_f$ (number of flavors) has
been inserted in the 
numerator  because a  $n_f$-flavor singlet meson
has $n_f$ contributions $C_P$, but $n_f^2$ contributions $D_P$.

In full QCD, the asymptotic time dependence of $G_P$ and $C_P$
is given by eq.(\ref{eq_mesprop_tdep}). Thus, it follows
\begin{equation}
R_{DC}(t) = 1 - C e^{(m_{\eta'} - m_8)t}
\label{eq_rdc_full}
\end{equation}
at large $t$.
Here we have identified  mass of the full singlet propagator ($G_P$)
with $m_{\eta'}$ and the mass of the `Goldstone part' ($C_P$) with
$m_8$. Note that $R_{DC}$ is sensitive to the difference of these
masses. The `Goldstone mass' $m_8$ vanishes in the chiral
limit. Therefore, one can extract $m_0$, see
eq. (\ref{eq_witten_veneziano}), from an extrapolation to this point.\\

The situation is slightly more complicated in quenched QCD. In this
case eq.(\ref{eq_mesprop_tdep}) does not describe the asymptotic
time dependence of $G_P$ correctly since higher quark loop 
contributions to the singlet propagator are not taken into account.
In the $1/N_c$
expansion of Witten and Veneziano \cite{witten_eta,veneziano_eta},
which in first order leads to eq.(\ref{eq_witten_veneziano}), the 
$\eta'$ propagator in momentum space, $\tilde{G}_P$, is described by
an infinite sum over virtual quark loops, connected by an effective gluonic
coupling $m_0^2$
\begin{equation}
\tilde{G}_P(p^2) \sim \frac{1}{p^2 + m_8^2} 
                     - \frac{m_0^2}{(p^2 + m_8^2)^2}  
                     + \frac{m_0^4}{(p^2 + m_8^2)^3} 
                     - \cdots \; = \frac{1}{p^2 + m_8^2 + m_0^2} \;.  
\label{eq_witten_geom}
\end{equation}
\begin{figure}
\epsfxsize=15cm
\epsfbox{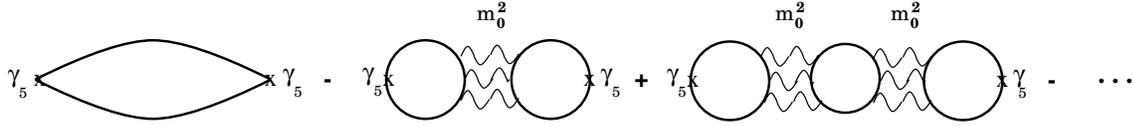}
\caption{\label{fig_schema_witten}{\it Contributions to the flavor
singlet propagator in the scheme of Witten and Veneziano.
 }} 
\end{figure}
This is shown in fig.\ref{fig_schema_witten}. The quenched
approximation neglects virtual quark loops, which corresponds to a
truncation of the geometric series in eq.(\ref{eq_witten_geom}) after
the second term. Obviously, the first term can be identified with the
connected contribution to the truncated propagator, whereas the second
term belongs to it's disconnected part. A Fourier transform of both
terms yields\footnote{Using the Witten-Veneziano expansion one can
also estimate the amplitude $C$ in eq.(\ref{eq_rdc_full}). One finds
$C=m_8/m_{\eta'}$.}
\begin{equation}
(R_{DC}(t))^{quen} = A + \frac{m_0^2}{2m_8}t\;,\,A=\frac{m_0^2}{2m_8^2}\;.
\label{eq_rdc_quen}
\end{equation}
Thus, under the assumption that the Witten-Veneziano expansion is
valid, one can determine $m_{\eta'}$ even in the quenched
approximation from a linear fit to  $(R_{DC}(t))^{quen}$.
It is
however by no means obvious, that quenched and full QCD calculations
should yield the same value for $m_{\eta'}$. Full QCD simulations with
reliable statistics are therefore urgently needed. 

A peculiar situation arises in the partially quenched approximation,
i.e. in the case where not all valence quarks have their counterparts
in the quark sea. Such a setting can be useful if one wants to exploit
gauge configurations with light sea quarks with respect to the
strange quark sector of QCD or if one works with $n_f=2$ dynamical
fermions in the Kogut-Susskind discretization. 
The form of the corresponding $\eta'$ propagator, i.e. the analogon
to eq.(\ref{eq_witten_geom}), has been worked out by Bernard and
Golterman  in the framework of the
partially quenched chiral perturbation theory \cite{bernard_golterman}.\\

The numerical techniques which are used to calculate the
connected and disconnected contributions to the $\eta'$ propagator,
eq. (\ref{eq_mesprop_qprop}), are quite similar to those already
explained in the previous sections.  
The connected part can be composed from solutions to
eq.(\ref{eq_inversion_simple}), and for the disconnected contribution
one can use the estimator techniques discussed in sections
\ref{sec_what_has} and \ref{sec_direct_meth}.  
We emphasize again that the computationally expensive and
statistically noisy part of this calculation is the disconnected
one. In order to gain
statistics one exploits the translational invariance of the
propagator, $G(0,t) = G(t_0,t_0+t)$, and performs the average over all
timeslices $N_t$, $D_P(0,t) = 1/N_t \sum_{t_0}D_P(t_0,t_0+t)$.\\ 
 
We mentioned at the beginning of this section that the calculation of
$m_{\eta'}$ and the proof that it does not vanish in the
chiral limit are not sufficient to proof that the 't Hooft scenario of
the $U(1)$ problem is realized in QCD. One
rather has to demonstrate that non-trivial topological excitations of the
vacuum really exist in QCD, and that those excitations are responsible
for the large mass of the $\eta'$ meson. 

One method to study the connection between topology and $m_{\eta'}$ is
to determine the topological charge $Q$, see
eq.(\ref{eq_topol_charge_def}), in addition to  $m_{\eta'}$ and $m_0$,
on a given set of gauge configurations\footnote{Lattice definitions
  and numerical methods to determine $Q$ have been discussed in 
section \ref{sec_topol_meth}}. With this information in hands
one can analyze the  dependence of $m_{\eta'}$ (or equivalently
of $m_0$)  on $Q$.
A finding $m_{\eta'}= m_8$, ($m_0=0$), at $Q=0$ and
$m_{\eta'}\neq m_8$, ($m_0\neq0$),    
at $Q\neq0$ would favor strongly the `topological solution' 
of the $U(1)$ problem.\\   
   
Additional information on this connection can be extracted from the
lowest eigenmodes of the Dirac operator.
The Atiyah-Singer index theorem \cite{atiyah_singer} 
\begin{equation}
Q = n^+ - n^- 
\label{eq_atiyah_singer}
\end{equation}
states that the topological charge $Q$ in the continuum is given
by the difference between the number of positive ($n^+$) and negative ($n^-$)
chirality eigenmodes $\Psi^{\pm}$ of the Dirac operator with zero eigenvalue
\begin{eqnarray}
D\!\!\!\!/ \,\Psi^{\pm}&=&
\gamma^{\mu}(\partial_{\mu} - i g
\frac{\lambda_a}{2}A^a_{\mu})\Psi^{\pm} =
i\lambda \Psi^{\pm}\quad,\quad \lambda=0 \\
\gamma_5 \Psi^{+} &=& + \Psi^{+}\;,\; \gamma_5 \Psi^{-} = - \Psi^{-}\;.
\end{eqnarray}
Thus , if $m_0$ is determined by topology, one would expect that
the related correlator, i.e. the disconnected part of
eq.(\ref{eq_mesprop_qprop}), could be calculated using only the zero
modes of the Dirac operator, whereas the latter should be insufficient
for a complete determination of the connected part.

Unfortunately, the Atiyah-Singer theorem does not hold (exactly) in a
discrete formulation (Wilson or Kogut-Susskind) of QCD on a finite
lattice. This is due to the fact that the fundamental property
$\gamma_5 D\!\!\!\!/ = -D\!\!\!\!/ \, \gamma_5$  of
the (anti hermitian) Dirac operator, is not fulfilled 
in the discretized case. It follows that, in principle, all eigenmodes of the
lattice Dirac operator can contribute to the topological charge.

 It has been shown however, by Smit and Vink \cite{q_smit_vink} and
Itoh, Iwasaki and Yoshi\'{e} \cite{q_itoh}, that the Atiyah-Singer theorem
holds still approximately on the lattice. Instead of taking into
account only the zero eigenmodes one now needs the `lowest'
eigenmodes to make up for the topological charge. Since  it is not
a priori known how many of these `lowest' eigenmodes are really
needed, one has to study  the influence of the number
of lowest eigenmodes on
the connected and disconnected contributions to the $\eta'$
propagator numerically.
 
We mention that the determination of the spectrum of the Dirac
operator on a (large) lattice represents a non-trivial numerical
problem. Algorithms suited for this task are explained in 
ref's. \cite{lanczos,bbunk,arnoldi_kstep,gattringer}.
 
\subsection{Lattice Results}

The `topological' scenario to explain the $U(1)$ puzzle
has been proposed more than 20 years ago, but it is  only in
these days that the lattice method may become powerful enough to prove
that the proposed solution is actually realized in QCD. This is mostly due
to the substantial computational effort which is necessary to
calculate the disconnected contributions to the $\eta'$ propagator
with reliable statistics.

Nevertheless there is already a handful of pioneering simulations
in quenched QCD and, to an even lesser extend, in full QCD,
which find promising evidence that the `t Hooft mechanism explains
the large mass of the $\eta'$ meson.\\ 

After early exploratory studies \cite{q_itoh,fukugita_early}
from the time where
estimator techniques for the disconnected part of the $\eta'$
propagator were not yet available, the first quenched simulation
with reliable statistics (200-300 gauge configurations) was performed
by the authors of ref. \cite{japan_eta}, using Wilson 
fermions\footnote{They chose quark masses corresponding to
$m_{\pi}/m_{\rho}= 0.74,0.70,0.59,0.42$. Unfortunately, the results 
for the lightest quark mass were achieved with only 40 gauge
configurations.}
on a
$12^3 \times 20$ lattice and a somewhat large lattice spacing,
$a \simeq 0.14$fm ($a^{-1}=1.45(3)$GeV, $\beta=5.7$).
They employed the volume source
technique, see eq's. (\ref{eq_walls_M}),(\ref{eq_walls_sumM}), to
calculate the disconnected contribution.
\begin{figure}[htb]
\begin{center}
\vskip -3.0cm
\noindent\parbox{15.0cm}{
\parbox{7.0cm}{\epsfxsize=7.0cm\epsfbox{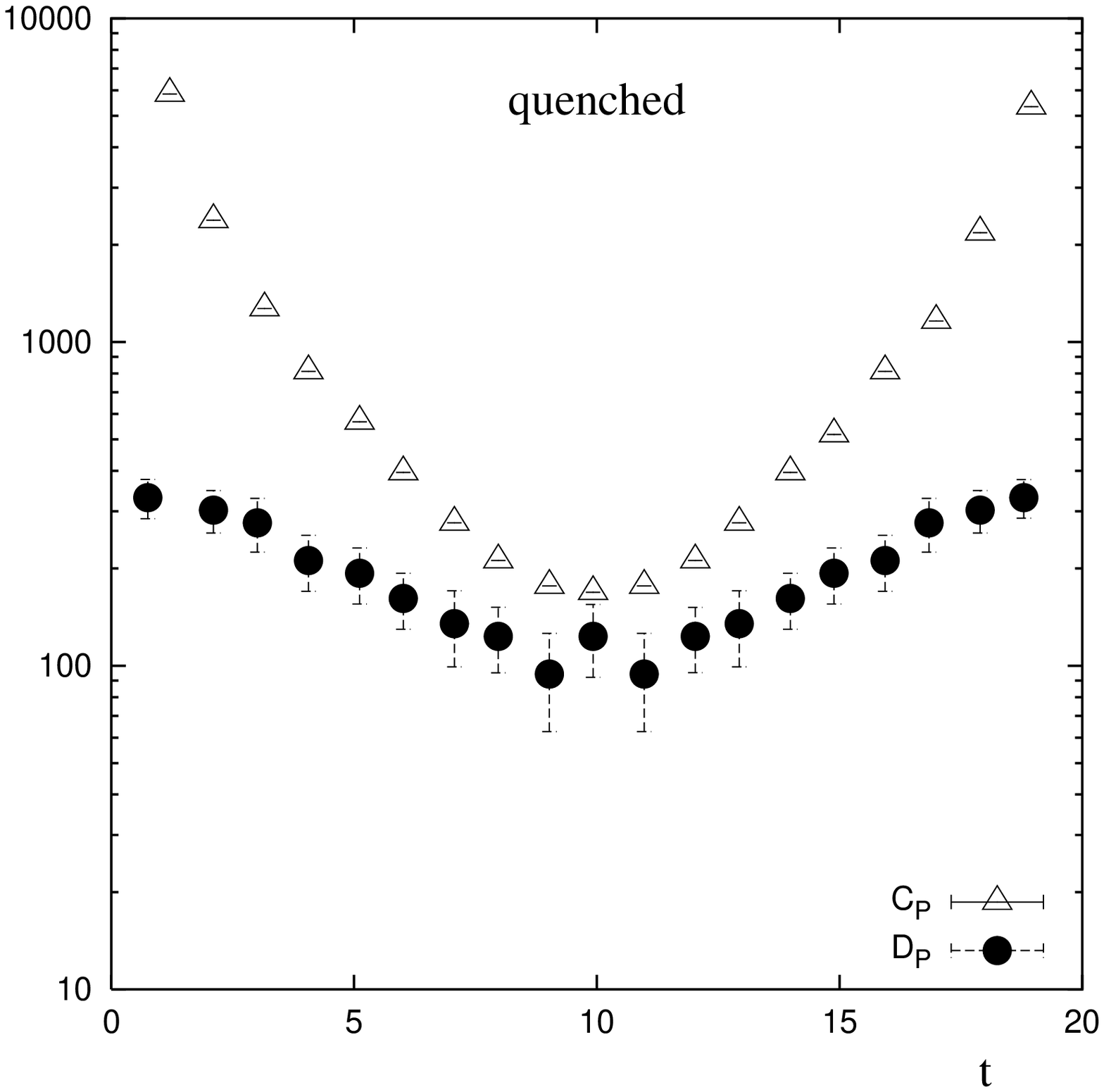}}
\parbox{7.0cm}{\epsfxsize=7.0cm\epsfbox{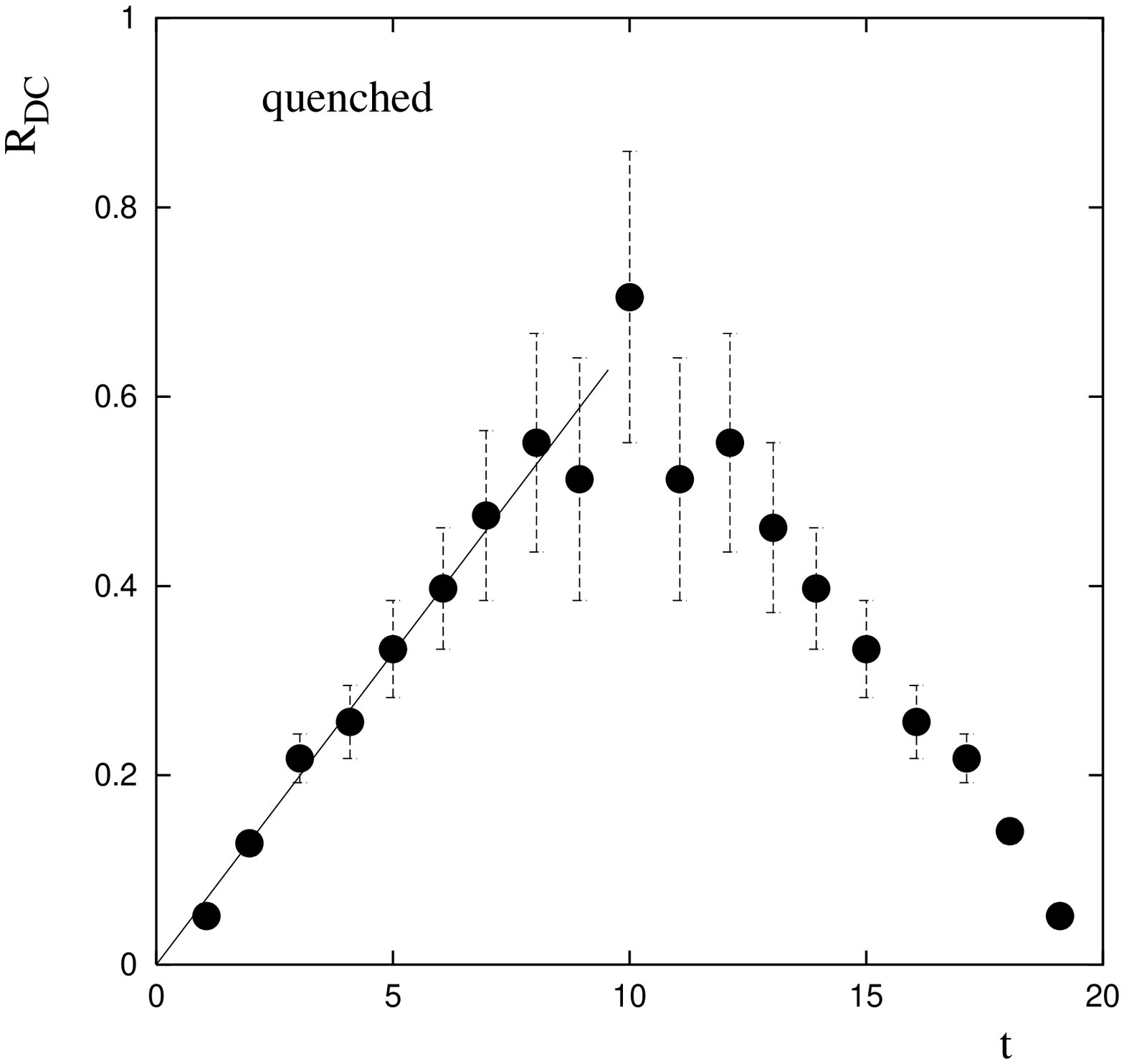}}
}
\caption{\label{fig_japan_eta_signals} {\it Left: Connected ($C_P$) and
disconnected ($D_P$) contributions to the $\eta'$ propagator at a 
quark mass corresponding to $m_{\pi}/m_{\rho}=0.59$. Right: The
ratio $R_{DC}$ at the same quark mass. The solid line indicates the
result of a linear fit to $R_{DC}$ in the range $4 \leq t \leq 8$.
The data has been extracted from ref.\cite{japan_eta}.
}}
\end{center}
\end{figure}

We illustrate the
statistical quality of the corresponding signals in
fig.\ref{fig_japan_eta_signals}, where we display connected and
disconnected contributions to the $\eta'$ propagator as well as 
the ratio $R_{DC}$. The disconnected part is much more noisy than the
connected one, as expected. The ratio $R_{DC}$ is reasonably
consistent with the predicted linear behavior in the quenched
approximation, see eq. (\ref{eq_rdc_quen}).
From this data, one can
extract $m_0$ at a given quark mass. To mimic the physical situation 
of $n_f=3$ quarks, one multiplies $m_0$ by $\sqrt{3}$. Clearly, such
a procedure is justified only in a flavor symmetric situation. 

The physical value for $m_0$ in this flavor symmetric world, i.e. it's
value at the light quark mass\footnote{The light quark mass
$m_l$ (in lattice units) 
can be determined by the condition $m_{\pi}/m_{\rho} = 0.1766$.
Alternatively one can extrapolate to the chiral point $m_l=0$. The
difference between these methods with respect to $m_0$ is negligible.
}
can be read off from the result of a fit to $m_0(m_q)$.

\begin{figure}[htb]
\begin{center}
\vskip -4cm
\epsfxsize=10.0cm
\centerline{\epsfbox{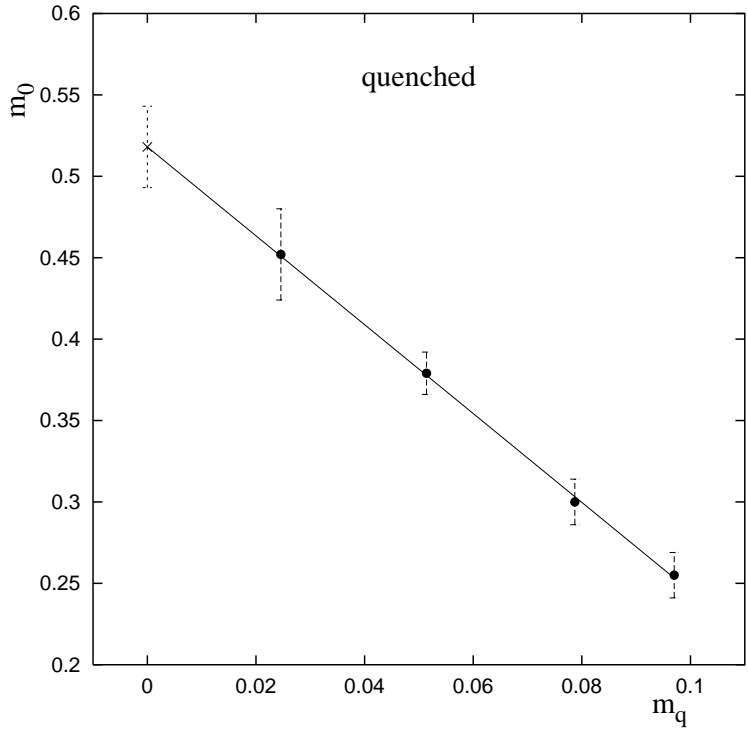}}
\caption{\label{fig_japan_eta_m0_extrap} {\it Ref. \cite{japan_eta}:
Dependence of $m_0$ on the bare quark mass $m_q$.
The solid line shows 
a linear fit to $m_0(m_q)$. The cross indicates the result of the fit 
at the chiral point. It's value in lattice units is $m_0=0.518(25)$,
which corresponds to $m_0 = 751(39)$MeV. Note that all $m_0$ data 
have been multiplied by $\sqrt{n_f}=\sqrt{3}$.
}}
\end{center}
\end{figure}
As shown in fig. \ref{fig_japan_eta_m0_extrap}, the authors of
ref. \cite{japan_eta} find a 
linear increase of $m_0(m_q)$ with decreasing $m_q$.
Note that this is in contrast
to the behavior of non-singlet pseudo scalar masses, where one 
obtains a linear decrease of $m_P^2(m_q)$. At the chiral point, $(m_q=0)$, 
they get $m_0 = 751(39)$MeV, where the number in brackets denotes the
statistical error. 

One should, for various reasons, not be worried too much about the
mismatch ($2-3\sigma$) of  this result with the physical estimate,
see eq.(\ref{eq_m0_phys}). As we mentioned above, the disconnected
part has been obtained with the volume source technique, which does
not allow to monitor the cancellation of the unwanted gauge variant
parts. Furthermore, the simulation has been performed  at a coarse
lattice spacing, but the influence of
finite cutoff effects has not been taken into account. 
Moreover, one has assumed flavor symmetry with respect to $m_0$,
which might not be realized in nature.
Last not least, the result has been achieved
in the quenched approximation. Since vacuum contributions are
essential in the determination of $m_0$, this could introduce a
systematic bias.
For all these reasons one should expect that the true error on $m_0$ 
is actually much larger.

The good news from the findings of ref.\cite{japan_eta}
is of course that $m_0$ is in the
right range. Thus, there is a good chance that, once all the
uncertainties listed above have been removed, one will indeed obtain
the experimental $\eta'$ mass in full QCD lattice calculations.\\

Given that, one can proceed to study the connection between vacuum
topology and $m_0$. The authors of ref.\cite{japan_eta_q} have
calculated the topological charge $Q$ according to the standard 
definition\footnote{The definition is applied to `cooled'
configurations. The cooling method is explained in 
ref. \cite{hoek_cooling}.}, see eq.(\ref{eq_q_latt_def}), on the same
set of gauge configurations as used in ref.\cite{japan_eta}, at a quark mass
corresponding to $m_{\pi}/m_{\rho}=0.59$. Thus one can analyze the
dependence of $m_0$ on $Q$ for this setup.  
\begin{figure}[htb]
\begin{center}
\vskip -3.0cm
\epsfxsize=11.0cm
\epsfbox{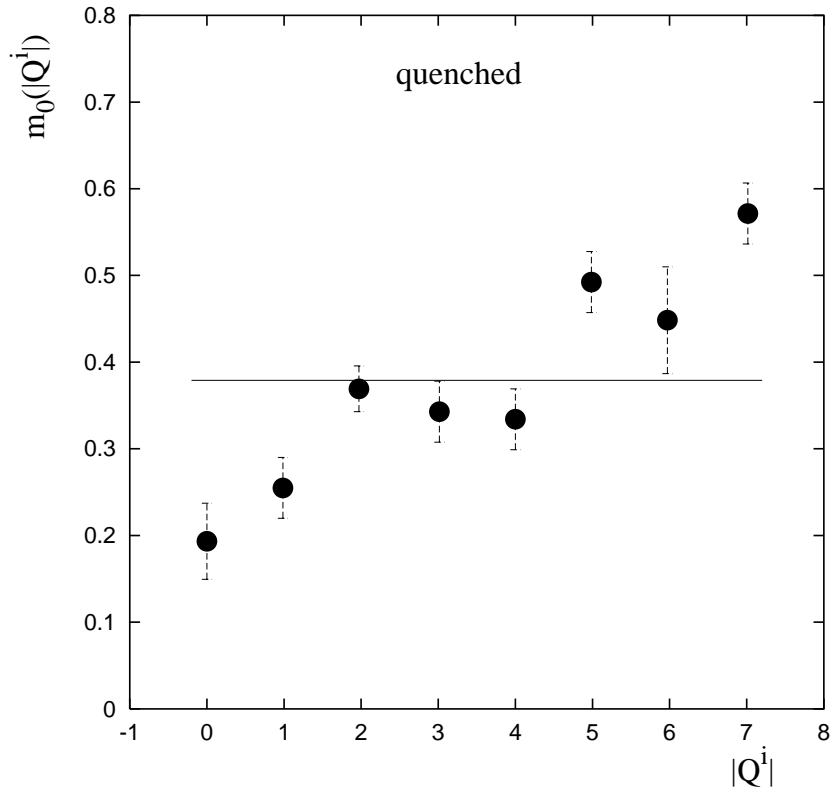}
\caption{\label{fig_japan_m0_q} {\it $m_0$ in lattice units as a function 
of the `integerized' topological charge $Q^i$, at a 
quark mass corresponding to $m_{\pi}/m_{\rho}=0.59$.
The solid line indicates the 
result for the entire ensemble of gauge configurations.
The data has been extracted from ref.\cite{japan_eta_q}.
}}
\end{center}
\end{figure}
We display the result of
ref. \cite{japan_eta_q} in fig.\ref{fig_japan_m0_q}. It appears that 
$m_0$ correlates to the value of $Q$:
$m_0$ increases as $Q$ increases. This indicates that the proposed
solution of the $U(1)$ problem is realized in QCD.
Unfortunately, the authors of ref.\cite{japan_eta_q} have used an 
`integerized' value of $Q$ instead of a renormalized one, i.e. they
set the lattice value of $Q$ to the nearest integer instead of
applying the renormalization procedure described in 
section \ref{sec_topol_meth}. However, it is unlikely that this
inaccuracy alters the conclusions of their work.  \\
  
In order to consolidate this result one needs to perform lattice
simulations which avoid the insufficiencies discussed above. As this
is, after all, a question of compute power and of the capacity of
stochastic algorithms, this goal can be achieved only in small steps.

Recently, Venkataraman and Kilcup \cite{kilcup} analyzed full
QCD gauge configurations with 2 flavors of Kogut-Susskind fermions
with respect to the $\eta'$ mass.
The configurations have been generated by the QCDSP
collaboration \cite{columbia,qcdsp} on $16^3 \times 32$ lattices, at 3
different values of the sea quark
mass, corresponding to $m_{\pi}/m_{\rho} = 0.704, 0.644$, and 0.572,
and at a lattice cutoff $a^{-1} \simeq 2$GeV. The disconnected part of
the $\eta'$ propagator has been calculated using a stochastic
estimator method, and `Wuppertal smearing' \cite{smearing_wtal} has
been employed to improve the ground state projection of the operators.
In addition, the authors of ref.\cite{kilcup} have performed a
quenched simulation at equal lattice spacing and volume. This allows
for a direct comparison of quenched and unquenched results.
Thus, this analysis has installed quite a number of improvements to
alleviate the deficiencies of ref's.\cite{japan_eta,japan_eta_q}.
Unfortunately, the size of the samples of gauge configurations used
here is rather limited (34-79 configurations for full QCD and 83
configurations for quenched QCD).   
\begin{figure}[htb]
\begin{center}
\vskip -3.0cm
\epsfxsize=11.0cm
\epsfbox{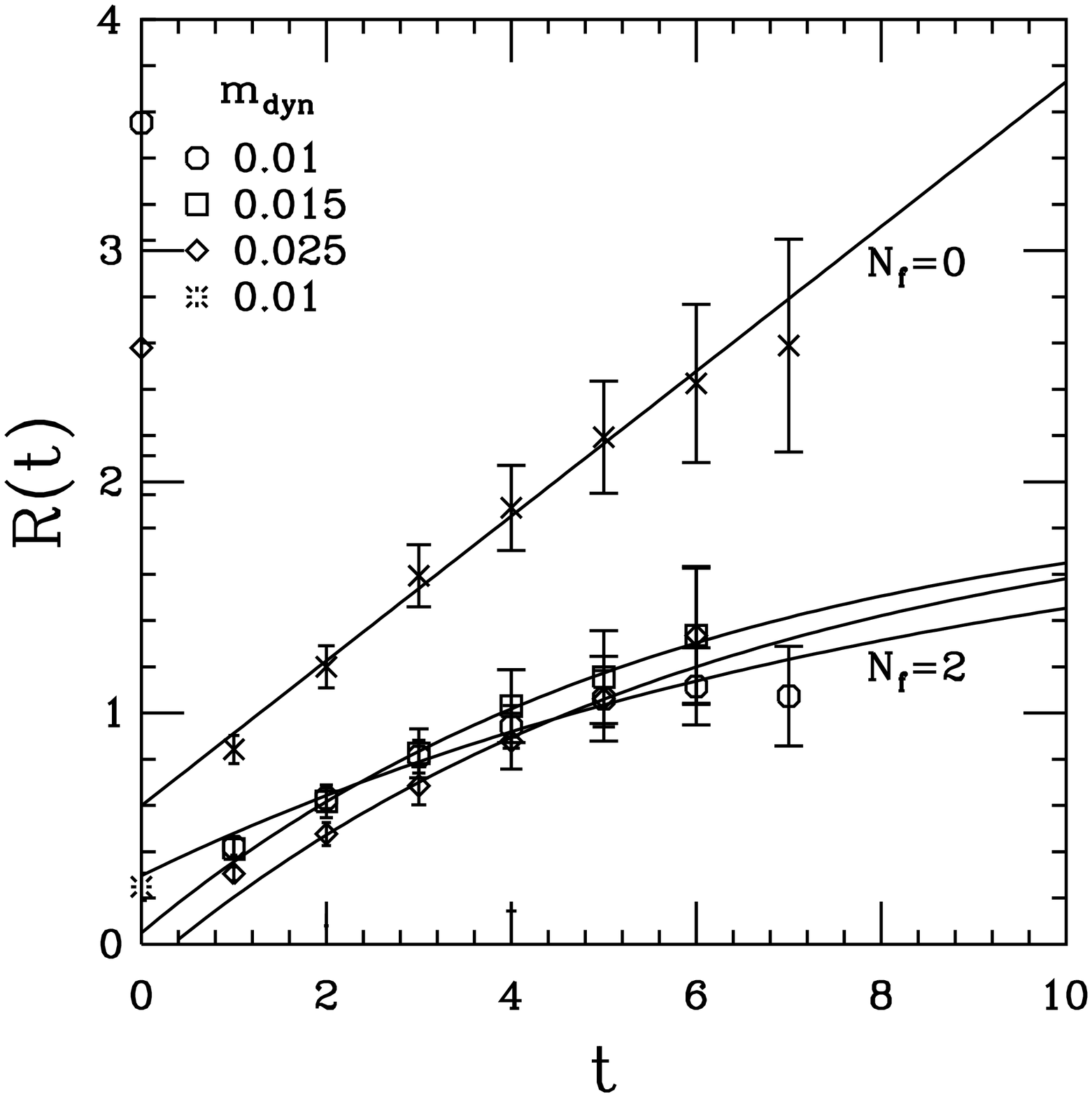}
\caption{\label{fig_kilcup_rdc} {\it The ratio $R_{DC}$ for quenched
$(N_f=0)$  and full $(N_f=2)$ QCD, as found by ref.\cite{kilcup}.
The solid lines correspond to fits according to eq. (\ref{eq_rdc_full}) and
eq. (\ref{eq_rdc_quen}).}
}
\end{center}
\end{figure}

In fig.\ref{fig_kilcup_rdc} we display the ratio $R_{DC}$ both for the
quenched and full QCD simulations. Obviously the expected $t$
dependence of $R_{DC}$, c.f. eq's. (\ref{eq_rdc_full}) and
(\ref{eq_rdc_quen}), is reasonably well satisfied by the data.

An extrapolation\footnote{The authors of ref.\cite{kilcup} 
assume a linear functional behavior of $m_0^2(m_q)$, in contrast to the 
finding of ref. \cite{japan_eta}, where $m_0$ (not $m_0^2$) increases
linearly with with decreasing $m_q$. Unfortunately, the systematic 
uncertainty arising from this difference has not been taken into
account.}
of the fit results at the various quark mass values
to the chiral limit yields $m_{\eta'} = 780(187)$MeV for $n_f=2$ full QCD,
and $m_{\eta'} = 891(101)$MeV in the quenched approximation. As in
ref. \cite{japan_eta}, these values are obtained by correcting  for the case of
3 quark flavors. Within large statistical errors both results,
quenched and unquenched, are consistent with experiment. 
This is similar to the situation found in lattice calculations
of the axial flavor singlet coupling $G_A^1$. It might
indicate, within the given level of accuracy, that the vacuum
contribution to $m_{\eta'}$ is determined largely  by gluonic properties,
rather than by fermion loops.\\

We mention that this interpretation is supported by the findings of 
lattice calculations using the `bermion' approach to
QCD\cite{bermion_app}. In this method one simulates QCD at negative
numbers of flavor, $n_f \leq 0$, where fermions are replaced by
`bermions', i.e. bosons with a fermion action, and then extrapolates
the results to $n_f > 0$. The authors of ref. \cite{bermion_eta} 
have applied this approach to the calculation of $m_0$. As a result 
they find no significant dependence of $m_0$ on $n_f$ in the
range $-8 \leq n_f \leq 0$.\\   
  
\begin{figure}[htb]
\begin{center}
\vskip -3.0cm
\noindent\parbox{15.0cm}{
\parbox{7.0cm}{\epsfxsize=7.0cm\epsfbox{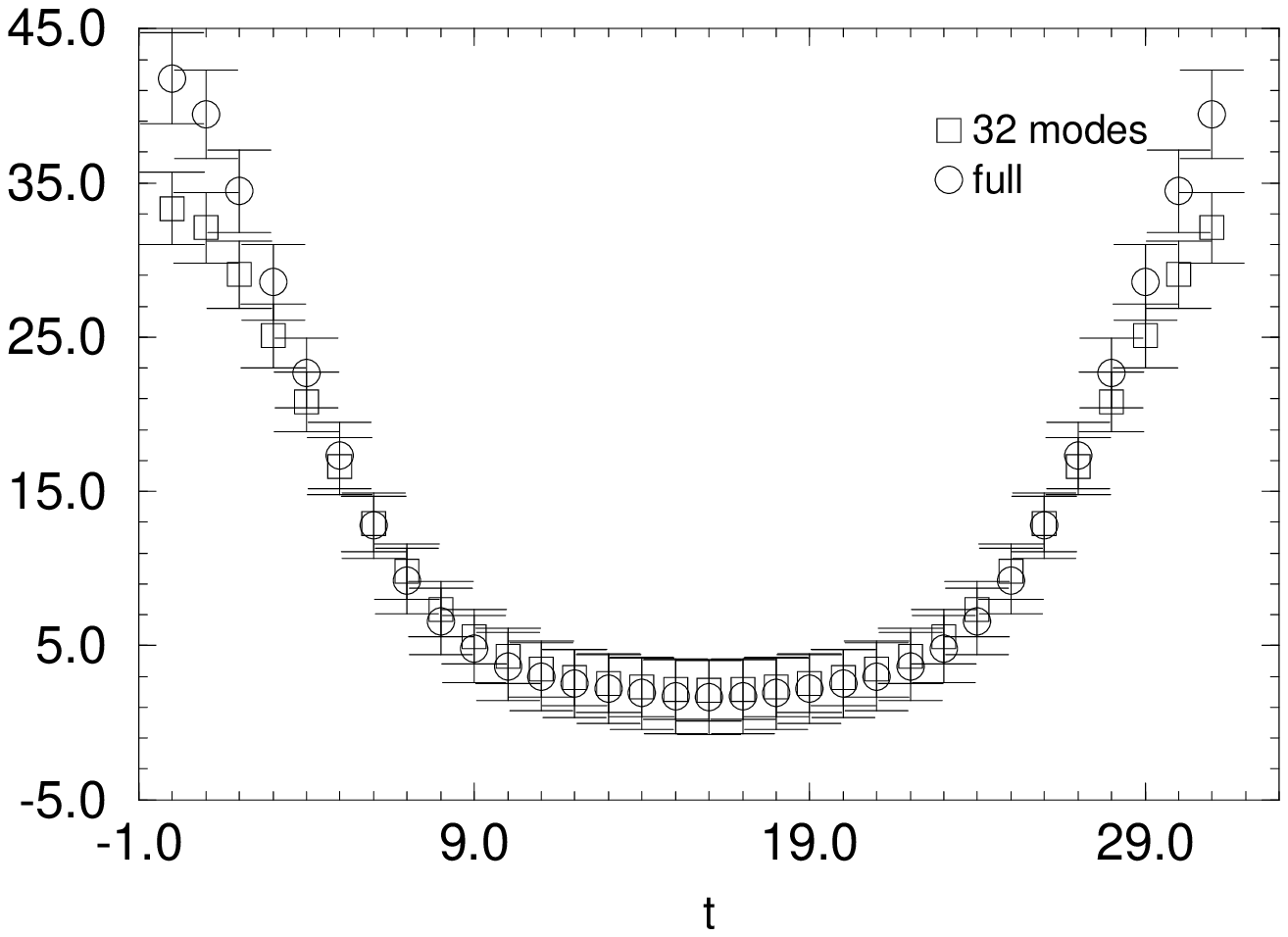}}
\parbox{7.55cm}{\epsfxsize=7.55cm\epsfbox{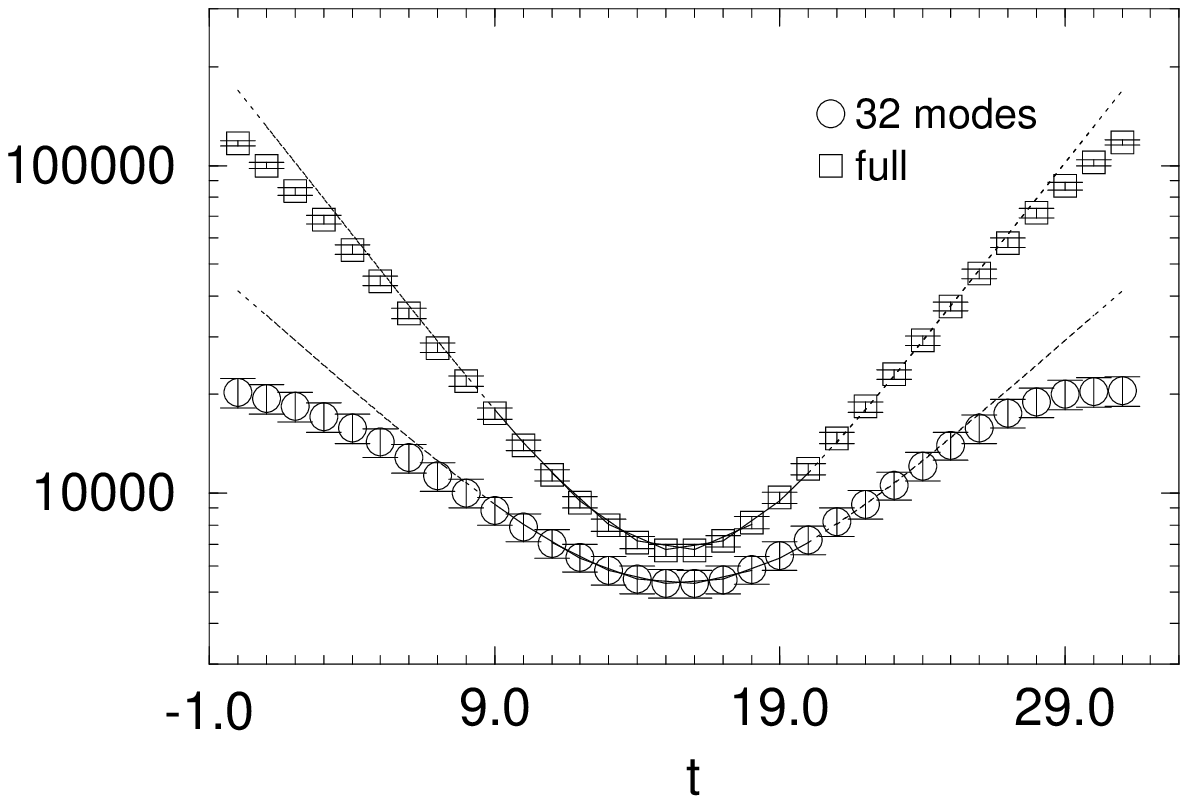}}
}
\caption{\label{fig_kilcup_eigenmodes} {\it Disconnected (left) and
connected (right) contributions to the $\eta'$ propagator 
as found by ref. \cite{kilcup} in a full QCD ($n_f=2$) simulation with
Kogut-Susskind fermions at a (lattice) quark mass $m_q a =0.01$. The
latter corresponds to $m_{\pi}/m_{\rho} = 0.572$. 
}}
\end{center}
\end{figure}
Finally, we comment on the numerical evidence for the
connection between topological vacuum excitations and the $\eta'$ mass
from a calculation of the low lying eigenmodes of the Dirac operator.
According to the approximate validity of the Atiyah-Singer theorem on
the lattice, the lowest eigenmodes reflect the topological
properties of the vacuum. Thus, if the disconnected part of the
$\eta'$ propagator in dominated by the latter, it should be possible
to compose it exclusively from these contributions. The connected
part, which represents the `normal' Goldstone mode,
should however not be obtainable from the lowest eigenmodes alone.

The authors of ref. \cite{kilcup} have compared the connected
and disconnected parts obtained from a full calculation with those
constructed from the lowest 32 eigenmodes. We display
their result in fig.\ref{fig_kilcup_eigenmodes}. Obviously the (full)
disconnected contribution is well represented by the eigenmodes, in
contrast to the connected part.

\subsection{Outlook}

The final goal to confirm  quantitatively that the `topological' solution
to the $U(1)$ problem is realized in QCD has not been reached yet.
There is however a lot of evidence that this solution is most likely.
All results achieved so far with lattice methods point in this
direction. 

From quenched and unquenched lattice simulations it appears,
within the given level of uncertainty, that the mass of the $\eta'$
meson calculated in QCD is comparable to it's experimental value. The
fact that both methods yield similar results may indicate that the
topological contributions are dominated by the gluonic properties of QCD. 

On top of this, the connection to topological vacuum excitations has been
established in quenched QCD by studying the dependence of $m_0$ on the
topological charge $Q$, and in full QCD by a comparison between the
$\eta'$ propagator calculated with all eigenmodes and the one composed
from only the low lying states.   \\

To consolidate these findings one has to perform lattice calculations
at high statistics and by application of (possibly improved)
stochastic estimator techniques. This, as a first step, would enable
for a reliable estimate of the statistical error. The next step is
then to vary lattice spacing and volume in order to perform a
continuum extrapolation of the lattice results. If this is done within
the framework of a `conventional' full QCD simulation with 2 flavors
and degenerate quark masses one could extract the QCD result for the
$\eta'$ mass under the assumption of flavor symmetry with respect to
$m_0$ and by neglecting $\eta-\eta'$ mixing effects. \\

In view of the
fact that the results discussed above have been achieved on
computers with a speed of less or about $10$Gflops this aim is
certainly within the range of upcoming TERAflops machines.
To include the
effects of flavor symmetry breaking and to calculate reliably the
influence of $\eta-\eta'$ mixing effects one would have to repeat the whole
procedure with $n_f \geq 3$ dynamical quark flavors and
non-degenerate masses. Without significantly improved stochastic
algorithms, such a program is currently beyond reach. It should
however be feasible to perform such a calculation with $n_f=2$ light
and $n_f=2$ heavy flavors at a fixed lattice spacing. This would help
to estimate the size of these effects.

Clearly, high statistics samples of QCD vacuum configurations will
allow for a more detailed study of the connection between topology
and $m_{\eta'}$. Especially the role of instantons\footnote{For a
recent review on instantons on the lattice see \cite{negele}.},
which we did not discuss here in detail, could be clarified.

\newpage   

\section{Final Remarks}

We said in the introduction that a major  goal in studying flavor
singlet phenomena is to learn about the properties  of the QCD
vacuum. So, what can we learn from the results of lattice QCD
achieved so far in this context?
The answer to this question embraces several aspects.

First of all, we have seen that disconnected (vacuum) amplitudes
indeed yield substantial contributions to flavor singlet processes.
It turns out that they are suited in sign and order of 
magnitude to explain `anomalous' phenomena like the small
value of $G_A^1$, the large mass of the $\eta'$ meson and the high value,
compared to the octet estimate $\sigma_0$, of $\sigma_{\pi N}$.

Secondly, the connection between topological excitations of the vacuum
and the $\eta'$ mass could be established. This is evidence, that the
$U(1)$ problem can be solved by the 't Hooft mechanism.  
Similarly, one finds that the influence of the axial anomaly, which
is (mostly) reflected in the size of the disconnected axial vector
insertion of the proton, lowers the value of $G_A^1$. 

Thirdly, comparing quenched and full QCD lattice results, one can 
estimate the effect of sea quarks on flavor singlet amplitudes.
They seem to have some influence on the scalar insertions,
e.g. on $\sigma_{\pi N}$ and $y$, as they lower the ratio of disconnected
to connected amplitudes. Axial vector and pseudo scalar insertions,
which are connected to the axial anomaly, do not seem to depend
largely on the presence of sea quarks.

Fourthly, flavor symmetry breaking effects in the disconnected
(vacuum) contributions appear to be small. This would explain 
the success of lattice determinations of flavor non-singlet 
quantities like $G_A^3$.\\
  
Unfortunately most of the findings mentioned here are still 
semi-quantitative results.
At the current level of statistical and systematical
reliability, the estimates of flavor singlet quantities from
lattice calculations cannot be used yet for a precise quantitative
comparison between experiment and theoretical QCD predictions.

Clearly, one would like to have an idea of what it would take
to reduce the uncertainties of the lattice results to the level
of about 20$\%$, which corresponds the accuracy currently achieved in
experimental measurements of $G_A^{1}$ and $\sigma_{\pi N}$. 

The various sources of uncertainty have been discussed in some detail
in this review. It can be seen from this discussion that the
error is dominated by the large statistical fluctuations of the 
disconnected amplitudes. These amount to 50 to $60\%$. To reduce
the fluctuations to the experimental level of $20\%$ just by `brute
force' one would need a factor of 9 in computer speed. 

The second largest error is due to cutoff effects and, closely
connected to this, uncertainties in the determination of
renormalization constants. From the lattice results of $G_A^3$
one estimates a $20$ to $30\%$ contribution to the error. Thus,
a rough finite cutoff analysis would be sufficient. Current lattice
calculations of flavor singlet quantities have been performed at 
a cutoff of $a^{-1} \simeq 2$GeV. An increase of $a^{-1}$ by a 
factor of 1.5  ($a^{-1} \simeq 2$GeV) would require an increase
of the number of lattice points by a factor of $1.5^4 \simeq 5$. On top
of this one has to take into account that it becomes more expensive
to generate statistically independent gauge configurations when going
closer to the continuum limit. For full QCD simulations, where the
generation of gauge configurations consumes the largest fraction of 
computer time, the dependence of the auto correlation length on the
cutoff is not well known. An additional factor of 2 in computer time
should be sufficient however for a moderate increase of $a^{-1}$.            

Thus, one estimates that, without any further algorithmic improvement,
a factor of $9\times 5 \times 2 = 90$ in computer power would be sufficient
to reduce the uncertainties on flavor singlet calculations to the
$20\%$ level.

According to the fact that the currently most advanced lattice 
calculations in this field, i.e. full QCD with $n_f=2$ dynamical
Wilson fermions, have been
performed on (APE) computers with a  sustained speed of 10 GigaFlops, one
would need  machines with a 1 TeraFlops performance to achieve this goal.      

Since computers with a speed of a few hundred Gigaflops are 
already available \cite{cp_pacs,qcdsp}, we would like to close
this review with the optimistic view that a $20\%$ uncertainty in full
QCD flavor singlet calculations can be realized within the next 5 
years. Quenched simulations with that accuracy could be done right now.

\paragraph{Acknowledgments}

I would like to thank Klaus Schilling, Apoorva Patel, Thomas Lippert,
and  Peer Ueberholz for many instructive and enlightening discussions
on the subject of flavor singlet processes.
I am also indepted to the Fachbereich Physik of the University
of Wuppertal which made it possible to complete this review.

\end{document}